\documentclass[%
 reprint,
superscriptaddress,
 amsmath,amssymb,
 aps,longbibliography,
]{revtex4-1}

\usepackage[colorlinks=true,linkcolor=blue,urlcolor=blue,citecolor=blue]{hyperref}
\usepackage{overpic}
\usepackage{physics}
\usepackage{graphicx}
\usepackage{dcolumn}
\usepackage{bm}
\usepackage{outlines}
\usepackage[dvipsnames]{xcolor}
\usepackage{upgreek, textcomp, gensymb}

\definecolor{matlabblue}{rgb}{0,0.4470,0.7410}
\definecolor{matlabred}{rgb}{0.6350,0.0780,0.1840}
\definecolor{matlabpurple}{rgb}{0.4940,0.1840,0.5560}
\definecolor{matlaborange}{rgb}{0.8500,0.3250,0.0980}

\definecolor{matlab1}{HTML}{0072BD}
\definecolor{matlab2}{HTML}{D95319}
\definecolor{matlab3}{HTML}{EDB120}
\definecolor{matlab4}{HTML}{7E2F8E}
\definecolor{matlab5}{HTML}{77AC30}
\definecolor{matlab6}{HTML}{4DBEEE}
\definecolor{matlab7}{HTML}{A2142F}

\definecolor{refA}{rgb}{0, 0, 0}
\definecolor{refB}{rgb}{0, 0, 0}
\definecolor{refAB}{rgb}{0, 0, 0}
\definecolor{otherRev}{rgb}{0, 0, 0}
\newcommand{\rthz}{$\sqrt{\text{Hz}}$}
\newcommand{\plotline}{\rule[0.75mm]{3mm}{.25mm}}
\newcommand{\plotdashline}{\rule[0.75mm]{1.1mm}{.25mm}~\rule[0.75mm]{1.1mm}{.35mm}}
\newcommand{\plotdotdashline}{\rule[0.75mm]{1.1mm}{.25mm}~\rule[0.75mm]{.2mm}{.25mm}~\rule[0.75mm]{1.1mm}{.25mm}}

\begin{document}

\title{Vector magnetometry using cavity-enhanced microwave readout in nitrogen-vacancy diamond}

\author{Reginald Wilcox}
\affiliation{Massachusetts Institute of Technology, Cambridge, MA 02139, USA}
\affiliation{MIT Lincoln Laboratory, Lexington, MA 02421, USA}
\author{David Phillips}
\affiliation{MIT Lincoln Laboratory, Lexington, MA 02421, USA}
\author{Matthew Steinecker}
\affiliation{MIT Lincoln Laboratory, Lexington, MA 02421, USA}
\author{Erik Eisenach}%
\affiliation{Massachusetts Institute of Technology, Cambridge, MA 02139, USA}
\affiliation{MIT Lincoln Laboratory, Lexington, MA 02421, USA}
\author{Corey Hawkins}
\affiliation{MIT Lincoln Laboratory, Lexington, MA 02421, USA}
\author{Linh Pham}
\affiliation{MIT Lincoln Laboratory, Lexington, MA 02421, USA}
\author{Jennifer Schloss}
\affiliation{MIT Lincoln Laboratory, Lexington, MA 02421, USA}
\author{Dirk Englund}
\affiliation{Massachusetts Institute of Technology, Cambridge, MA 02139, USA}
\author{Danielle Braje}
\affiliation{MIT Lincoln Laboratory, Lexington, MA 02421, USA}

\date{\today}

\begin{abstract}
We demonstrate $4\pi$-steradian vector magnetic field sensing using an ensemble of nitrogen-vacancy (NV) centers in a single-crystal diamond coupled to a microwave (MW) cavity.
The MW cavity enhances the spin-photon coupling which enables efficient, high-contrast spin-state readout via MW interrogation and removes the need for bulky optical collection components.
An applied AC bias magnetic field lifts the zero-field degeneracy of the four crystallographic NV orientations, allowing each orientation to be individually addressed and used for vector reconstruction of the magnetic field.
The resulting magnetometer has a 40\% contrast (20x higher than typical for optical spin-ensemble readout) and achieves a single-axis sensitivity of 250~pT/\rthz{} which is flat from DC to 1~kHz.
Noise models of the composite spin-cavity system establish MW amplitude noise as the dominant noise source and predict a thermal noise limit of 2~pT/\rthz{}.


\vspace{5mm}
\end{abstract}


\clearpage

\maketitle

\section{Introduction}
Over the past decade, solid-state spin systems have undergone a period of rapid research and development, progressing from novel physics demonstrations to compelling quantum sensors \cite{karadas_feasibility_2018, sturner_integrated_2021, barry_optical_2016, davis_mapping_2018, arai_millimetre-scale_2022, webb_detection_2021, fescenko_diamond_2019, shi_single-protein_2015, zhou_imaging_2021, turner_magnetic_2020, patel_subnanotesla_2020, bertelli_magnetic_2020, lenz_imaging_2021, jenkins_imaging_2020, hsieh_imaging_2019, lovchinsky_nuclear_2016}, driven in part by advances in materials engineering \cite{achard_chemical_2020,tallaire_high_2020, alsid_photoluminescence_2019, ashfold_nitrogen_2020, bauch_decoherence_2020, wolfowicz_quantum_2021, edmonds_characterisation_2021, watanabe_shallow_2021, bluvstein_identifying_2019} and coherent control \cite{barry_sensitive_2024, aslam_nanoscale_2017, waeber_pulse_2019, okeeffe_hamiltonian_2019, glenn_high-resolution_2018, bauch_ultralong_2018, smits_two-dimensional_2019, childress_coherent_2006, bar-gill_solid-state_2013, dreau_avoiding_2011, hart_n-vdiamond_2021, pham_enhanced_2012, dwyer_probing_2022}.
Nitrogen-vacancy (NV) centers in diamond have become the most extensively studied solid-state defect for sensing applications \cite{barry_sensitive_2024} and are promising candidates for high-sensitivity room-temperature magnetometry owing to their exceptionally long coherence times \cite{jarmola_temperature-_2012, rosskopf_investigation_2014, balasubramanian_ultralong_2009}.
Though spin-dependent fluorescence readout of the spin state has long been the dominant technique, its poor contrast and typically low collection efficiency have prompted the development of alternative readout modalities \cite{neumann_single-shot_2010, shields_efficient_2015, eisenach_cavity-enhanced_2021, ebel_dispersive_2021, bourgeois_photoelectric_2015, hopper_near-infrared-assisted_2016, hopper_spin_2018, niethammer_coherent_2019}.
Microwave (MW) cavity readout -- a newly developed, non-optical readout method -- offers promising sensitivities in diamond \cite{eisenach_cavity-enhanced_2021, ebel_dispersive_2021} and other solid-state defects \cite{wilcox_thermally_2022}.


Readout is accomplished by coupling the NV spins to a small dielectric resonator, eliminating bulky fluorescence collection and measurement components and enabling compact designs.
High readout contrast combined with efficient MW coupling structures produce competitive sensitivities rivaling state-of-the-art optical techniques \cite{wang_spin-refrigerated_2024, barry_sensitive_2024}.
But because the dielectric resonator frequency is fixed and confines the addressable spins to a narrow frequency band, existing vector sensing methods which resolve individual NV orientations by splitting the spin resonance frequencies with a bias field are not directly applicable \cite{schloss_simultaneous_2018}.

Full vector magnetometry allows complete reconstruction of all three cartesian components of the magnetic field, offering advantages over scalar and single-axis vector magnetometers in application spaces including magnetic navigation \cite{canciani_analysis_2020} and magnetocardiography \cite{su_vector_2024, yang_new_2021, budker_optical_2013}.
The face-centered cubic lattice of diamond confines the NV center axes to fixed orientations, providing a natural basis with which to make vector magnetic field measurements, thereby making NV diamond an ideal platform for vector magnetometry which is free from heading errors \cite{schloss_simultaneous_2018, clevenson_robust_2018, eisenach_vector_2022}.
By using an AC magnetic field to successively tune each NV orientation's resonance near the MW cavity resonance, this work extends the MW cavity readout technique from single-axis vector magnetometry to $4\pi$-steradian magnetic field sensing.

\begin{figure*}[t] 
\hspace{-2mm}
\begin{minipage}[b]{0.3\textwidth}
\begin{overpic}[width=2.1in]{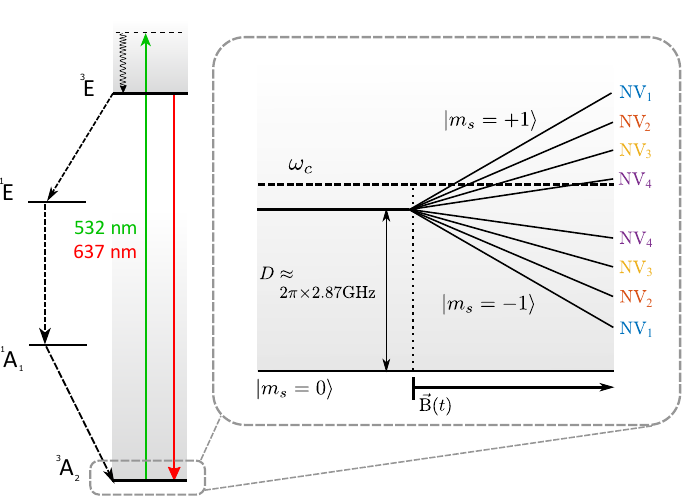} \put(0,75){\textbf{a)}}
\end{overpic}
\end{minipage}
\;
\begin{minipage}[b]{0.3\textwidth}
\begin{Overpic}{{\includegraphics[width=2.1in]{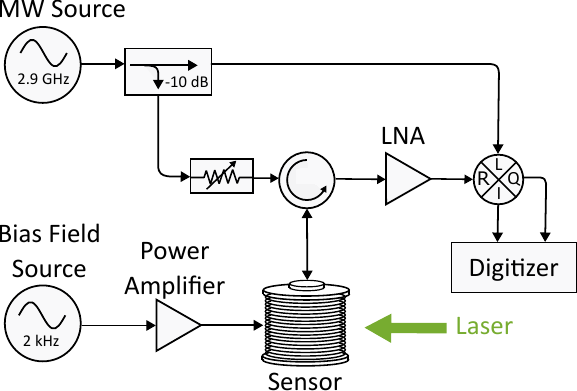}}} \put(0,75){\textbf{b)}} 
\end{Overpic}
\end{minipage}
\;
\begin{minipage}[b]{0.3\textwidth}
\begin{overpic}[width=1.9in]{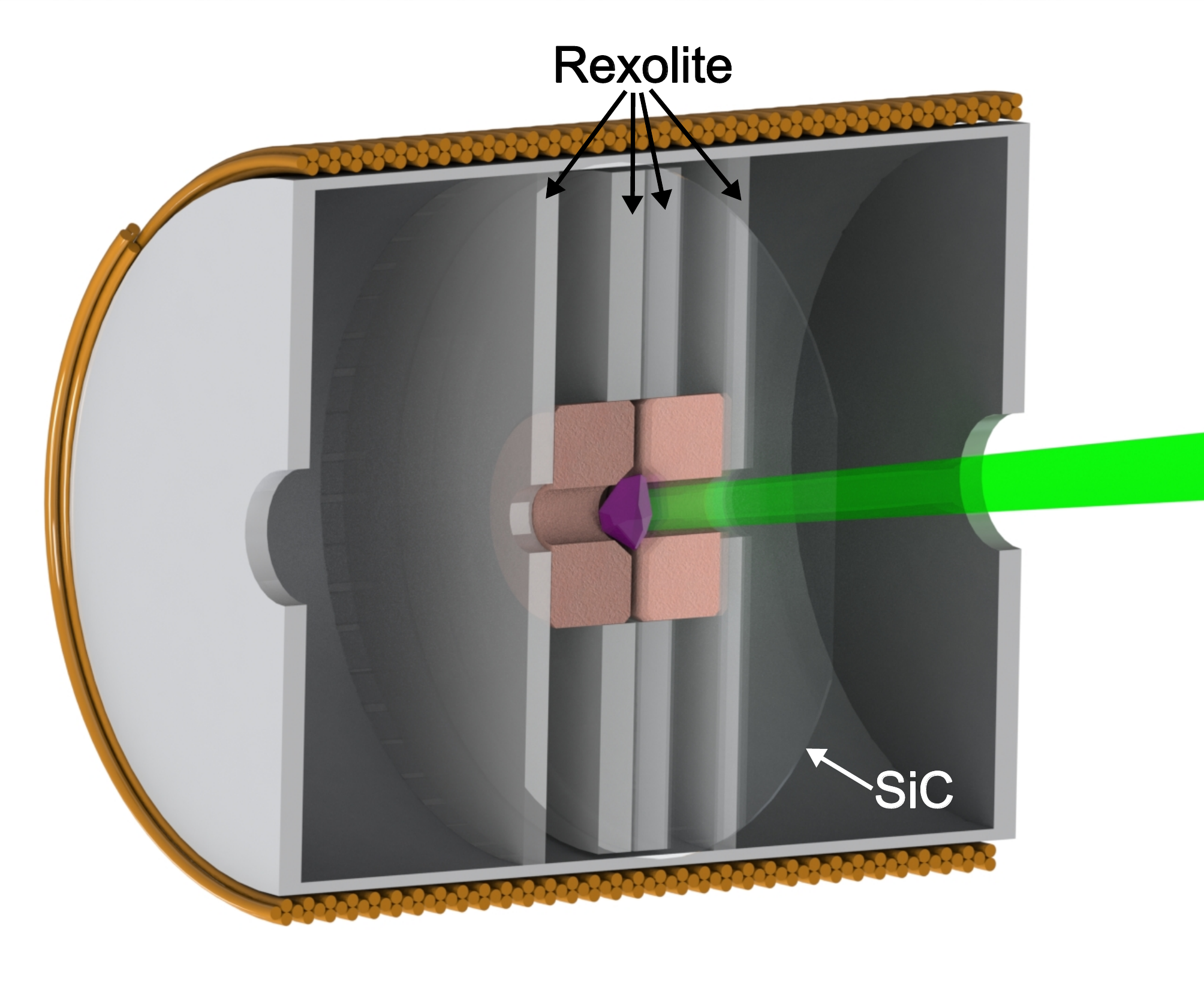} \put(-5,75){\textbf{c)}} 
\end{overpic}
\end{minipage}
\caption{\textbf{Cavity design and schematic.}
\textbf{(a) NV diamond energy levels.}
The NV diamond exhibits a $D = 2\pi\times2.87$~GHz zero-field-splitting for the $|m_s = 0\rangle\leftrightarrow|m_s=\pm 1\rangle$ spin transition.
Applied magnetic fields linearly shift this transition frequency in proportion to the projection of the applied field along the NV orientation.
The applied field shown here has highest projection along NV orientation 1 and the smallest projection along NV orientation 4.
\textbf{(b) Schematic of sensor.}
A 2~kHz bias field is delivered to the sensor using a signal generator and amplifier.
Microwaves from a signal generator are split using a directional coupler, with the through port being used to drive the local oscillator (LO) port of an IQ mixer.
A variable attenuator adjusts the power from the -10~dB arm of the coupler which is then delivered to the cavity through a circulator.
The reflected signal from the cavity is then amplified with a low-noise amplifier (LNA) before driving the RF port of an IQ mixer, whose in-phase (I) and quadrature (Q) ports are then digitized.
\textbf{(c) Rendering of cavity.}
The diamond (purple) is mounted on a SiC wafer (transparent) and suspended between two dielectric resonators (pink) with a center bore hole to allow for laser delivery (green).
The diamond resonator assembly is held in place inside a ceramic cylinder using Rexolite supports.
The interior of the ceramic cylinder is silver plated to prevent MW radiative losses and magnet wire is wrapped around the exterior to apply the bias magnetic field.}
\label{vcr:fig:cavity}
\end{figure*}

\section{NV Background}
Nitrogen-vacancy centers in diamond are defects in the diamond lattice that consist of substitutional nitrogen bound to an adjacent lattice vacancy \cite{barry_sensitivity_2020, doherty_nitrogen-vacancy_2013}.
An NV center's symmetry axis is confined to be along one of the diamond's [111] crystallographic directions, resulting in four unique NV axes \cite{barry_sensitivity_2020}.
Within the lattice, NV centers behave as isolated quantum systems, with their ground state forming a spin-1 system featuring long coherence times at room temperature \cite{jarmola_temperature-_2012, rosskopf_investigation_2014, balasubramanian_ultralong_2009}.
The $|m_s = 0\rangle \leftrightarrow |m_s = \pm 1\rangle$ transition in the NV ground state experiences a Zeeman shift due to the magnetic field.
Its transition frequency $\omega_{s,i,\pm}$ can be expressed to first order in $\gamma B/D$ as
\begin{equation}
\omega_{s, i, \pm} = D \mp \gamma \vec{B}\cdot{\hat{n}}_i,
\end{equation}
where $i = 1, 2, 3, 4$ indexes the NV orientation, $D = 2\pi\times2.87$~GHz is the NV diamond zero-field-splitting, $\gamma \approx 2\pi\times 28$~GHz/T is the electron gyromagnetic ratio, $\vec{B}$ is the applied magnetic field, and $\hat{n}_i$ is the symmetry axis of the NV orientation \cite{barry_sensitivity_2020} (see Fig.~\ref{vcr:fig:cavity}a).
The vector magnetic field can be reconstructed by independently measuring these spin transition frequencies \cite{schloss_simultaneous_2018}.

NV centers must be polarized into a known spin state to perform magnetometry \cite{barry_sensitivity_2020}.
This is done by applying 532~nm light which drives the NV center into an excited state which then preferentially decays into the $|m_s = 0\rangle$ ground state through a non-radiative inter-system crossing \cite{goldman_state-selective_2015, goldman_phonon-induced_2015}.
NV centers feature spin-state-dependent fluorescence, which enables most readout techniques for sensing applications \cite{barry_sensitivity_2020}.
However, in this work we employ an alternative readout scheme using MWs to probe the spin transition frequencies \cite{eisenach_cavity-enhanced_2021, wilcox_thermally_2022}, avoiding optical readout and its associated challenges, including bulky components and poor contrast.

\section{Theory of Operation}
\label{vcr:sec:setup}
\begin{figure*}
    \centering
    \includegraphics[width=5.5in]{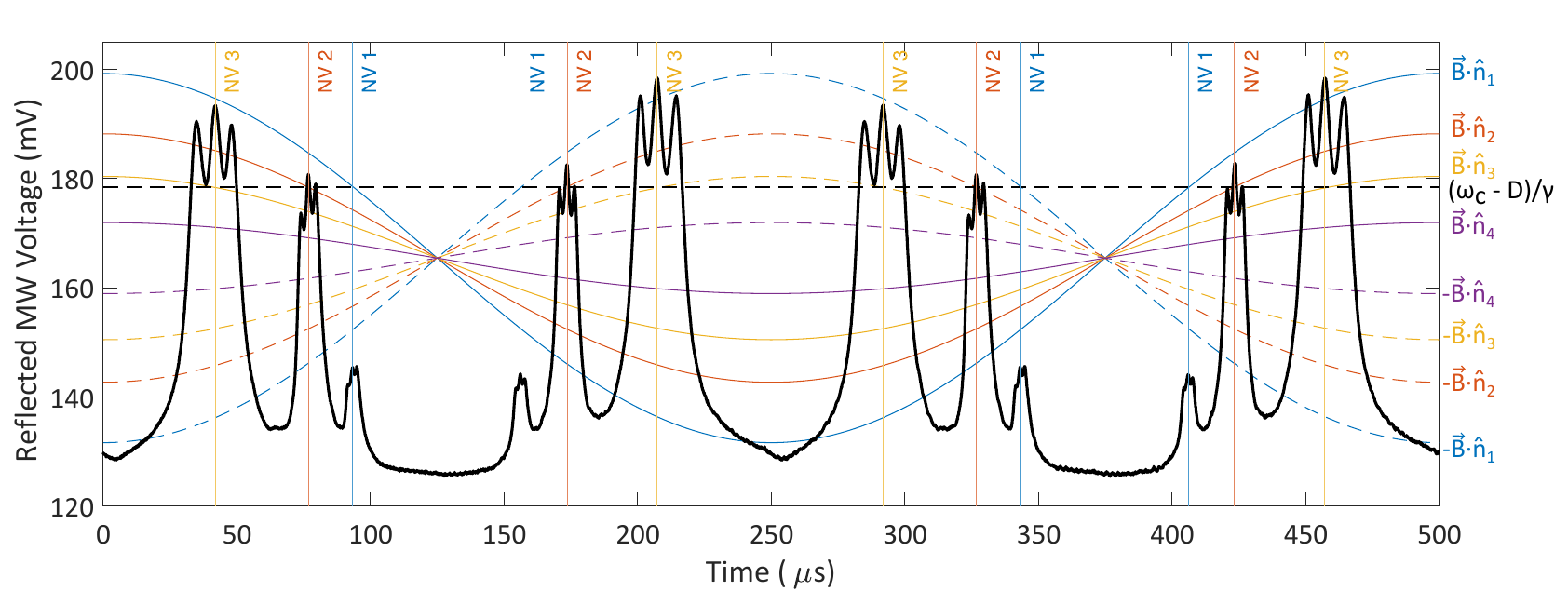}
    \caption{\textbf{Reflected signal from cavity.}
    The reflected signal from the MW cavity for a single cycle of AC bias field.
    The overlaid family of sinusoids correspond to the bias field projection onto each of the 4 NV axes.
    The dashed horizontal line marks $(\omega_c - D)/\gamma$, the required bias field projection to make the spin transition frequency resonant with the cavity.
    When the sinusoids intersect the dashed horizontal line, a peak in the reflected signal is observed, marked by the vertical lines.
    The peaks on either side of the vertical lines correspond to hyperfine splitting.
    The first half period and second half period are near perfect repetitions of each other.
    Asymmetry between the first and second quarter periods is attributed to time dynamics related to depolarization of the NV orientation as it is swept -- decreasing the sweep frequency or increasing laser power improves the symmetry.}
    \label{vcr:fig:timeseries}
\end{figure*}
Cavity readout of a quantum ensemble relies on the coupling between spins and a resonant cavity, which causes the spin system to change the cavity properties.
The cavity properties are then measured by a MW signal that probes the cavity.
Due to the spin-cavity coupling, information about the spin system becomes encoded in the MWs, allowing measurement of the spin system by monitoring the MW signal.

To perform MW cavity readout, a polarized spin system is placed within a high-Q dielectric resonator and a continuous-wave (CW) MW source is coupled into the resonator to probe the cavity resonance \cite{eisenach_cavity-enhanced_2021, wilcox_thermally_2022}.
A magnetic bias field lifts degeneracies between the $|m_s = \pm 1\rangle$ states and tunes the spin transition frequency, $\omega_{s, i, +}$, to $\omega_c$.
Setting the MW frequency $\omega_d$ to $\omega_c$ enhances the spin-photon coupling through the high Q of the dielectric resonator.
External magnetic fields then shift $\omega_{s,i,+}$, which produces a change in the reflected MW signal due to the cavity-enhanced spin-photon coupling.
By monitoring the reflected signal from the composite system, the external magnetic field can be determined.

In this work, an AC bias field projects unequally on all 4 NV axes, splitting the transition frequencies as shown in Fig.~\ref{vcr:fig:cavity}a and enabling vector sensing.
The time-varying bias field sweeps the NV orientations through the cavity resonance sequentially, producing peaks in the reflected signal at different times, as seen in Fig.~\ref{vcr:fig:timeseries}.
Within a single bias field period, an NV orientation is interrogated 4 times: twice via the $|m_s = 0\rangle \leftrightarrow |m_s = + 1\rangle$ transition and twice via the $|m_s = 0\rangle \leftrightarrow |m_s = - 1\rangle$ transition \footnote{A different quantization axis is chosen for each NV orientation along its NV axis. These quantization axes remain fixed and do not change even when the bias field changes direction.}.
The central peak for each NV orientation, marked with vertical lines, occurs when the bias field projection onto the NV axis shifts the spin transition frequency to be resonant with the dielectric resonator.
Hyperfine splitting  due to the nuclear spin of $^{14}$N shows distinct triplet structure.
When an external magnetic field is applied, the temporal locations of the resonances shift.
By measuring these changes, we can calculate the magnetic field projection onto each NV axis and reconstruct the applied magnetic field vector.

\section{Experimental Setup}
The experimental setup is comprised of two main parts: the sensor head and the readout.
The sensor head, which consists of a diamond mounted in a dielectric resonator, orients the bias field relative to the diamond, couples the MWs into the resonator, and provides thermal management.
Meanwhile, the readout consists of a homodyne architecture to measure the reflected signal from the sensor head.

A rendering of the sensor head is shown in Fig.~\ref{vcr:fig:cavity}c.
A brilliant-cut, cleaved NV diamond is adhered to a 300~\micro m-thick silicon carbide (SiC) wafer which provides thermal management for the laser heat load and mechanical support of the diamond within the cavity.
The diamond has an approximate volume of 6~mm$^3$ and an estimated laser-broadened spin dephasing time of $\kappa_s = 2\pi\times 2$~MHz at the laser power used in this experiment.
Two separate cylindrical, ceramic resonators with a center hole (Q~$\approx 20,000$, with relative dielectric $\epsilon_r \approx 34$) enclose the diamond, forming a composite dielectric resonator.
Threaded Rexolite fixtures hold the resonators, allowing for $\omega_c$ to be adjusted by varying the spacing between the resonators.
A patch antenna (not pictured) is placed near the composite resonator to deliver MWs to the diamond, also held by a threaded Rexolite fixture which allows for tuning the coupling rate $\kappa_{c1}$ by varying its distance from resonators.

The resonator assembly is placed inside an alumina ceramic shield with 50.8~mm inner diameter.
To provide the AC bias field, 151 turns of 20-gauge magnet wire are wound around the ceramic shield and made resonant at $\omega_m = 2\pi\times 2$~kHz with a $Q = 8$ matching circuit.
The coil is driven with $\approx 1.25$~A$_{\text{rms}}$ of current, resulting in an applied bias field of $\approx 25$~G$_{\text{rms}}$, sufficient to sweep three of the four NV orientations through the cavity resonance.
Full vector magnetometry necessitates three NV orientations, providing sufficient information to fully reconstruct the magnetic field.

A 532~nm CW laser is used to polarize the NV centers into the $|m_s = 0\rangle$ spin state.
Data are taken with 2.1~W of laser power, limited by thermal dissipation in the sensor head.

\begin{figure*}[t]
\begin{tabular}{ccc}
a) NV Orientation 1 & & d) $x$-direction magnetic field \\
\includegraphics[width=3.2in]{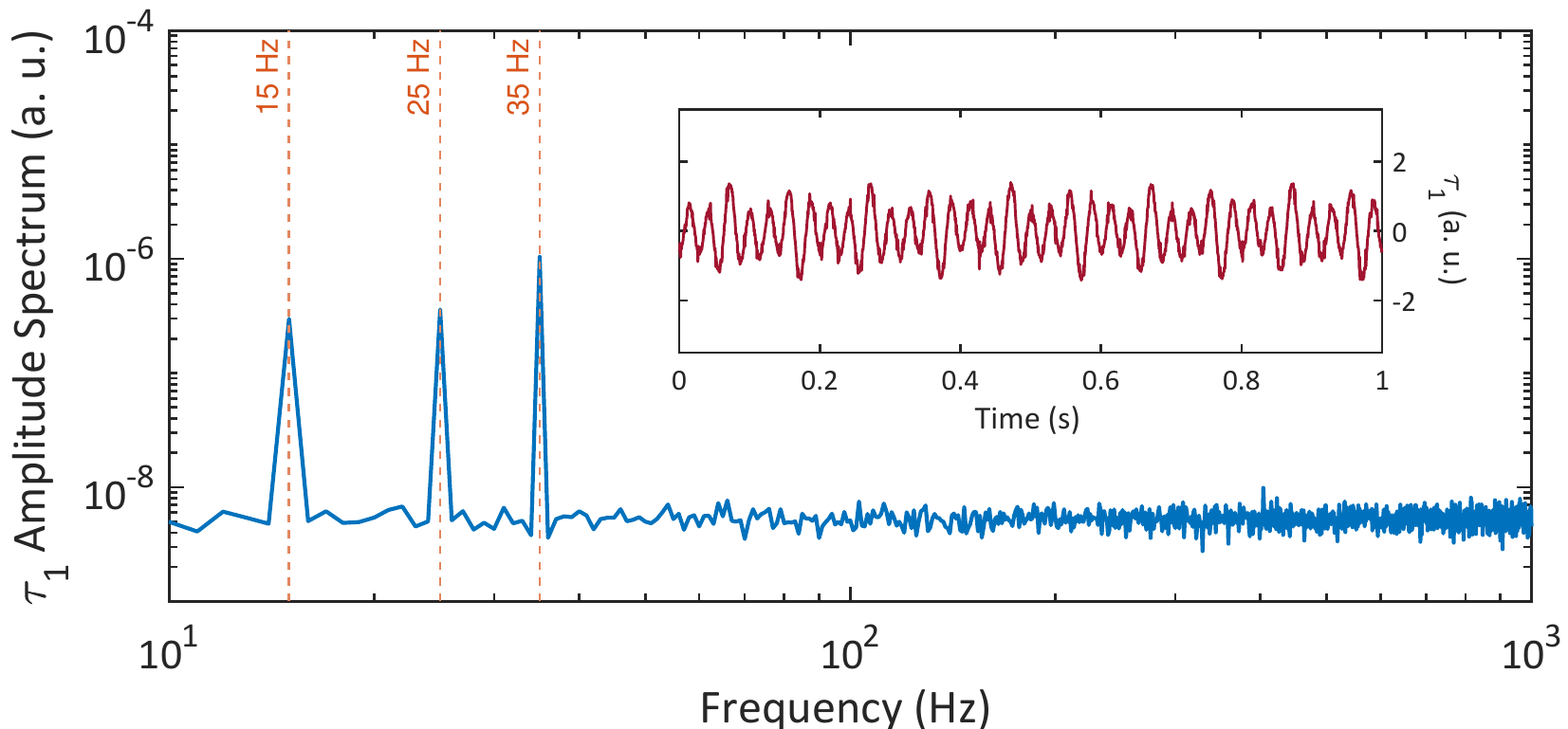} & \hspace{8pt} & \includegraphics[width=3.2in]{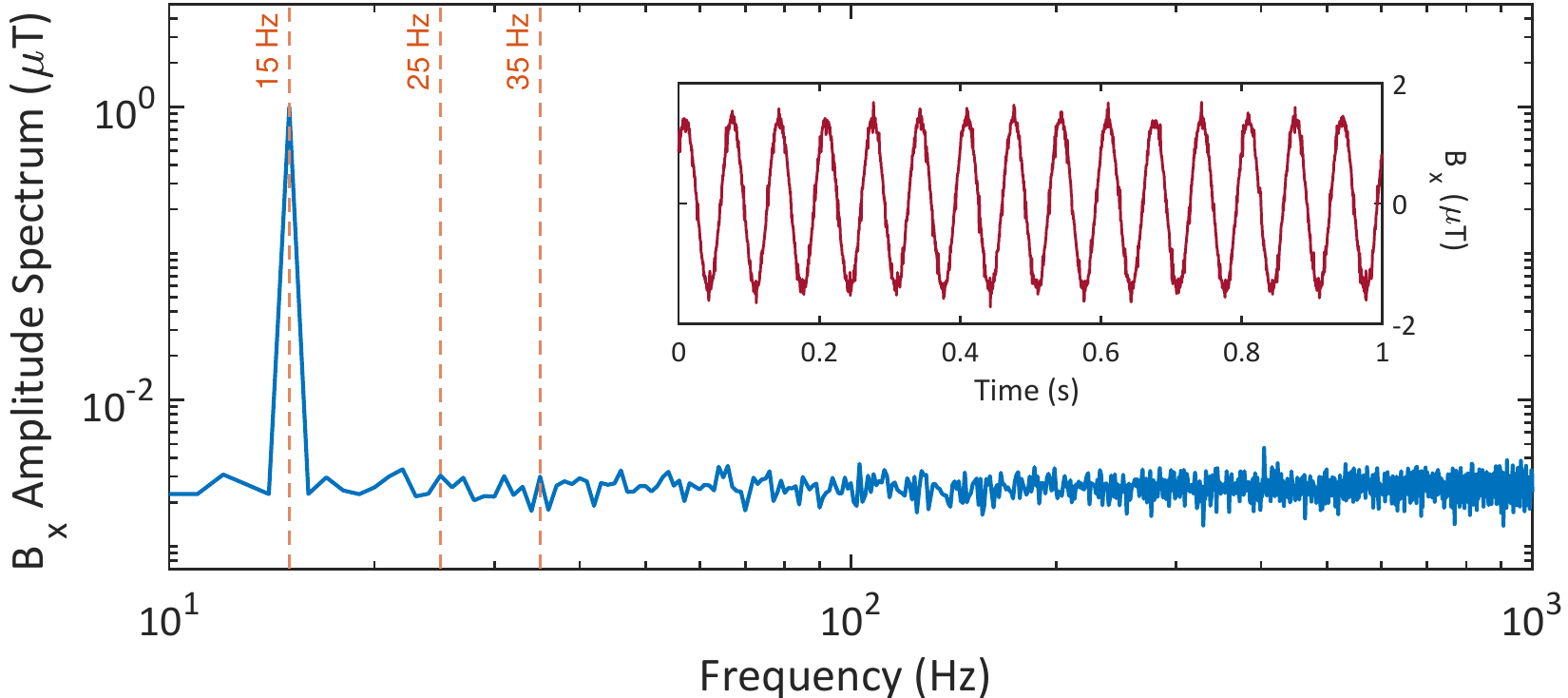} \\
\\
b) NV Orientation 2 & & e) $y$-direction magnetic field \\ 
\includegraphics[width=3.2in]{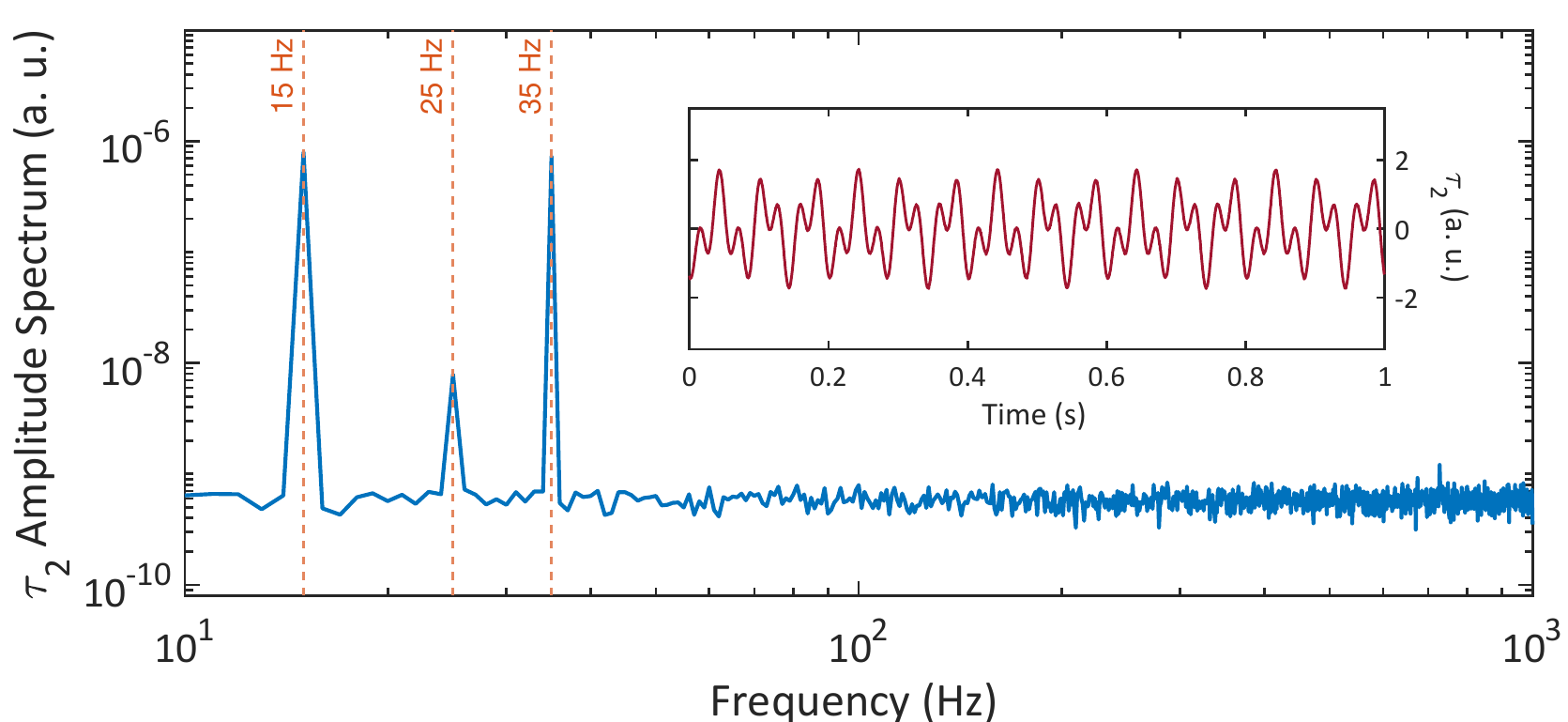} &   & \includegraphics[width=3.2in]{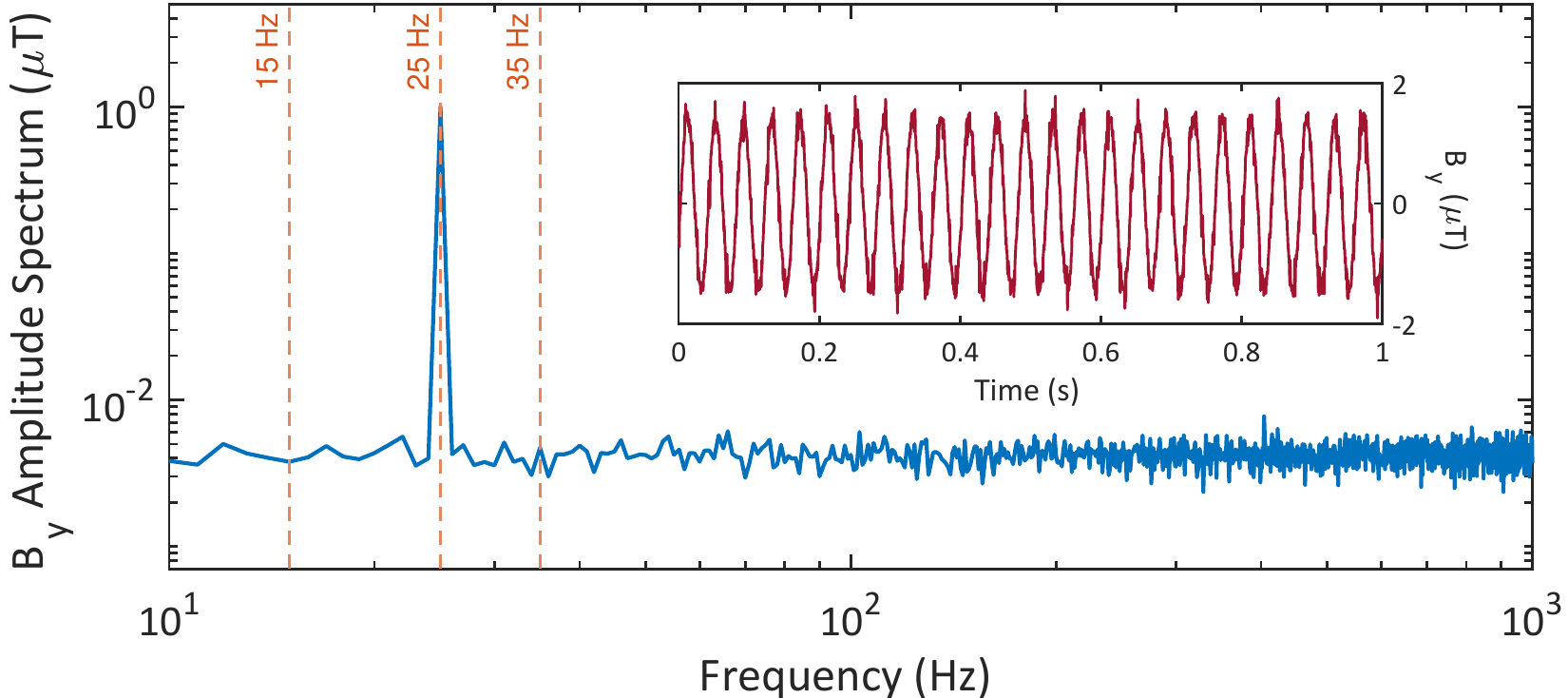} \\
\\
c) NV Orientation 3 & & f) $z$-direction magnetic field \\
\includegraphics[width=3.2in]{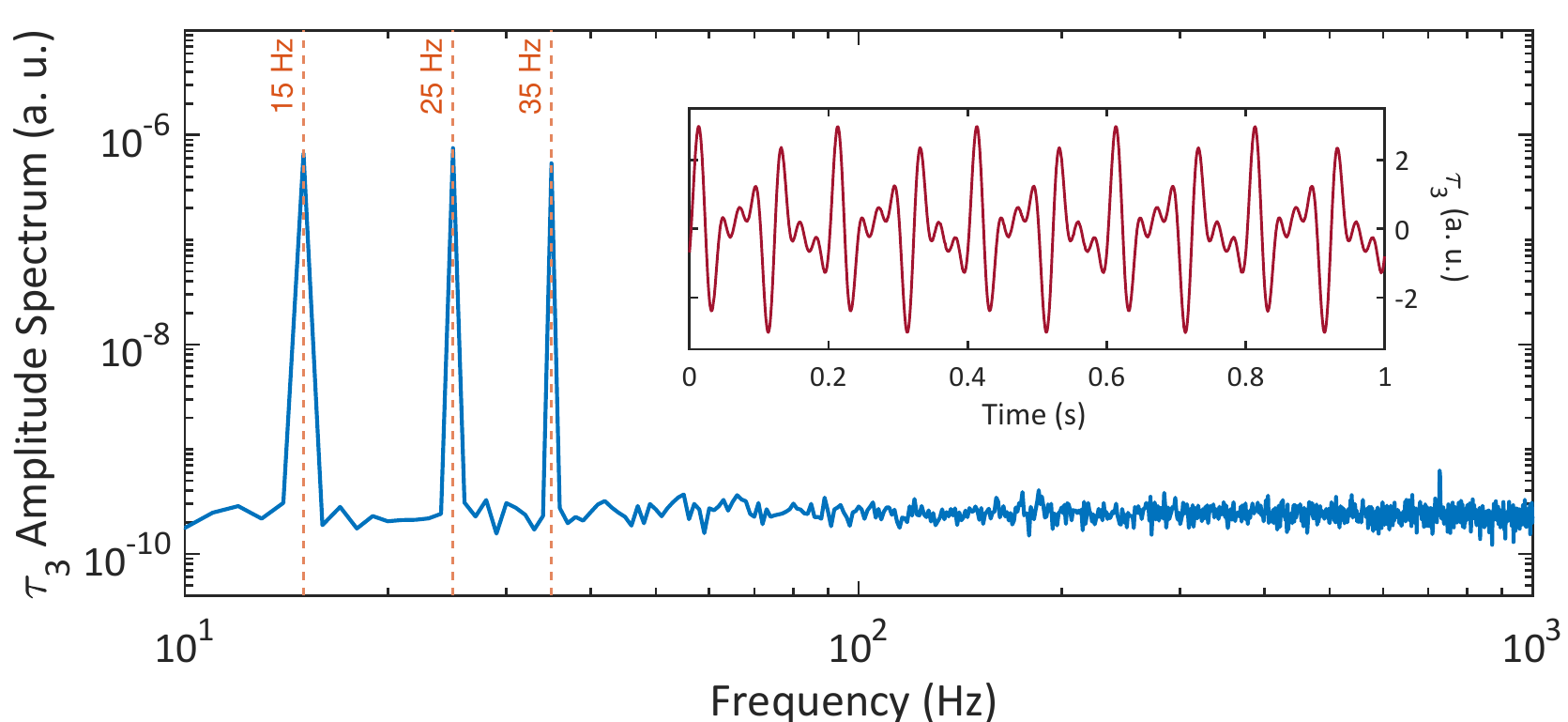} &   & \includegraphics[width=3.2in]{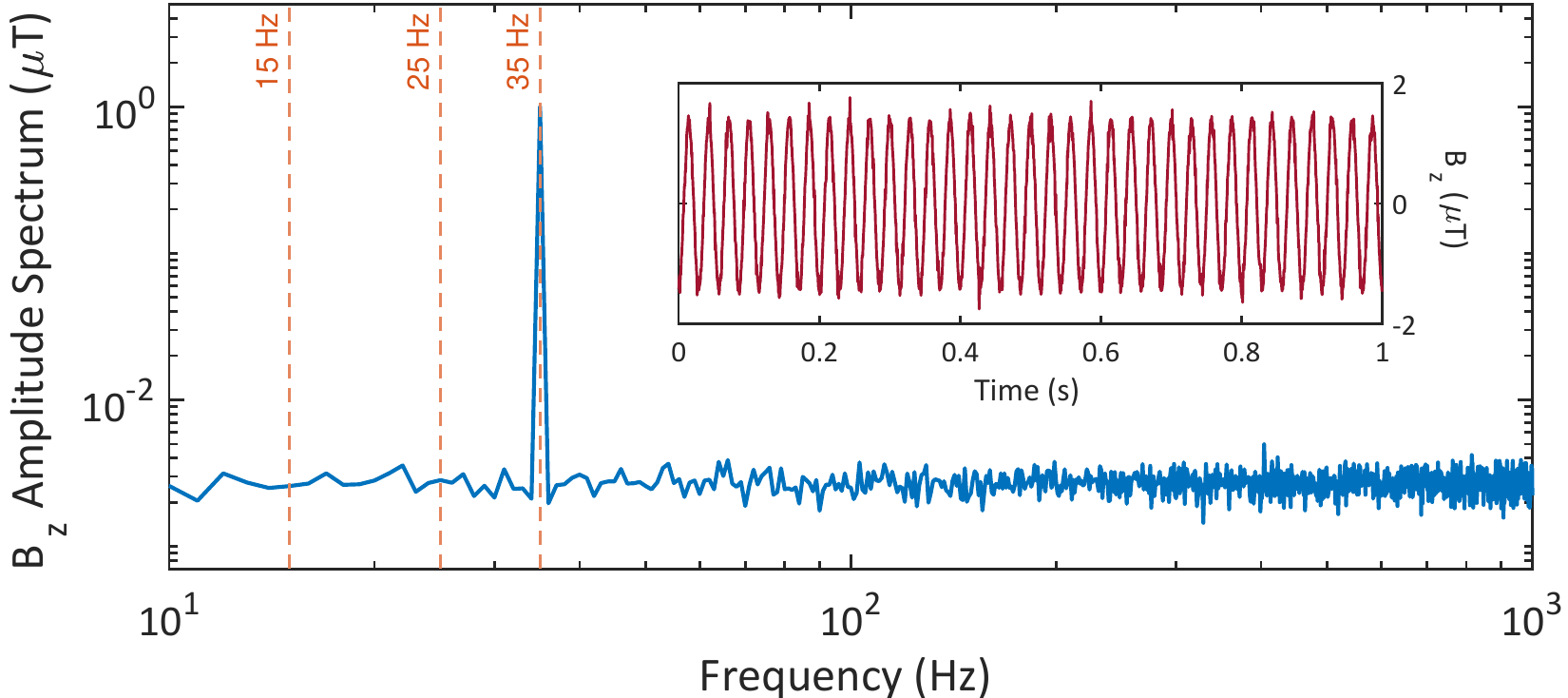} \\
\end{tabular}
\caption{\textbf{Full vector magnetometry demonstration.}
\textbf{(a)-(c) NV orientation response to vector fields.}
For each NV orientation, the inset shows $\tau_i$ (\textcolor{matlab7}{\rule[0.75mm]{3mm}{.25mm}}), with its amplitude spectrum displayed on the main plot (\textcolor{matlab1}{\rule[0.75mm]{3mm}{.25mm}}).
Three orthogonal magnetic test field signals at distinct frequencies (15~Hz, 25~Hz, and 35~Hz in the $x$-, $y$-, and $z$-directions respectively) are applied to the magnetometer, with the frequencies marked by a vertical dashed lines (\textcolor{matlab2}{\rule[0.75mm]{1.1mm}{.25mm}~\rule[0.75mm]{1.1mm}{.35mm}}).
Multiple signals are present on each $\tau_i$ spectrum as the NV axes are not orthogonal and do not coincide with the cartesian basis.
\textbf{(d)-(e) Vector reconstruction of fields.} Using the calibration, the $x$-, $y$-, and $z$-components of the magnetic field are determined from $\tau_i$ and plotted in the inset (\textcolor{matlab7}{\rule[0.75mm]{3mm}{.25mm}}), with their corresponding amplitude spectra in the main plot (\textcolor{matlab1}{\rule[0.75mm]{3mm}{.25mm}}).
The three orthogonal signals which appeared in multiple $\tau_i$ appear only along a single cartesian direction.}
\label{vcr:fig:vector}
\end{figure*}

Sensor readout, illustrated in Fig.~\ref{vcr:fig:cavity}b, is designed to measure the MW reflection from the sensor head.
A signal generator produces the MW signal at frequency $\omega_d$ which is split into a signal and reference arm using a directional coupler.
The reference arm  is used to drive the local oscillator (LO) port of an IQ mixer while the signal arm probes the sensor head.
A circulator in the signal arm isolates the incident MWs from the reflected MWs, delivering the reflected signal to the radio frequency (RF) port of the mixer.
The resulting in-phase (I) and quadrature (Q) outputs of the mixer are then digitized for analysis.
The resulting signal has a contrast of 40\% which compares favorably to typical spin-ensemble readout contrasts of 1\% - 2\%. \cite{poulsen_optimal_2022}.

Precise control of the power delivered to the diamond is enabled with a variable attenuator.
Under the experimental conditions here, we found that -22~dBm is the highest amount of power which can be applied before deleterious power broadening effects occur.

\section{Experimental Results}
\subsection{Vector Magnetometry}
\label{vcr:sec:vector}
To convert time-domain signals to magnetic sensor readout, the reflected time trace is parsed in single-cycle increments of 500~\micro s as shown in Fig.~\ref{vcr:fig:timeseries}.
Each frame is then independently processed to calculate the external magnetic field during that frame, resulting in a 2~ksps measurement of the magnetic field.
Within a single frame period, individual NV orientations are interrogated four times, producing four resonances in the reflected signal which encodes magnetometry information in their time domain locations.
To optimize processing, a matched filter is applied to each NV orientation using previously recorded data which incorporates the shape of the resonance, including the hyperfine peaks \cite{proakis_digital_2008}.
For each NV orientation, the frame data is filtered with the corresponding matched filter and the locations of the four peaks are found by fitting the filtered data to a second order polynomial.
The matched filter extracts information from all the hyperfine peaks while still being robust to noise sources that can deform them.

Individually, the temporal locations of the four peaks are susceptible to amplitude and phase noise from the bias field.
However, external magnetic fields, bias field amplitude noise, and bias field phase noise each produce a unique signature in the movement of the peaks relative to each other.
It is therefore possible to combine these four peak locations into a single quantity, denoted $\tau_i$ for NV orientation $i$, which retains the magnetometry information while suppressing bias field noise (see Supplemental Material Sec.~\ref{vcr:supp:biasnoise}).
$\tau_i$ is expected to vary linearly with the projection of the magnetic field along $\hat{n}_i$, allowing the applied magnetic field $\vec{B}_T(t)$ to be computed by
\begin{equation}
\label{vcr:eq:A}
\vec{B}_T = (I + C)A\left[\begin{array}{c}\tau_1 \\ \tau_2 \\ \tau_3 \end{array} \right],
\end{equation}
where $I$ is the identity matrix, $A$ is a $3\times 3$ geometric calibration matrix, which transforms the NV orientation basis to the lab and accounts for the bias field strength, and $C$ is a $3\times 3$ correction matrix which accounts for non-idealities introduced by orientation cross-talk and sensitivity to transverse magnetic fields, both of which are discussed in more detail later.

To perform calibration, 3 orthogonal fields with amplitude 1~\micro T$_{\text{rms}}$ and varying frequencies are applied: 15~Hz in the $x$-direction, 25~Hz in the $y$-direction, and 35~Hz in the $z$-direction.
From the measured data, the $A$ matrix is obtained by solving for the NV orientations and bias field strength that best map the observed peak locations $\tau_i$ to the applied test field.
The magnetic fields calculated by $A$ using the calibration data are not orthogonal due to two main effects.
In the first effect, cross-talk between orientations can occur when the shift in one NV orientation's peak causes an apparent shift in another NV orientation's peak due to overlapping resonances as shown in Fig.~\ref{vcr:fig:timeseries}.
In the second effect, a strong bias magnetic field transverse to the NV axis causes the NV resonance frequency to become sensitive to transverse magnetic fields \cite{barry_sensitivity_2020}.
A correction matrix $C$ is thus fit from the calibration data to make the calculated magnetic fields orthogonal.
Corrections from $C$ are of order 16\%, which is commensurate with estimates of the magnitude of the effects of the non-idealities (see Supplemental Material Sec.~\ref{vcr:supp:calibration}).
Figure~\ref{vcr:fig:vector} shows the results of this procedure: each $\tau_i$ contains multiple applied fields, but upon transformation to the $x$-, $y$-, and $z$-bases using the calibration, only a single applied field is present in each direction.


\subsection{Sensitivity}
\begin{figure}[t]
\hspace{-2mm}
\begin{minipage}[b]{0.45\textwidth}
\begin{overpic}[width=3.2in]{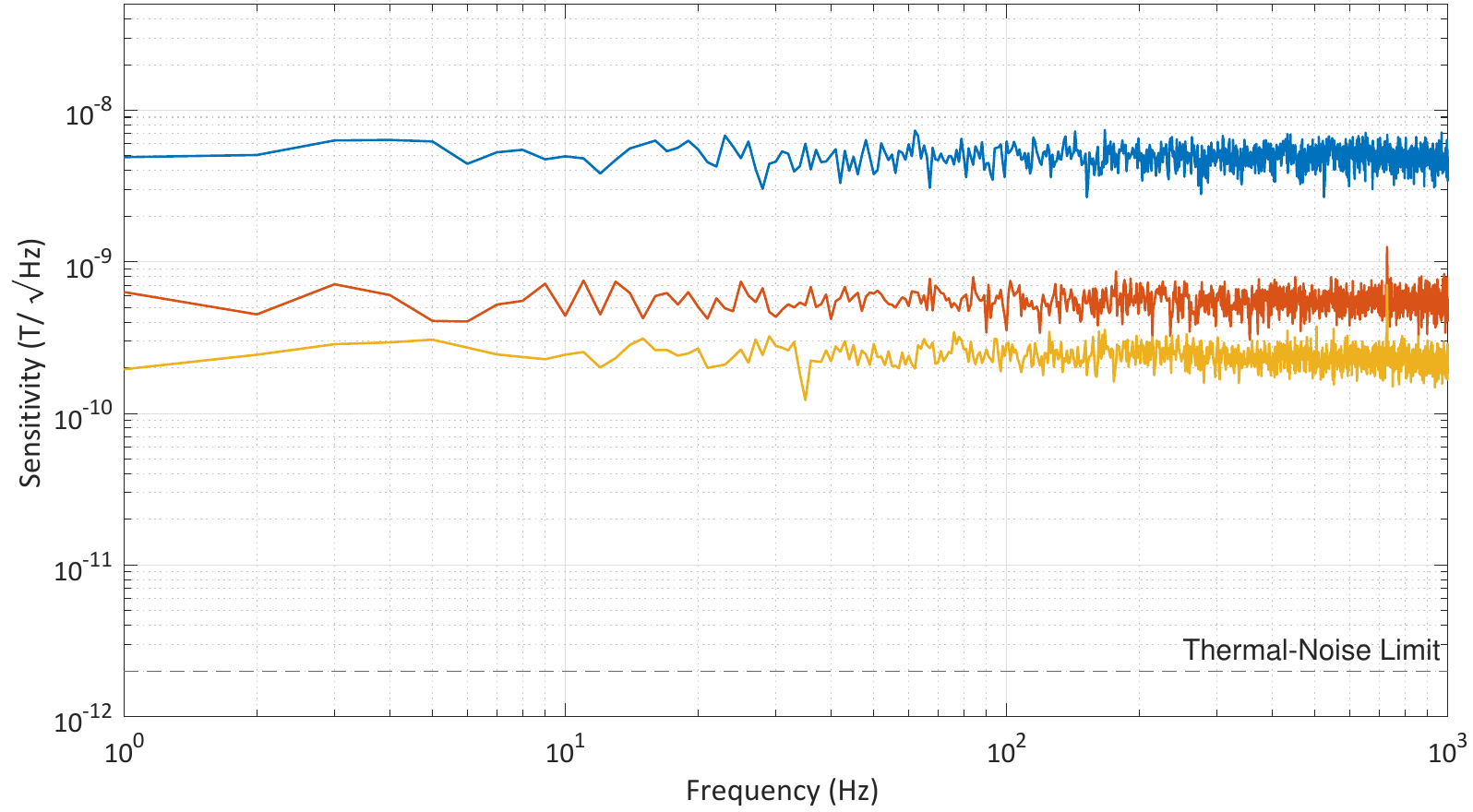} \put(0,55){\textbf{a)}}
\end{overpic}
\end{minipage}
\vspace{10pt}
\;
\begin{minipage}[b]{0.45\textwidth}
\begin{Overpic}{{
\put(-116,0){\includegraphics[width=3.15in]{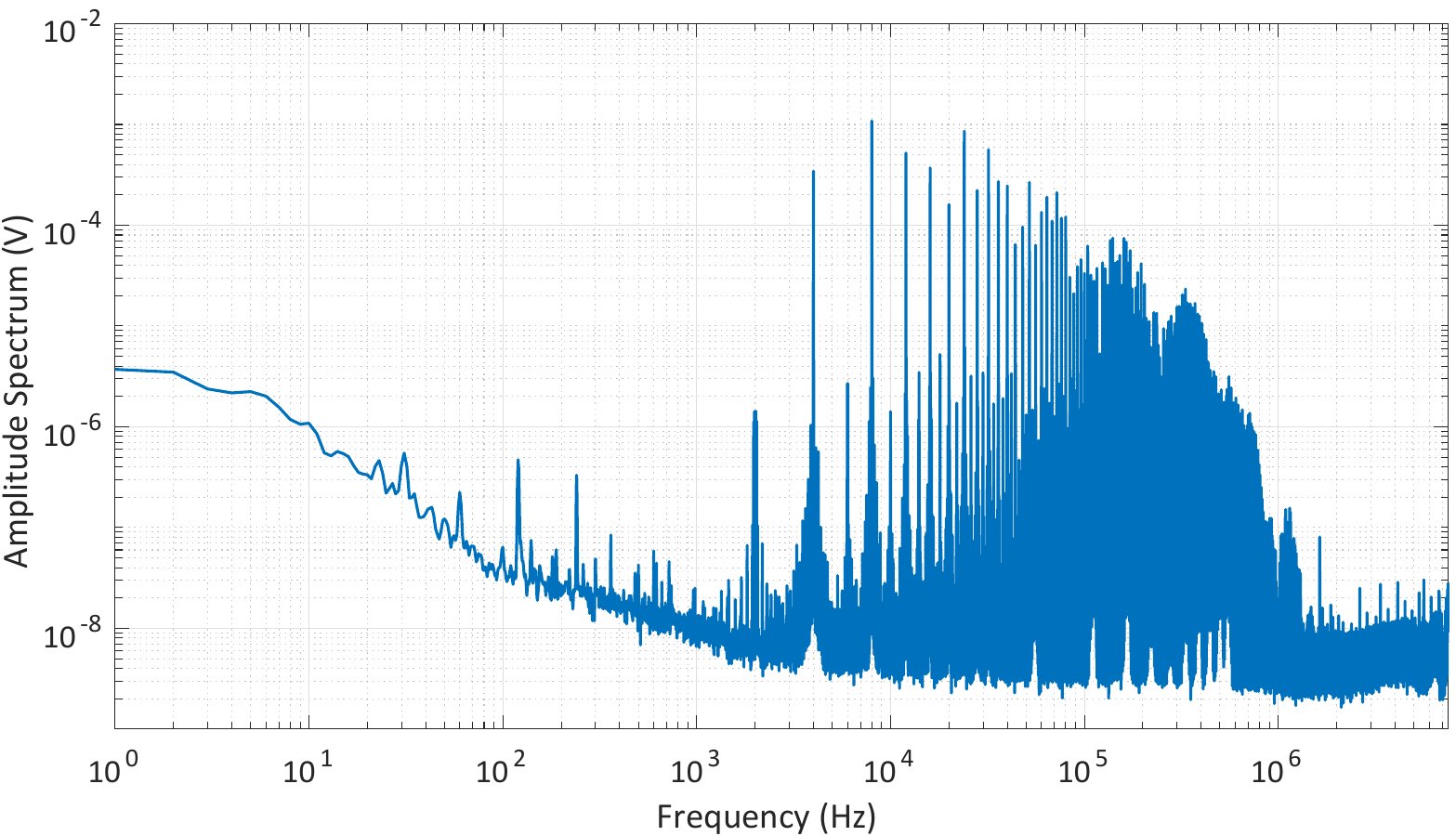}}}} \put(-92,100){\textbf{b)}}
\end{Overpic}
\end{minipage}
\caption{\textbf{Device sensitivity and reflected spectrum.}
\textbf{(a) Device sensitivity} The single-axis sensitivity for NV orientation 1 (\textcolor{matlab1}{\rule[0.75mm]{3mm}{.25mm}}), NV orientation 2 (\textcolor{matlab2}{\rule[0.75mm]{3mm}{.25mm}}), and NV orientation 3 (\textcolor{matlab3}{\rule[0.75mm]{3mm}{.25mm}}). Each orientation is flat over the 1~kHz sensing band, with a measured rms sensitivity of 5.3~nT/\rthz{}, 580~pT/\rthz{}, and 250~pT/\rthz{} for NV orientations 1, 2, and 3, respectively.
Simulations predict a thermal noise limit of 2~pT/\rthz{} (\plotdashline).
\textbf{(b) Spectrum of the reflected signal.} Amplitude spectral density of the MW signal after reflecting off of the sensor head. The sensed magnetic field is encoded at even harmonics of 2~kHz (2~kHz, 6~kHz, 10~kHz, etc.) up to $\approx 785$~kHz, making the system susceptible to MW noise in this bandwidth.}
\label{vcr:fig:sensitivity}
\end{figure}

Precision measurement of device sensitivity requires eliminating ambient noise.
The sensor is thus operated inside a 5-layer MuMETAL shield to observe the noise floor.
Using the known NV geometry, the measured $\tau_i$ are scaled so that each orientation behaves as an independent single-axis vector magnetometer (see Supplemental Material Sec.~\ref{vcr:supp:calibration} for details).
The resulting noise floors are 5.3~nT/\rthz{}, 580~pT/\rthz{}, and 250~pT/\rthz{} for NV orientations 1, 2, and 3, respectively, as shown in Fig.~\ref{vcr:fig:sensitivity}a.
This sensitivity remains flat over the 1~kHz bandwidth of the sensor, which is set by the bias field frequency, as the device samples the full magnetic field once per bias field sweep.
When the calibration in Eq.~\eqref{vcr:eq:A} is applied to bring the measurement to an orthogonal basis, the sensitivity varies from 210~pT/\rthz{} to 5.6~nT/\rthz{} depending on the orientation (see Supplemental Material Sec.~\ref{vcr:supp:orthnoise} for conversion from single-axis sensitivity to orthogonal sensitivity).

The reduced performance of NV orientation 1 can be attributed to two main factors.
First, having the largest bias field projection, it is swept through the quickest and therefore the digitizer samples fewer points  during its resonance peak for analysis.
Second, this NV orientation's peak is significantly smaller than the others, due to a decreased interaction between the spin resonance and the MW photons.
In this device, the MW resonator is oriented so that its electric field is parallel to the bias magnetic field. Because the spin-photon coupling is proportional to the MW electric field transverse to the NV axis, NV orientations with a high bias field projection, such as NV orientation 1, have low MW coupling.


The sensitivity noise floor is limited by amplitude noise on the MW source.
Because the reflected signal from the sensor head is a non-linear function of the applied magnetic field (see Supplemental Sec.\ref{vcr:supp:refcoeff} for the equation describing the reflected signal), the spectrum of the reflected signal, shown in Fig.~\ref{vcr:fig:sensitivity}b, contains harmonics of the 2~kHz bias field up to $\approx$~785~kHz.
The bandwidth over which harmonics appear is set by the width of the peaks in Fig.~\ref{vcr:fig:timeseries}.
An externally applied magnetic field, even one small enough to be treated linearly, is upmodulated by these harmonics, encoding the magnetometry information over that same 785~kHz bandwidth (see Supplemental Material Sec.~\ref{vcr:supp:spectrum} for analysis of the reflected signal spectrum).
By spreading the magnetometry information across this band, the system  now becomes susceptible to noise across this entire bandwidth, effectively aliasing MW noise from this entire 785~kHz band onto the 1~kHz bandwidth of the sensor (see Supplemental Material Sec.~\ref{vcr:supp:mwnoise} for supporting simulation results).
As a result of this broadband noise dependence, even the low-noise MW source used in this work introduces sufficient amplitude noise to set the noise floor.

\section{Discussion}
At present, the cartesian vector sensitivity of the device is limited by NV orientation 1.
This could be improved by changing the dielectric resonator alignment relative to the diamond so that the MW photons couple equally to all NV orientations, thus increasing the size of the NV orientation 1 peak.
The new coupling would be similar to the current coupling of NV orientation 2, suggesting that we could achieve full vector sensitivity which approaches 580~pT/\rthz, the single-axis sensitivity of NV orientation 2.

Furthermore, an improved bias coil design could increase the bias field strength, allowing interrogation of all four NV orientations.
By sampling the fourth orientation, the magnetic field vector becomes over defined, improving the magnetic field reconstruction and sensitivity.

The above improvements can push the sensitivity to the floor set by amplitude noise on the MW source.
Additional improvements could be realized by lowering the MW noise floor, which can be accomplished either through amplitude noise cancellation techniques \cite{depellette_amplitude_2025} or by switching to a lower noise, fixed-frequency MW source, with an expected improvement of $\approx 3$x in sensitivity.
If using a fixed-frequency source, the cavity resonance must be controlled to ensure the MWs are resonant.
This can be accomplished through techniques including cavity temperature control or mechanical adjustment of the separation between the two halves of the dielectric resonator.
With a thermal-noise-limited MW source, the device sensitivity could be pushed below 2~pT/\rthz{} (see Supplemental Material Sec.~\ref{vcr:supp:mwnoise} for supporting simulation details).


By careful application of an AC bias field, we have extended the MW readout technique from a single-axis measurement to $4\pi$-steradian vector magnetic field sensing.
With the suggested improvements, the sensor would offer sensitivities of $\approx 200$~pT/\rthz, making it well-suited to applications which are enhanced by vector capabilities, including magnetocardiography \cite{su_vector_2024, yang_new_2021}.
Paired with a larger diamond, thermal polarization from the NV zero-field splitting \cite{wilcox_thermally_2022} could present a path towards a laser-free, diamond-based vector magnetometer.


\section*{Acknowledgements}
We acknowledge Vescent Technologies for hardware contribution, particularly in early phases of device construction.

DISTRIBUTION STATEMENT A. Approved for public release. Distribution is unlimited.

This material is based upon work supported by the Vescent Technologies, Inc. and Under Secretary of War for Research and Engineering under Air Force Contract No. FA8702-15-D-0001 or FA8702-25-D-B002. Any opinions, findings, conclusions or recommendations expressed in this material are those of the author(s) and do not necessarily reflect the views of the Vescent Technologies, Inc. and Under Secretary of War for Research and Engineering.

\copyright 2025 Massachusetts Institute of Technology.

Delivered to the U.S. Government with Unlimited Rights, as defined in DFARS Part 252.227-7013 or 7014 (Feb 2014). Notwithstanding any copyright notice, U.S. Government rights in this work are defined by DFARS 252.227-7013 or DFARS 252.227-7014 as detailed above. Use of this work other than as specifically authorized by the U.S. Government may violate any copyrights that exist in this work.

\bibliography{thebib.bib}

@article{barry_optical_2016,
	title = {Optical magnetic detection of single-neuron action potentials using quantum defects in diamond},
	volume = {113},
	url = {https://www.pnas.org/content/113/49/14133.short},
	number = {49},
	journal = {Proc. Natl. Acad. Sci.},
	author = {Barry, John F and Turner, Matthew J and Schloss, Jennifer M and Glenn, David R and Song, Yuyu and Lukin, Mikhail D and Park, Hongkun and Walsworth, Ronald L},
	year = {2016},
	note = {Publisher: National Acad Sciences},
	pages = {14133--14138},
}

@article{eisenach_cavity-enhanced_2021,
	title = {Cavity-enhanced microwave readout of a solid-state spin sensor},
	volume = {12},
	url = {https://www.nature.com/articles/s41467-021-21256-7},
	number = {1},
	journal = {Nat. Commun.},
	author = {Eisenach, Erik R and Barry, John F and O'Keeffe, Michael F and Schloss, Jennifer M and Steinecker, Matthew H and Englund, Dirk R and Braje, Danielle A},
	year = {2021},
	note = {Publisher: Nature Publishing Group},
	pages = {1357},
}

@article{ashfold_nitrogen_2020,
	title = {Nitrogen in diamond},
	volume = {120},
	url = {https://pubs.acs.org/doi/10.1021/acs.chemrev.9b00518?ref=pdf},
	number = {12},
	journal = {Chem. Rev.},
	author = {Ashfold, Michael NR and Goss, Jonathan P and Green, Ben L and May, Paul W and Newton, Mark E and Peaker, Chloe V},
	year = {2020},
	note = {Publisher: ACS Publications},
	pages = {5745--5794},
}

@article{loubser_electron_1978,
	title = {Electron spin resonance in the study of diamond},
	volume = {41},
	url = {https://doi.org/10.1088/0034-4885/41/8/002},
	doi = {10.1088/0034-4885/41/8/002},
	number = {8},
	journal = {Rep. Prog. Phys.},
	author = {Loubser, J. H. N. and Wyk, J. A. van},
	month = aug,
	year = {1978},
	note = {Publisher: IOP Publishing},
	pages = {1201--1248},
}

@article{karadas_feasibility_2018,
	title = {Feasibility and resolution limits of opto-magnetic imaging of neural network activity in brain slices using color centers in diamond},
	volume = {8},
	url = {https://www.nature.com/articles/s41598-018-22793-w},
	number = {1},
	journal = {Sci. Rep.},
	author = {Karadas, M{\"u}rsel and Wojciechowski, Adam M and Huck, Alexander and Dalby, Nils Ole and Andersen, Ulrik Lund and Thielscher, Axel},
	year = {2018},
	note = {Publisher: Nature Publishing Group},
	pages = {1--14},
}

@article{okeeffe_hamiltonian_2019,
	title = {Hamiltonian engineering with constrained optimization for quantum sensing and control},
	volume = {21},
	url = {https://iopscience.iop.org/article/10.1088/1367-2630/ab00be/meta},
	number = {2},
	journal = {New J. Phys.},
	author = {O'Keeffe, Michael F and Horesh, Lior and Barry, John F and Braje, Danielle A and Chuang, Isaac L},
	year = {2019},
	note = {Publisher: IOP Publishing},
	pages = {023015},
}

@article{alsid_photoluminescence_2019,
	title = {Photoluminescence {Decomposition} {Analysis}: {A} {Technique} to {Characterize} {N}-{V} {Creation} in {Diamond}},
	volume = {12},
	url = {https://journals.aps.org/prapplied/abstract/10.1103/PhysRevApplied.12.044003},
	number = {4},
	journal = {Phys. Rev. Appl.},
	author = {Alsid, Scott T and Barry, John F and Pham, Linh M and Schloss, Jennifer M and O'Keeffe, Michael F and Cappellaro, Paola and Braje, Danielle A},
	year = {2019},
	note = {Publisher: APS},
	pages = {044003},
}

@article{hopper_near-infrared-assisted_2016,
	title = {Near-infrared-assisted charge control and spin readout of the nitrogen-vacancy center in diamond},
	volume = {94},
	url = {https://journals.aps.org/prb/abstract/10.1103/PhysRevB.94.241201},
	number = {24},
	journal = {Phys. Rev. B},
	author = {Hopper, David A and Grote, Richard R and Exarhos, Annemarie L and Bassett, Lee C},
	year = {2016},
	note = {Publisher: APS},
	pages = {241201},
}

@article{waeber_pulse_2019,
	title = {Pulse control protocols for preserving coherence in dipolar-coupled nuclear spin baths},
	volume = {10},
	url = {https://www.nature.com/articles/s41467-019-11160-6},
	number = {1},
	journal = {Nat. Commun.},
	author = {Waeber, AM and Gillard, G and Ragunathan, G and Hopkinson, M and Spencer, P and Ritchie, DA and Skolnick, MS and Chekhovich, EA},
	year = {2019},
	note = {Publisher: Nature Publishing Group},
	pages = {3157},
}

@article{ebel_dispersive_2021,
	title = {Dispersive readout of room-temperature ensemble spin sensors},
	volume = {6},
	url = {https://doi.org/10.1088/2058-9565/abfaaf},
	doi = {10.1088/2058-9565/abfaaf},
	number = {3},
	journal = {Quantum Sci. Technol.},
	author = {Ebel, J. and Joas, T. and Schalk, M. and Weinbrenner, P. and Angerer, A. and Majer, J. and Reinhard, F.},
	month = jun,
	year = {2021},
	note = {Publisher: IOP Publishing},
	pages = {03LT01},
}

@article{bourgeois_photoelectric_2015,
	title = {Photoelectric detection of electron spin resonance of nitrogen-vacancy centres in diamond},
	volume = {6},
	url = {https://www.nature.com/articles/ncomms9577},
	number = {1},
	journal = {Nat. Commun.},
	author = {Bourgeois, E and Jarmola, A and Siyushev, P and Gulka, M and Hruby, J and Jelezko, Fedor and Budker, D and Nesladek, M},
	year = {2015},
	note = {Publisher: Nature Publishing Group},
	pages = {1--8},
}

@incollection{rubiola_phase_2009,
	series = {The {Cambridge} {RF} and {Microwave} {Engineering} {Series}},
	title = {Phase noise and frequency stability},
	booktitle = {Phase {Noise} and {Frequency} {Stability} in {Oscillators}},
	publisher = {Cambridge University Press},
	author = {Rubiola, Enrico},
	year = {2009},
	doi = {10.1017/CBO9780511812798.005},
}

@article{dreau_avoiding_2011,
	title = {Avoiding power broadening in optically detected magnetic resonance of single {NV} defects for enhanced dc magnetic field sensitivity},
	volume = {84},
	url = {https://link.aps.org/doi/10.1103/PhysRevB.84.195204},
	doi = {10.1103/PhysRevB.84.195204},
	number = {19},
	journal = {Phys. Rev. B},
	author = {Dr{\'e}au, A. and Lesik, M. and Rondin, L. and Spinicelli, P. and Arcizet, O. and Roch, J.-F. and Jacques, V.},
	month = nov,
	year = {2011},
	note = {Publisher: American Physical Society},
	pages = {195204},
}

@article{barry_sensitivity_2020,
	title = {Sensitivity optimization for {NV}-diamond magnetometry},
	volume = {92},
	url = {https://link.aps.org/doi/10.1103/RevModPhys.92.015004},
	doi = {10.1103/RevModPhys.92.015004},
	number = {1},
	journal = {Rev. Mod. Phys.},
	author = {Barry, John F. and Schloss, Jennifer M. and Bauch, Erik and Turner, Matthew J. and Hart, Connor A. and Pham, Linh M. and Walsworth, Ronald L.},
	month = mar,
	year = {2020},
	note = {Publisher: American Physical Society},
	pages = {015004},
}

@article{pound_electronic_1946,
	title = {Electronic {Frequency} {Stabilization} of {Microwave} {Oscillators}},
	volume = {17},
	doi = {10.1063/1.1770414},
	number = {11},
	journal = {Review of Scientific Instruments},
	author = {Pound, R. V.},
	year = {1946},
	pages = {490--505},
}

@article{fescenko_diamond_2019,
	title = {Diamond {Magnetic} {Microscopy} of {Malarial} {Hemozoin} {Nanocrystals}},
	volume = {11},
	url = {https://link.aps.org/doi/10.1103/PhysRevApplied.11.034029},
	doi = {10.1103/PhysRevApplied.11.034029},
	number = {3},
	journal = {Phys. Rev. Appl.},
	author = {Fescenko, Ilja and Laraoui, Abdelghani and Smits, Janis and Mosavian, Nazanin and Kehayias, Pauli and Seto, Jong and Bougas, Lykourgos and Jarmola, Andrey and Acosta, Victor M.},
	month = mar,
	year = {2019},
	note = {Publisher: American Physical Society},
	pages = {034029},
}

@article{lovchinsky_nuclear_2016,
	title = {Nuclear magnetic resonance detection and spectroscopy of single proteins using quantum logic},
	volume = {351},
	issn = {0036-8075},
	url = {http://science.sciencemag.org/content/351/6275/836},
	doi = {10.1126/science.aad8022},
	number = {6275},
	journal = {Science},
	author = {Lovchinsky, I. and Sushkov, A. O. and Urbach, E. and de Leon, N. P. and Choi, S. and De Greve, K. and Evans, R. and Gertner, R. and Bersin, E. and M{\"u}ller, C. and McGuinness, L. and Jelezko, F. and Walsworth, R. L. and Park, H. and Lukin, M. D.},
	year = {2016},
	note = {Publisher: American Association for the Advancement of Science},
	pages = {836--841},
}

@article{glenn_high-resolution_2018,
	title = {High-resolution magnetic resonance spectroscopy using a solid-state spin sensor},
	volume = {555},
	issn = {1476-4687},
	url = {https://doi.org/10.1038/nature25781},
	doi = {10.1038/nature25781},
	number = {7696},
	journal = {Nature},
	author = {Glenn, David R. and Bucher, Dominik B. and Lee, Junghyun and Lukin, Mikhail D. and Park, Hongkun and Walsworth, Ronald L.},
	month = mar,
	year = {2018},
	pages = {351--354},
}

@article{schloss_simultaneous_2018,
	title = {Simultaneous {Broadband} {Vector} {Magnetometry} {Using} {Solid}-{State} {Spins}},
	volume = {10},
	url = {https://link.aps.org/doi/10.1103/PhysRevApplied.10.034044},
	doi = {10.1103/PhysRevApplied.10.034044},
	number = {3},
	journal = {Phys. Rev. Applied},
	author = {Schloss, Jennifer M. and Barry, John F. and Turner, Matthew J. and Walsworth, Ronald L.},
	month = sep,
	year = {2018},
	note = {Publisher: American Physical Society},
	pages = {034044},
}

@article{hopper_spin_2018,
	title = {Spin {Readout} {Techniques} of the {Nitrogen}-{Vacancy} {Center} in {Diamond}},
	volume = {9},
	issn = {2072-666X},
	url = {http://www.mdpi.com/2072-666X/9/9/437},
	number = {9},
	journal = {Micromachines},
	author = {Hopper, David A. and Shulevitz, Henry J. and Bassett, Lee C.},
	year = {2018},
	pages = {437},
}

@article{davis_mapping_2018,
	title = {Mapping the microscale origins of magnetic resonance image contrast with subcellular diamond magnetometry},
	volume = {9},
	url = {https://doi.org/10.1038/s41467-017-02471-7},
	doi = {10.1038/s41467-017-02471-7},
	number = {1},
	journal = {Nat. Commun.},
	author = {Davis, Hunter C. and Ramesh, Pradeep and Bhatnagar, Aadyot and Lee-Gosselin, Audrey and Barry, John F. and Glenn, David R. and Walsworth, Ronald L. and Shapiro, Mikhail G.},
	year = {2018},
	note = {ISBN: 2041-1723},
	pages = {131},
}

@article{jenkins_imaging_2020,
	title = {Imaging the breakdown of ohmic transport in graphene},
	url = {https://arxiv.org/abs/2002.05065},
	journal = {arXiv:2002.05065},
	author = {Jenkins, A. and Baumann, S. and Zhou, H. and Meynell, S. A. and Yang, D. and Watanabe, K. and Taniguchi, T. and Lucas, A. and Young, A. F. and Bleszynski Jayich, A. C.},
	year = {2020},
}

@article{shi_single-protein_2015,
	title = {Single-protein spin resonance spectroscopy under ambient conditions},
	volume = {347},
	issn = {0036-8075},
	url = {http://science.sciencemag.org/content/347/6226/1135},
	doi = {10.1126/science.aaa2253},
	number = {6226},
	journal = {Science},
	author = {Shi, Fazhan and Zhang, Qi and Wang, Pengfei and Sun, Hongbin and Wang, Jiarong and Rong, Xing and Chen, Ming and Ju, Chenyong and Reinhard, Friedemann and Chen, Hongwei and Wrachtrup, J{\"o}rg and Wang, Junfeng and Du, Jiangfeng},
	year = {2015},
	note = {Publisher: American Association for the Advancement of Science},
	pages = {1135--1138},
}

@article{smits_two-dimensional_2019,
	title = {Two-dimensional nuclear magnetic resonance spectroscopy with a microfluidic diamond quantum sensor},
	volume = {5},
	url = {https://advances.sciencemag.org/content/5/7/eaaw7895},
	number = {7},
	journal = {Sci. Adv.},
	author = {Smits, Janis and Damron, Joshua T. and Kehayias, Pauli and McDowell, Andrew F. and Mosavian, Nazanin and Fescenko, Ilja and Ristoff, Nathaniel and Laraoui, Abdelghani and Jarmola, Andrey and Acosta, Victor M.},
	year = {2019},
	note = {Publisher: American Association for the Advancement of Science},
	pages = {eaaw7895},
}

@article{aslam_nanoscale_2017,
	title = {Nanoscale nuclear magnetic resonance with chemical resolution},
	volume = {357},
	issn = {0036-8075},
	url = {http://science.sciencemag.org/content/357/6346/67},
	number = {6346},
	journal = {Science},
	author = {Aslam, Nabeel and Pfender, Matthias and Neumann, Philipp and Reuter, Rolf and Zappe, Andrea and F{\'a}varo de Oliveira, Felipe and Denisenko, Andrej and Sumiya, Hitoshi and Onoda, Shinobu and Isoya, Junichi and Wrachtrup, J{\"o}rg},
	year = {2017},
	note = {Publisher: American Association for the Advancement of Science},
	pages = {67--71},
}

@article{achard_chemical_2020,
	title = {Chemical vapour deposition diamond single crystals with nitrogen-vacancy centres: a review of material synthesis and technology for quantum sensing applications},
	volume = {53},
	doi = {10.1088/1361-6463/ab81d1},
	number = {31},
	journal = {J. Phys. D: Appl. Phys.},
	author = {Achard, J and Jacques, Vincent and Tallaire, A},
	year = {2020},
	note = {Publisher: IOP Publishing},
	pages = {313001},
}

@article{bar-gill_solid-state_2013,
	title = {Solid-state electronic spin coherence time approaching one second},
	volume = {4},
	url = {http://dx.doi.org/10.1038/ncomms2771},
	journal = {Nat. Commun.},
	author = {Bar-Gill, N. and Pham, L. M. and Jarmola, A. and Budker, D. and Walsworth, R. L.},
	year = {2013},
	pages = {1743},
}

@article{bauch_ultralong_2018,
	title = {Ultralong {Dephasing} {Times} in {Solid}-{State} {Spin} {Ensembles} via {Quantum} {Control}},
	volume = {8},
	url = {https://link.aps.org/doi/10.1103/PhysRevX.8.031025},
	doi = {10.1103/PhysRevX.8.031025},
	number = {3},
	journal = {Phys. Rev. X},
	author = {Bauch, Erik and Hart, Connor A. and Schloss, Jennifer M. and Turner, Matthew J. and Barry, John F. and Kehayias, Pauli and Singh, Swati and Walsworth, Ronald L.},
	month = jul,
	year = {2018},
	note = {Publisher: American Physical Society},
	pages = {031025},
}

@article{bauch_decoherence_2020,
	title = {Decoherence of ensembles of nitrogen-vacancy centers in diamond},
	volume = {102},
	url = {https://link.aps.org/doi/10.1103/PhysRevB.102.134210},
	doi = {10.1103/PhysRevB.102.134210},
	number = {13},
	journal = {Phys. Rev. B},
	author = {Bauch, Erik and Singh, Swati and Lee, Junghyun and Hart, Connor A. and Schloss, Jennifer M. and Turner, Matthew J. and Barry, John F. and Pham, Linh M. and Bar-Gill, Nir and Yelin, Susanne F. and Walsworth, Ronald L.},
	month = oct,
	year = {2020},
	note = {Publisher: American Physical Society},
	pages = {134210},
}

@article{niethammer_coherent_2019,
	title = {Coherent electrical readout of defect spins in silicon carbide by photo-ionization at ambient conditions},
	volume = {10},
	url = {https://doi.org/10.1038/s41467-019-13545-z},
	journal = {Nat. Commun.},
	author = {Niethammer, Matthias and Widmann, Matthias and Rendler, Torsten and Morioka, Naoya and Chen, Yu-Chen and St{\"o}hr, Rainer and Hassan, Jawad Ul and Onoda, Shinobu and Ohshima, Takeshi and Lee, Sang-Yun and Mukherjee, Amlan and Isoya, Junichi and Son, Nguyen Tien and Wrachtrup, J{\"o}rg},
	month = dec,
	year = {2019},
	pages = {5569},
}

@article{shields_efficient_2015,
	title = {Efficient {Readout} of a {Single} {Spin} {State} in {Diamond} via {Spin}-to-{Charge} {Conversion}},
	volume = {114},
	url = {https://link.aps.org/doi/10.1103/PhysRevLett.114.136402},
	doi = {10.1103/PhysRevLett.114.136402},
	number = {13},
	journal = {Phys. Rev. Lett.},
	author = {Shields, B. J. and Unterreithmeier, Q. P. and de Leon, N. P. and Park, H. and Lukin, M. D.},
	month = mar,
	year = {2015},
	note = {Publisher: American Physical Society},
	pages = {136402},
}

@article{neumann_single-shot_2010,
	title = {Single-{Shot} {Readout} of a {Single} {Nuclear} {Spin}},
	volume = {329},
	issn = {0036-8075},
	url = {https://science.sciencemag.org/content/329/5991/542},
	doi = {10.1126/science.1189075},
	number = {5991},
	journal = {Science},
	author = {Neumann, Philipp and Beck, Johannes and Steiner, Matthias and Rempp, Florian and Fedder, Helmut and Hemmer, Philip R. and Wrachtrup, J{\"o}rg and Jelezko, Fedor},
	year = {2010},
	note = {Publisher: American Association for the Advancement of Science},
	pages = {542--544},
}

@article{bluvstein_identifying_2019,
	title = {Identifying and {Mitigating} {Charge} {Instabilities} in {Shallow} {Diamond} {Nitrogen}-{Vacancy} {Centers}},
	volume = {122},
	url = {https://link.aps.org/doi/10.1103/PhysRevLett.122.076101},
	doi = {10.1103/PhysRevLett.122.076101},
	number = {7},
	journal = {Phys. Rev. Lett.},
	author = {Bluvstein, Dolev and Zhang, Zhiran and Jayich, Ania C. Bleszynski},
	month = feb,
	year = {2019},
	note = {Publisher: American Physical Society},
	pages = {076101},
}

@article{steiner_universal_2010,
	title = {Universal enhancement of the optical readout fidelity of single electron spins at nitrogen-vacancy centers in diamond},
	volume = {81},
	url = {https://link.aps.org/doi/10.1103/PhysRevB.81.035205},
	doi = {10.1103/PhysRevB.81.035205},
	number = {3},
	journal = {Phys. Rev. B},
	author = {Steiner, M. and Neumann, P. and Beck, J. and Jelezko, F. and Wrachtrup, J.},
	month = jan,
	year = {2010},
	note = {Publisher: American Physical Society},
	pages = {035205},
}

@article{webb_detection_2021,
	title = {Detection of biological signals from a live mammalian muscle using an early stage diamond quantum sensor},
	volume = {11},
	issn = {2045-2322},
	url = {https://doi.org/10.1038/s41598-021-81828-x},
	doi = {10.1038/s41598-021-81828-x},
	number = {1},
	journal = {Sci. Rep.},
	author = {Webb, James Luke and Troise, Luca and Hansen, Nikolaj Winther and Olsson, Christoffer and Wojciechowski, Adam M. and Achard, Jocelyn and Brinza, Ovidiu and Staacke, Robert and Kieschnick, Michael and Meijer, Jan and Thielscher, Axel and Perrier, Jean-Fran{\c c}ois and Berg-S{\o}rensen, Kirstine and Huck, Alexander and Andersen, Ulrik Lund},
	month = jan,
	year = {2021},
	pages = {2412},
}

@article{zhou_imaging_2021,
	title = {Imaging {Damage} in {Steel} {Using} a {Diamond} {Magnetometer}},
	volume = {15},
	url = {https://link.aps.org/doi/10.1103/PhysRevApplied.15.024015},
	doi = {10.1103/PhysRevApplied.15.024015},
	number = {2},
	journal = {Phys. Rev. Appl.},
	author = {Zhou, L. Q. and Patel, R. L. and Frangeskou, A. C. and Nikitin, A. and Green, B. L. and Breeze, B. G. and Onoda, S. and Isoya, J. and Morley, G. W.},
	month = feb,
	year = {2021},
	note = {Publisher: American Physical Society},
	pages = {024015},
}

@article{hart_n-vdiamond_2021,
	title = {N-{V}{\textendash}{Diamond} {Magnetic} {Microscopy} {Using} a {Double} {Quantum} 4-{Ramsey} {Protocol}},
	volume = {15},
	url = {https://link.aps.org/doi/10.1103/PhysRevApplied.15.044020},
	doi = {10.1103/PhysRevApplied.15.044020},
	number = {4},
	journal = {Phys. Rev. Applied},
	author = {Hart, Connor A. and Schloss, Jennifer M. and Turner, Matthew J. and Scheidegger, Patrick J. and Bauch, Erik and Walsworth, Ronald L.},
	month = apr,
	year = {2021},
	note = {Publisher: American Physical Society},
	pages = {044020},
}

@article{wolfowicz_quantum_2021,
	title = {Quantum guidelines for solid-state spin defects},
	url = {https://www.nature.com/articles/s41578-021-00306-y},
	journal = {Nat. Rev. Mater.},
	author = {Wolfowicz, Gary and Heremans, F. Joseph and Anderson, Christopher P. and Kanai, Shun and Seo, Hosung and Gali, Adam and Galli, Giulia and Awschalom, David D.},
	year = {2021},
}

@article{edmonds_characterisation_2021,
	title = {Characterisation of {CVD} diamond with high concentrations of nitrogen for magnetic-field sensing applications},
	volume = {1},
	url = {https://doi.org/10.1088/2633-4356/abd88a},
	journal = {Mater. Quantum Technol.},
	author = {Edmonds, Andrew Mark and Hart, Connor A. and Turner, Matthew J. and Colard, Pierre-Olivier and Schloss, Jennifer M. and Olsson, Kevin and Trubko, Raisa and Markham, Matthew L. and Rathmill, Adam and Horne-Smith, Ben and Lew, Wilbur and Manickam, Arul and Bruce, Scott and Kaup, Peter G. and Russo, Jon C. and DiMario, Michael J. and South, Joseph T. and Hansen, Jay T. and Twitchen, Daniel J. and Walsworth, Ronald},
	month = jan,
	year = {2021},
	note = {Publisher: IOP Publishing},
	pages = {025001},
}

@article{sturner_integrated_2021,
	title = {Integrated and {Portable} {Magnetometer} {Based} on {Nitrogen}-{Vacancy} {Ensembles} in {Diamond}},
	volume = {4},
	doi = {https://doi.org/10.1002/qute.202000111},
	number = {4},
	journal = {Adv. Quantum Technol.},
	author = {St{\"u}rner, Felix M. and Brenneis, Andreas and Buck, Thomas and Kassel, Julian and R{\"o}lver, Robert and Fuchs, Tino and Savitsky, Anton and Suter, Dieter and Grimmel, Jens and Hengesbach, Stefan and Fortsch, Michael and Nakamura, Kazuo and Sumiya, Hitoshi and Onoda, Shinobu and Isoya, Junichi and Jelezko, Fedor},
	year = {2021},
	pages = {2000111},
}

@article{turner_magnetic_2020,
	title = {Magnetic {Field} {Fingerprinting} of {Integrated}-{Circuit} {Activity} with a {Quantum} {Diamond} {Microscope}},
	volume = {14},
	url = {https://link.aps.org/doi/10.1103/PhysRevApplied.14.014097},
	doi = {10.1103/PhysRevApplied.14.014097},
	number = {1},
	journal = {Phys. Rev. Appl.},
	author = {Turner, Matthew J. and Langellier, Nicholas and Bainbridge, Rachel and Walters, Dan and Meesala, Srujan and Babinec, Thomas M. and Kehayias, Pauli and Yacoby, Amir and Hu, Evelyn and Loncar, Marko and Walsworth, Ronald L. and Levine, Edlyn V.},
	month = jul,
	year = {2020},
	note = {Publisher: American Physical Society},
	pages = {014097},
}

@article{patel_subnanotesla_2020,
	title = {Subnanotesla {Magnetometry} with a {Fiber}-{Coupled} {Diamond} {Sensor}},
	volume = {14},
	url = {https://link.aps.org/doi/10.1103/PhysRevApplied.14.044058},
	doi = {10.1103/PhysRevApplied.14.044058},
	number = {4},
	journal = {Phys. Rev. Appl.},
	author = {Patel, R.L. and Zhou, L.Q. and Frangeskou, A.C. and Stimpson, G.A. and Breeze, B.G. and Nikitin, A. and Dale, M.W. and Nichols, E.C. and Thornley, W. and Green, B.L. and Newton, M.E. and Edmonds, A.M. and Markham, M.L. and Twitchen, D.J. and Morley, G.W.},
	month = oct,
	year = {2020},
	note = {Publisher: American Physical Society},
	pages = {044058},
}

@article{pham_enhanced_2012,
	title = {Enhanced solid-state multispin metrology using dynamical decoupling},
	volume = {86},
	url = {https://link.aps.org/doi/10.1103/PhysRevB.86.045214},
	doi = {10.1103/PhysRevB.86.045214},
	number = {4},
	journal = {Phys. Rev. B},
	author = {Pham, L. M. and Bar-Gill, N. and Belthangady, C. and Le Sage, D. and Cappellaro, P. and Lukin, M. D. and Yacoby, A. and Walsworth, R. L.},
	month = jul,
	year = {2012},
	note = {Publisher: American Physical Society},
	pages = {045214},
}

@article{bertelli_magnetic_2020,
	title = {Magnetic resonance imaging of spin-wave transport and interference in a magnetic insulator},
	volume = {6},
	url = {https://advances.sciencemag.org/content/6/46/eabd3556},
	number = {46},
	journal = {Sci. Adv.},
	author = {Bertelli, Iacopo and Carmiggelt, Joris J. and Yu, Tao and Simon, Brecht G. and Pothoven, Coosje C. and Bauer, Gerrit E. W. and Blanter, Yaroslav M. and Aarts, Jan and van der Sar, Toeno},
	year = {2020},
	note = {Publisher: American Association for the Advancement of Science},
	pages = {eabd3556},
}

@article{lenz_imaging_2021,
	title = {Imaging {Topological} {Spin} {Structures} {Using} {Light}-{Polarization} and {Magnetic} {Microscopy}},
	volume = {15},
	url = {https://link.aps.org/doi/10.1103/PhysRevApplied.15.024040},
	number = {2},
	journal = {Phys. Rev. Appl.},
	author = {Lenz, Till and Chatzidrosos, Georgios and Wang, Zhiyuan and Bougas, Lykourgos and Dumeige, Yannick and Wickenbrock, Arne and Kerber, Nico and Z{\'a}zvorka, Jakub and Kammerbauer, Fabian and Kl{\"a}ui, Mathias and al, et},
	year = {2021},
	pages = {024040},
}

@article{watanabe_shallow_2021,
	title = {Shallow {NV} centers augmented by exploiting n-type diamond},
	volume = {178},
	issn = {0008-6223},
	url = {https://www.sciencedirect.com/science/article/pii/S0008622321003110},
	doi = {https://doi.org/10.1016/j.carbon.2021.03.010},
	journal = {Carbon},
	author = {Watanabe, A. and Nishikawa, T. and Kato, H. and Fujie, M. and Fujiwara, M. and Makino, T. and Yamasaki, S. and Herbschleb, E. D. and Mizuochi, N.},
	year = {2021},
	pages = {294--300},
}

@article{hsieh_imaging_2019,
	title = {Imaging stress and magnetism at high pressures using a nanoscale quantum sensor},
	volume = {366},
	issn = {0036-8075},
	url = {https://science.sciencemag.org/content/366/6471/1349},
	doi = {10.1126/science.aaw4352},
	number = {6471},
	journal = {Science},
	author = {Hsieh, S. and Bhattacharyya, P. and Zu, C. and Mittiga, T. and Smart, T. J. and Machado, F. and Kobrin, B. and H{\"o}hn, T. O. and Rui, N. Z. and Kamrani, M. and Chatterjee, S. and Choi, S. and Zaletel, M. and Struzhkin, V. V. and Moore, J. E. and Levitas, V. I. and Jeanloz, R. and Yao, N. Y.},
	year = {2019},
	note = {Publisher: American Association for the Advancement of Science},
	pages = {1349--1354},
}

@article{tallaire_high_2020,
	title = {High {NV} density in a pink {CVD} diamond grown with {N}$_{2}${O} addition},
	volume = {170},
	issn = {0008-6223},
	url = {https://www.sciencedirect.com/science/article/pii/S0008622320308149},
	doi = {https://doi.org/10.1016/j.carbon.2020.08.048},
	journal = {Carbon},
	author = {Tallaire, Alexandre and Brinza, Ovidiu and Huillery, Paul and Delord, Tom and Pellet-Mary, Cl{\'e}ment and Staacke, Robert and Abel, Bernd and Pezzagna, S{\'e}bastien and Meijer, Jan and Touati, Nadia and Binet, Laurent and Ferrier, Alban and Goldner, Philippe and Hetet, Gabriel and Achard, Jocelyn},
	year = {2020},
	pages = {421--429},
}

@article{balasubramanian_ultralong_2009,
	title = {Ultralong spin coherence time in isotopically engineered diamond},
	volume = {8},
	issn = {1476-4660},
	url = {https://doi.org/10.1038/nmat2420},
	doi = {10.1038/nmat2420},
	number = {5},
	journal = {Nature Materials},
	author = {Balasubramanian, Gopalakrishnan and Neumann, Philipp and Twitchen, Daniel and Markham, Matthew and Kolesov, Roman and Mizuochi, Norikazu and Isoya, Junichi and Achard, Jocelyn and Beck, Johannes and Tissler, Julia and Jacques, Vincent and Hemmer, Philip R. and Jelezko, Fedor and Wrachtrup, J{\"o}rg},
	month = may,
	year = {2009},
	pages = {383--387},
}

@article{clevenson_robust_2018,
	title = {Robust high-dynamic-range vector magnetometry with nitrogen-vacancy centers in diamond},
	volume = {112},
	issn = {0003-6951},
	url = {https://doi.org/10.1063/1.5034216},
	doi = {10.1063/1.5034216},
	number = {25},
	journal = {Applied Physics Letters},
	author = {Clevenson, Hannah and Pham, Linh M. and Teale, Carson and Johnson, Kerry and Englund, Dirk and Braje, Danielle},
	month = jun,
	year = {2018},
	note = {\_eprint: https://pubs.aip.org/aip/apl/article-pdf/doi/10.1063/1.5034216/14514188/252406\_1\_online.pdf},
	pages = {252406},
}

@article{arai_millimetre-scale_2022,
	title = {Millimetre-scale magnetocardiography of living rats with thoracotomy},
	volume = {5},
	issn = {2399-3650},
	url = {https://doi.org/10.1038/s42005-022-00978-0},
	doi = {10.1038/s42005-022-00978-0},
	number = {1},
	journal = {Communications Physics},
	author = {Arai, Keigo and Kuwahata, Akihiro and Nishitani, Daisuke and Fujisaki, Ikuya and Matsuki, Ryoma and Nishio, Yuki and Xin, Zonghao and Cao, Xinyu and Hatano, Yuji and Onoda, Shinobu and Shinei, Chikara and Miyakawa, Masashi and Taniguchi, Takashi and Yamazaki, Masatoshi and Teraji, Tokuyuki and Ohshima, Takeshi and Hatano, Mutsuko and Sekino, Masaki and Iwasaki, Takayuki},
	month = aug,
	year = {2022},
	pages = {200},
}

@article{wang_spin-refrigerated_2024,
	title = {A spin-refrigerated cavity quantum electrodynamic sensor},
	volume = {15},
	issn = {2041-1723},
	url = {https://www.nature.com/articles/s41467-024-54333-8},
	doi = {10.1038/s41467-024-54333-8},
	number = {1},
	urldate = {2025-06-22},
	journal = {Nature Communications},
	author = {Wang, Hanfeng and Tiwari, Kunal L. and Jacobs, Kurt and Judy, Michael and Zhang, Xin and Englund, Dirk R. and Trusheim, Matthew E.},
	month = nov,
	year = {2024},
	pages = {10320},
}

@article{wilcox_thermally_2022,
	title = {Thermally {Polarized} {Solid}-{State} {Spin} {Sensor}},
	volume = {17},
	issn = {2331-7019},
	url = {https://link.aps.org/doi/10.1103/PhysRevApplied.17.044004},
	doi = {10.1103/PhysRevApplied.17.044004},
	number = {4},
	urldate = {2025-06-22},
	journal = {Physical Review Applied},
	author = {Wilcox, Reginald and Eisenach, Erik and Barry, John and Steinecker, Matthew and O{\textquoteright}Keeffe, Michael and Englund, Dirk and Braje, Danielle},
	month = apr,
	year = {2022},
	pages = {044004},
}

@phdthesis{eisenach_vector_2022,
	title = {Vector magnetometry using cavity-enhanced microwave readout of solid-state spin sensors},
	url = {https://hdl.handle.net/1721.1/144924},
	school = {MIT},
	author = {Eisenach, Erik},
	year = {2022},
}

@article{doherty_nitrogen-vacancy_2013,
	title = {The nitrogen-vacancy colour centre in diamond},
	volume = {528},
	issn = {03701573},
	url = {https://linkinghub.elsevier.com/retrieve/pii/S0370157313000562},
	doi = {10.1016/j.physrep.2013.02.001},
	number = {1},
	urldate = {2025-06-23},
	journal = {Physics Reports},
	author = {Doherty, Marcus W. and Manson, Neil B. and Delaney, Paul and Jelezko, Fedor and Wrachtrup, J{\"o}rg and Hollenberg, Lloyd C.L.},
	month = jul,
	year = {2013},
	pages = {1--45},
}

@article{rosskopf_investigation_2014,
	title = {Investigation of {Surface} {Magnetic} {Noise} by {Shallow} {Spins} in {Diamond}},
	volume = {112},
	copyright = {http://link.aps.org/licenses/aps-default-license},
	issn = {0031-9007, 1079-7114},
	url = {https://link.aps.org/doi/10.1103/PhysRevLett.112.147602},
	doi = {10.1103/PhysRevLett.112.147602},
	number = {14},
	urldate = {2025-07-06},
	journal = {Physical Review Letters},
	author = {Rosskopf, T. and Dussaux, A. and Ohashi, K. and Loretz, M. and Schirhagl, R. and Watanabe, H. and Shikata, S. and Itoh, K. M. and Degen, C. L.},
	month = apr,
	year = {2014},
	pages = {147602},
}

@article{jarmola_temperature-_2012,
	title = {Temperature- and {Magnetic}-{Field}-{Dependent} {Longitudinal} {Spin} {Relaxation} in {Nitrogen}-{Vacancy} {Ensembles} in {Diamond}},
	volume = {108},
	copyright = {http://link.aps.org/licenses/aps-default-license},
	issn = {0031-9007, 1079-7114},
	url = {https://link.aps.org/doi/10.1103/PhysRevLett.108.197601},
	doi = {10.1103/PhysRevLett.108.197601},
	number = {19},
	urldate = {2025-07-06},
	journal = {Physical Review Letters},
	author = {Jarmola, A. and Acosta, V. M. and Jensen, K. and Chemerisov, S. and Budker, D.},
	month = may,
	year = {2012},
	pages = {197601},
}

@article{goldman_state-selective_2015,
	title = {State-selective intersystem crossing in nitrogen-vacancy centers},
	volume = {91},
	copyright = {http://link.aps.org/licenses/aps-default-license},
	issn = {1098-0121, 1550-235X},
	url = {https://link.aps.org/doi/10.1103/PhysRevB.91.165201},
	doi = {10.1103/PhysRevB.91.165201},
	number = {16},
	urldate = {2025-07-06},
	journal = {Physical Review B},
	author = {Goldman, M. L. and Doherty, M. W. and Sipahigil, A. and Yao, N. Y. and Bennett, S. D. and Manson, N. B. and Kubanek, A. and Lukin, M. D.},
	month = apr,
	year = {2015},
	pages = {165201},
}

@article{goldman_phonon-induced_2015,
	title = {Phonon-{Induced} {Population} {Dynamics} and {Intersystem} {Crossing} in {Nitrogen}-{Vacancy} {Centers}},
	volume = {114},
	copyright = {http://link.aps.org/licenses/aps-default-license},
	issn = {0031-9007, 1079-7114},
	url = {https://link.aps.org/doi/10.1103/PhysRevLett.114.145502},
	doi = {10.1103/PhysRevLett.114.145502},
	number = {14},
	urldate = {2025-07-06},
	journal = {Physical Review Letters},
	author = {Goldman, M. L. and Sipahigil, A. and Doherty, M. W. and Yao, N. Y. and Bennett, S. D. and Markham, M. and Twitchen, D. J. and Manson, N. B. and Kubanek, A. and Lukin, M. D.},
	month = apr,
	year = {2015},
	pages = {145502},
}

@article{berzins_impact_2024,
	title = {Impact of microwave phase noise on diamond quantum sensing},
	volume = {6},
	issn = {2643-1564},
	url = {https://link.aps.org/doi/10.1103/PhysRevResearch.6.043148},
	doi = {10.1103/PhysRevResearch.6.043148},
	number = {4},
	urldate = {2025-07-07},
	journal = {Physical Review Research},
	author = {Berzins, Andris and Saleh Ziabari, Maziar and Silani, Yaser and Fescenko, Ilja and Damron, Joshua T. and Barry, John F. and Jarmola, Andrey and Kehayias, Pauli and Richards, Bryan A. and Smits, Janis and Acosta, Victor M.},
	month = nov,
	year = {2024},
	pages = {043148},
}

@article{barry_sensitive_2024,
	title = {Sensitive ac and dc magnetometry with nitrogen-vacancy-center ensembles in diamond},
	volume = {22},
	copyright = {https://link.aps.org/licenses/aps-default-license},
	issn = {2331-7019},
	url = {https://link.aps.org/doi/10.1103/PhysRevApplied.22.044069},
	doi = {10.1103/physrevapplied.22.044069},
	number = {4},
	urldate = {2025-07-22},
	journal = {Physical Review Applied},
	author = {Barry, John F. and Steinecker, Matthew H. and Alsid, Scott T. and Majumder, Jonah and Pham, Linh M. and O{\textquoteright}Keeffe, Michael F. and Braje, Danielle A.},
	month = oct,
	year = {2024},
	note = {Publisher: American Physical Society (APS)},
}

@article{he_paramagnetic_1993,
	title = {Paramagnetic resonance of photoexcited {N}-\textit{{V}}defects in diamond. {II}. {Hyperfine} interaction with {theN14nucleus}},
	volume = {47},
	copyright = {http://link.aps.org/licenses/aps-default-license},
	issn = {0163-1829, 1095-3795},
	url = {https://link.aps.org/doi/10.1103/PhysRevB.47.8816},
	doi = {10.1103/physrevb.47.8816},
	number = {14},
	urldate = {2025-07-22},
	journal = {Physical Review B},
	author = {He, Xing-Fei and Manson, Neil B. and Fisk, Peter T. H.},
	month = apr,
	year = {1993},
	note = {Publisher: American Physical Society (APS)},
	pages = {8816--8822},
}

@article{smeltzer_robust_2009,
	title = {Robust control of individual nuclear spins in diamond},
	volume = {80},
	copyright = {http://link.aps.org/licenses/aps-default-license},
	issn = {1050-2947, 1094-1622},
	url = {https://link.aps.org/doi/10.1103/PhysRevA.80.050302},
	doi = {10.1103/physreva.80.050302},
	number = {5},
	urldate = {2025-07-22},
	journal = {Physical Review A},
	author = {Smeltzer, Benjamin and McIntyre, Jean and Childress, Lilian},
	month = nov,
	year = {2009},
	note = {Publisher: American Physical Society (APS)},
}

@article{felton_electron_2008,
	title = {Electron paramagnetic resonance studies of the neutral nitrogen vacancy in diamond},
	volume = {77},
	copyright = {http://link.aps.org/licenses/aps-default-license},
	issn = {1098-0121, 1550-235X},
	url = {https://link.aps.org/doi/10.1103/PhysRevB.77.081201},
	doi = {10.1103/physrevb.77.081201},
	number = {8},
	urldate = {2025-07-22},
	journal = {Physical Review B},
	author = {Felton, S. and Edmonds, A. M. and Newton, M. E. and Martineau, P. M. and Fisher, D. and Twitchen, D. J.},
	month = feb,
	year = {2008},
	note = {Publisher: American Physical Society (APS)},
}

@book{bracewell_fourier_1985,
	address = {Auckland},
	edition = {2. ed., 4. pr},
	series = {{McGraw}-{Hill} series in electrical engineering},
	title = {The {Fourier} transform and its applications},
	isbn = {978-0-07-007013-4 978-0-07-066196-7},
	publisher = {McGraw-Hill},
	author = {Bracewell, Ronald N.},
	year = {1985},
}

@article{canciani_analysis_2020,
	title = {An {Analysis} of the {Benefits} and {Difficulties} of {Aerial} {Magnetic} {Vector} {Navigation}},
	volume = {56},
	copyright = {https://creativecommons.org/licenses/by/4.0/legalcode},
	issn = {0018-9251, 1557-9603, 2371-9877},
	url = {https://ieeexplore.ieee.org/document/9076056/},
	doi = {10.1109/taes.2020.2987475},
	number = {6},
	urldate = {2025-07-28},
	journal = {IEEE Transactions on Aerospace and Electronic Systems},
	author = {Canciani, Aaron Joseph and Brennan, Christopher J.},
	month = dec,
	year = {2020},
	note = {Publisher: Institute of Electrical and Electronics Engineers (IEEE)},
	pages = {4161--4176},
}

@book{oppenheim_discrete-time_2010,
	address = {Upper Saddle River},
	edition = {3rd ed},
	title = {Discrete-time signal processing},
	isbn = {978-0-13-198842-2},
	publisher = {Pearson},
	author = {Oppenheim, Alan V. and Schafer, Ronald W.},
	year = {2010},
}

@book{horowitz_art_2024,
	address = {Cambridge, New York},
	edition = {Third edition, 20th printing with corrections},
	title = {The art of electronics},
	isbn = {978-0-521-80926-9},
	publisher = {Cambridge University Press},
	author = {Horowitz, Paul and Hill, Winfield},
	year = {2024},
}

@book{proakis_digital_2008,
	address = {Boston, Mass.},
	edition = {5. ed},
	title = {Digital communications},
	isbn = {978-0-07-295716-7},
	publisher = {McGraw-Hill},
	author = {Proakis, John G. and Salehi, Masoud},
	year = {2008},
}

@article{depellette_amplitude_2025,
	title = {Amplitude noise cancellation of microwave tones},
	volume = {96},
	issn = {0034-6748, 1089-7623},
	url = {https://pubs.aip.org/rsi/article/96/8/084705/3360903/Amplitude-noise-cancellation-of-microwave-tones},
	doi = {10.1063/5.0283567},
	number = {8},
	urldate = {2025-10-19},
	journal = {Review of Scientific Instruments},
	author = {Depellette, Joe and Rej, Ewa and Herbst, Matthew and Cutting, Richa and Liu, Yulong and Sillanp{\"a}{\"a}, Mika A.},
	month = aug,
	year = {2025},
	pages = {084705},
}

@book{axler_linear_2024,
	address = {Cham},
	series = {Undergraduate {Texts} in {Mathematics}},
	title = {Linear {Algebra} {Done} {Right}},
	isbn = {978-3-031-41025-3 978-3-031-41026-0},
	publisher = {Springer Nature},
	author = {Axler, Sheldon Jay},
	year = {2024},
}

@article{su_vector_2024,
	title = {Vector magnetocardiography using compact optically-pumped magnetometers},
	volume = {10},
	issn = {24058440},
	url = {https://linkinghub.elsevier.com/retrieve/pii/S2405844024051235},
	doi = {10.1016/j.heliyon.2024.e29092},
	number = {7},
	urldate = {2025-10-23},
	journal = {Heliyon},
	author = {Su, Shengran and Xu, Zhenyuan and He, Xiang and Zhang, Guoyi and Wu, Haijun and Gao, Yalan and Ma, Yueliang and Yin, Chanling and Ruan, Yi and Li, Kan and Lin, Qiang},
	month = apr,
	year = {2024},
	pages = {e29092},
}

@article{yang_new_2021,
	title = {A new wearable multichannel magnetocardiogram system with a {SERF} atomic magnetometer array},
	volume = {11},
	issn = {2045-2322},
	url = {https://www.nature.com/articles/s41598-021-84971-7},
	doi = {10.1038/s41598-021-84971-7},
	number = {1},
	urldate = {2025-10-23},
	journal = {Scientific Reports},
	author = {Yang, Yanfei and Xu, Mingzhu and Liang, Aimin and Yin, Yan and Ma, Xin and Gao, Yang and Ning, Xiaolin},
	month = mar,
	year = {2021},
	pages = {5564},
}

@book{budker_optical_2013,
	edition = {1},
	title = {Optical {Magnetometry}},
	copyright = {https://www.cambridge.org/core/terms},
	isbn = {978-1-107-01035-2 978-0-511-84638-0},
	url = {https://www.cambridge.org/core/product/identifier/9780511846380/type/book},
	urldate = {2025-10-23},
	publisher = {Cambridge University Press},
	editor = {Budker, Dmitry and Jackson Kimball, Derek F.},
	month = mar,
	year = {2013},
	doi = {10.1017/CBO9780511846380},
}

@article{poulsen_optimal_2022,
	title = {Optimal control of a nitrogen-vacancy spin ensemble in diamond for sensing in the pulsed domain},
	volume = {106},
	issn = {2469-9950, 2469-9969},
	url = {https://link.aps.org/doi/10.1103/PhysRevB.106.014202},
	doi = {10.1103/PhysRevB.106.014202},
	number = {1},
	urldate = {2025-11-04},
	journal = {Physical Review B},
	author = {Poulsen, Andreas F. L. and Clement, Joshua D. and Webb, James L. and Jensen, Rasmus H. and Troise, Luca and Berg-S{\o}rensen, Kirstine and Huck, Alexander and Andersen, Ulrik Lund},
	month = jul,
	year = {2022},
	pages = {014202},
}

@article{childress_coherent_2006,
	title = {Coherent {Dynamics} of {Coupled} {Electron} and {Nuclear} {Spin} {Qubits} in {Diamond}},
	volume = {314},
	issn = {0036-8075, 1095-9203},
	url = {https://www.science.org/doi/10.1126/science.1131871},
	doi = {10.1126/science.1131871},
	number = {5797},
	urldate = {2025-11-10},
	journal = {Science},
	author = {Childress, L. and Gurudev Dutt, M. V. and Taylor, J. M. and Zibrov, A. S. and Jelezko, F. and Wrachtrup, J. and Hemmer, P. R. and Lukin, M. D.},
	month = oct,
	year = {2006},
	pages = {281--285},
}

@article{dwyer_probing_2022,
	title = {Probing {Spin} {Dynamics} on {Diamond} {Surfaces} {Using} a {Single} {Quantum} {Sensor}},
	volume = {3},
	issn = {2691-3399},
	url = {https://link.aps.org/doi/10.1103/PRXQuantum.3.040328},
	doi = {10.1103/PRXQuantum.3.040328},
	number = {4},
	urldate = {2025-11-10},
	journal = {PRX Quantum},
	author = {Dwyer, Bo L. and Rodgers, Lila V.H. and Urbach, Elana K. and Bluvstein, Dolev and Sangtawesin, Sorawis and Zhou, Hengyun and Nassab, Yahia and Fitzpatrick, Mattias and Yuan, Zhiyang and De Greve, Kristiaan and Peterson, Eric L. and Knowles, Helena and Sumarac, Tamara and Chou, Jyh-Pin and Gali, Adam and Dobrovitski, V.V. and Lukin, Mikhail D. and De Leon, Nathalie P.},
	month = dec,
	year = {2022},
	pages = {040328},
}

\cleardoublepage

\onecolumngrid

\section*{Supplemental Material}
\setcounter{page}{1}
\subsection{Properties of NV Axes}
\label{vcr:supp:nvaxes}
While the NV axes form a basis, they are not orthogonal, which requires an additional layer of analysis to perform vector magnetometry.
As the spin resonance frequency is, to first order, sensitive to the projection of magnetic fields along its NV axis, we provide here an overview of the consequences of measuring a signal in a non-orthogonal basis.
Within the diamond coordinate system, the NV unit vectors $\hat{n}_i$, $i = 1,...,4$ can be chosen as any set of 4 non-parallel vectors from $V = \{[\begin{array}{ccc}\pm 1 & \pm 1 & \pm 1\end{array}]^T/\sqrt{3}\}$.
However, it is convenient to choose a set of vectors such that $\hat{n}_i\cdot\hat{n}_j = \tfrac{4}{3}\delta_{ij} - \frac{1}{3}$, such as
\begin{equation}
\hat{n}_1 = \frac{1}{\sqrt{3}}\left[\begin{array}{c}1 \\ 1\\ 1\end{array}\right],\;\; \hat{n}_2 = \frac{1}{\sqrt{3}}\left[\begin{array}{c}1 \\ -1\\ -1\end{array}\right], \;\; \hat{n}_3 = \frac{1}{\sqrt{3}}\left[\begin{array}{c}-1 \\ 1\\ -1\end{array}\right], \;\; \hat{n}_4 = \frac{1}{\sqrt{3}}\left[\begin{array}{c}-1 \\ -1\\ 1\end{array}\right].
\end{equation}
Throughout, we use the notation 
\begin{equation}
N_3 = \left[\begin{array}{ccc}\hat{n}_1 & \hat{n}_2 & \hat{n}_3\end{array}\right],\;\;N_4 = \left[\begin{array}{cccc}\hat{n}_1 & \hat{n}_2 & \hat{n}_3 & \hat{n}_4\end{array}\right].
\end{equation}
Because any 3 NV axes are linearly independent, $N_3$ is full rank.
Direct computation gives the important relation
\begin{equation}
    N_4 N_4^T = \frac{4}{3}I.
\end{equation}
This is true for any set of valid basis vectors $\hat{n}_i$, and can be verified checking all 16 possible sets of $\hat{n}_i$ from $V$.
Changing basis to another orthogonal coordinate system results in the transformation $N_4 \rightarrow M N_4$, where $M$ is an orthogonal matrix, so the product
\begin{equation}
N_4 N_4^T \rightarrow M^T N_4 N_4^T M^T = \frac{4}{3} M^T M = \frac{4}{3}I
\end{equation}
remains unchanged, meaning that this property is universal for selection of NV axes in any coordinate system.

Let $\vec{p}_4 = N_4^T \vec{B}$ be the vector of projections of $\vec{B}$ onto the NV axes. Computing $\vec{p}_4^T\vec{p}_4 = \vec{B}^TN_4N_4^T\vec{B} = \tfrac{4}{3}\vec{B}^T\vec{B}$ gives the important relation
\begin{equation}
\label{vcr:eq:4proj}
||\vec{B}||^2 = \frac{3}{4}||\vec{p}_4||^2,
\end{equation}
which allows the computation of the magnitude of applied fields by its projection onto each NV orientation.

In this work, however, we only measure the projection onto 3 NV orientations, requiring a slight modification to Eq.~\eqref{vcr:eq:4proj}. Consistent with the NV orientation labeling in this work, we enforce
\begin{equation}
\label{vcr:eq:order}
    |\hat{n}_1\cdot\vec{B}| \geq |\hat{n}_2\cdot\vec{B}| \geq |\hat{n}_3\cdot\vec{B}| > |\hat{n}_4\cdot\vec{B}|
\end{equation}
In this case, let $\vec{p}_3 = N_3^T\vec{B}$.
Because $N_3$ is full rank, $\hat{n}_4 = N_3\vec{a}$, for some vector $\vec{a}$.
It can be verified for all $\hat{n}_i$ which form a valid basis from $V$ that $\vec{a} = [\begin{array}{ccc} \pm 1 & \pm 1 & \pm 1 \end{array}]^T$.
$\vec{a}$ clearly does not change under rotations of the NV axis, so this must be true for all NV bases.
The final projection is then $p_4 = \vec{a}\cdot\vec{p}_3$.

In the laboratory, we do not know the orientation of the axes relative to each other and so when we measure $\vec{p}_3$ in the laboratory, the sign choices for each component are arbitrary.
This means that we need to derive an equation for $p_4$ which depends only on the absolute values of each component of $\vec{p}_3$.
It is clear for a given $\vec{p}_3$ and $\vec{a}$ that $\vec{a}\cdot\vec{p}_3$ can be written as a sum or difference of $|p_1|$, $|p_2|$, and $|p_3|$.
We thus write out all such possible combinations and choose the one consistent with our ordering in Eq.~\eqref{vcr:eq:order}.
From From Eq.~\eqref{vcr:eq:4proj}, $||\vec{B}||^2$ does not depend on an overall sign change in $\vec{p}_4$, allowing us to consider only a couple choices for $p_4$:
\begin{itemize}
    \item Case 1: $p_4 = ||p_1| + |p_2| + |p_3||$. This is obviously larger than $|p_1|$ and violates Eq.~\eqref{vcr:eq:order}.   \\
    \item Case 2: $p_4 = ||p_1| + |p_2| - |p_3||$. Since $|p_2| \geq |p_3|$, $|p_4| \geq |p_1|$, violating Eq.~\eqref{vcr:eq:order}.   \\
    \item Case 3: $p_4 = ||p_1| - |p_2| + |p_3||$. Since $|p_1| \geq |p_2|$, $|p_4| \geq |p_3|$, again violating Eq.~\eqref{vcr:eq:order}.   \\
    \item Case 4: $p_4 = ||p_1| - |p_2| - |p_3||$. The only remaining option for $p_4$ must be true.
\end{itemize}
From this, we arrive at the final result
\begin{equation}
\label{vcr:eq:3proj}
||\vec{B}||^2 = \frac{3}{4} (p_1^2 + p_2^2 + p_3^2 + (|p_1| - |p_2| - |p_3|)^2).
\end{equation}

\subsection{Reflection Coefficient Discussion}
\label{vcr:supp:refcoeff}
The reflection coefficient of the composite NV diamond and dielectric resonator structure is given by \cite{eisenach_cavity-enhanced_2021, wilcox_thermally_2022}
\begin{equation}
\label{vcr:eq:ref}
\Gamma = -1 + \frac{\kappa_{c1}}{\frac{\kappa_{c0} + \kappa_{c1}}{2} + i(\omega_d - \omega_c) + \Pi},
\end{equation}
where $\kappa_{c0}$ and $\kappa_{c1}$ are the intrinsic dielectric resonator linewidth and the input couplings rates, $\kappa_c = \kappa_{c0} + \kappa_{c1}$ is the loaded linewidth, $\omega_c$ is the dielectric resonator resonant frequency, $\omega_d$ is the MW drive frequency, and $\Pi$ is the spin-photon interaction term. The interaction term $\Pi$ may be written as
\begin{equation} \label{vcr:eqn:spinphoton}
    \Pi = \frac{g_s^2 N}{\frac{\kappa_s}{2} + i(\omega_d - \omega_s) + \frac{g_s^2 n_{\mathrm{cav}}\kappa_s/(2\kappa_{\mathrm{op}})}{\frac{\kappa_s}{2} - i(\omega_d - \omega_s)}},
\end{equation}
where $\kappa_s$ is the spin resonance linewidth, $\kappa_{\mathrm{op}}$ is the optical pumping rate, $n_{\mathrm{cav}}$ is the average number of MW photons in the cavity, and $N$ is effective number of polarized spins.
$g_s = \tfrac{\gamma n_{\perp}}{2}\sqrt{\tfrac{\hbar\omega_c \mu_0}{V_{\mathrm{cav}}}}$ is the single spin-photon coupling, where $\gamma$ is the electron gyromagnetic ratio, $V_{\mathrm{cav}}$ is the resonator modal field volume, $\mu_0$ is the vacuum permeability, and $0\leq n_{\perp} \leq 1$ is a geometric factor, which occurs because only fields transverse to the spin quantization axis can drive transitions.

\begin{figure}
    \centering
    \includegraphics[width=2.0in]{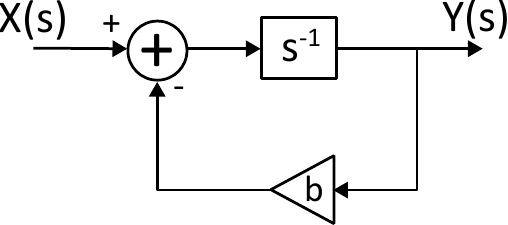}
    \caption{\textbf{Simple integrator.} Example of a simple integrator described by $Y(s) = \frac{1}{s + b}X(s)$.}
    \label{vcr:fig:integrator}
\end{figure}

\begin{figure}
    \centering
    \includegraphics[width=3.2in]{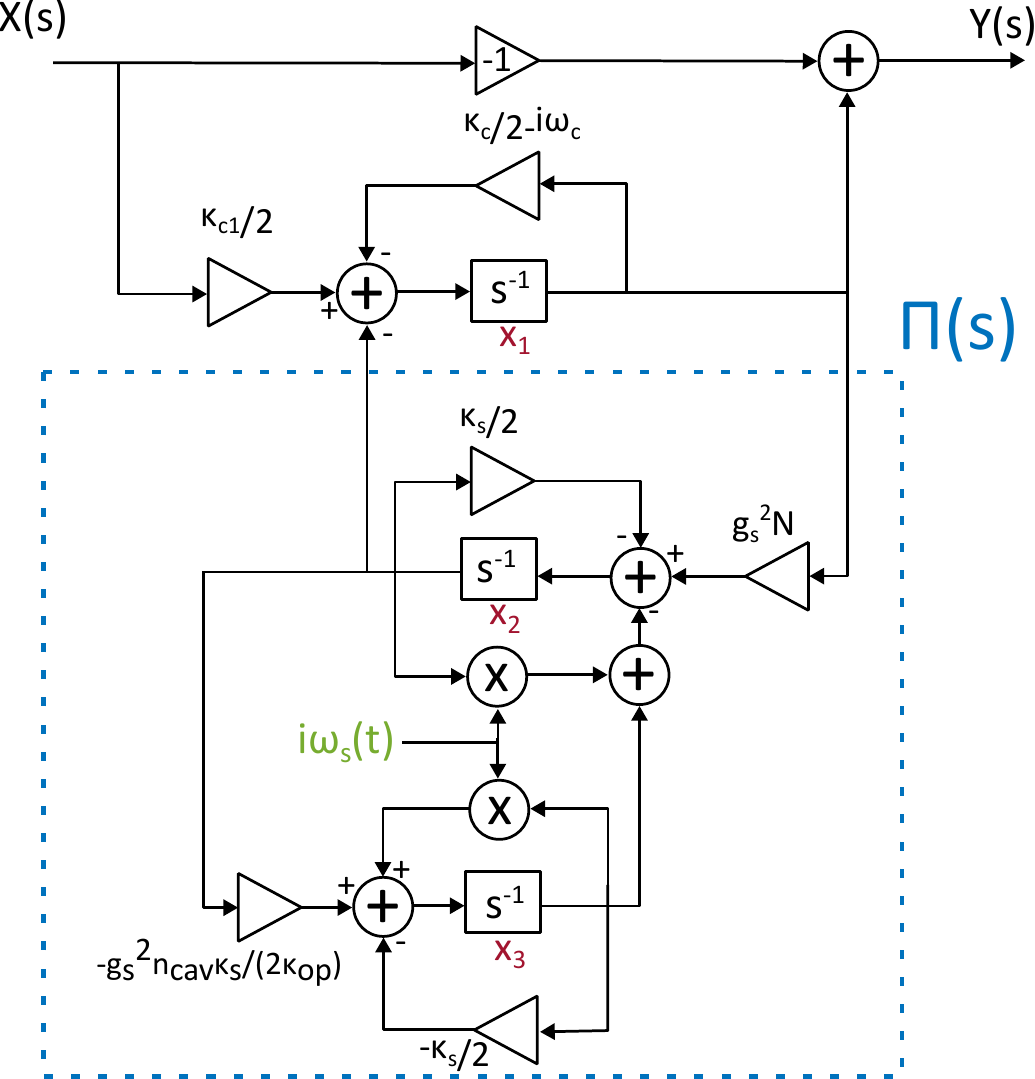}
    \caption{\textbf{State space representation of the reflection coefficient.} The reflection coefficient can be represented by a series of integrators.}
    \label{vcr:fig:refblockdiagram}
\end{figure}

Drawing inspiration from control theory, we replace $i\omega_d \rightarrow s$.
The presence of an $s$ term in the denominator indicates the presence of an integrator.
From Eqs.~\eqref{vcr:eq:ref} and \eqref{vcr:eqn:spinphoton}, we see that there must be 3 integrators to fully describe the state of the reflection coefficient.
As a simple example, the system $Y(s) = \frac{1}{s + b(s)}X(s)$ can be written $Y(s) = s^{-1}(X(s) - b Y(s))$, which is represented by the block diagram in Fig.~\ref{vcr:fig:integrator}.
This representation can be generalized to represent the full reflection coefficient of the cavity, as in Fig.~\ref{vcr:fig:refblockdiagram}.
The dashed portion represents the spin-photon interaction term $\Pi$, while the external input, $\omega_s(t)$ is the time-dependent spin resonance frequency.

If we were to take some input signal $x(t)$ and apply it to the input of Fig.~\ref{vcr:fig:refblockdiagram}, then the output $y(t)$ will be signal reflected off the cavity.
In the special case of applying a sinusoid $x(t) = e^{i\omega_d t}$ and holding $\omega_s(t)$ constant, then $y(t) = \Gamma x(t)$, where $\Gamma$ is evaluated using Eq.~\eqref{vcr:eq:ref}.
For time varying $\omega_s(t)$, Eq.~\eqref{vcr:eq:ref} can no longer be used to predict the reflected signal, but the representation in Fig.~\ref{vcr:fig:refblockdiagram} remains a valid way to simulate the system.
This allows for certain time-dependent effects, such as spin depolarization by the applied MWs, to be investigated.
It should be noted, however, that the incorporation of $\omega_s(t)$ through multiplication means that the system is not linear, so simulation can be time consuming.

\begin{figure}
    \centering
    \includegraphics[width=3.2in]{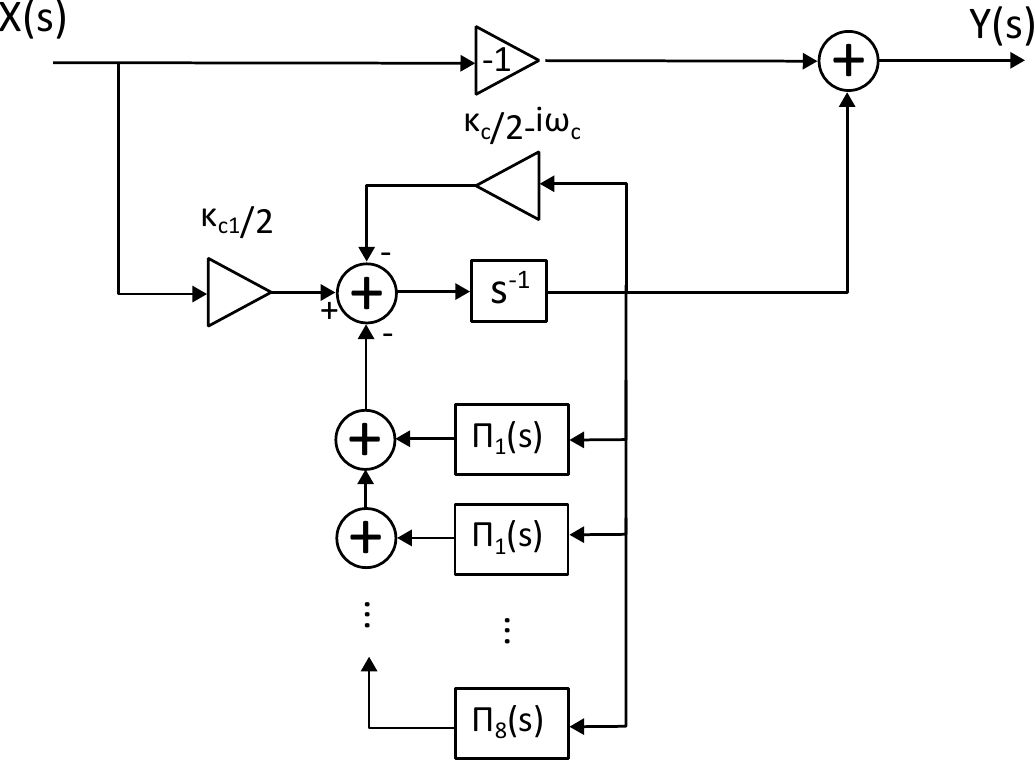}
    \caption{\textbf{Including multiple orientations in the reflection coefficient.} To account for multiple NV orientations, the spin interaction terms, $\Pi_i$ are summed together in the feedback term.}
    \label{vcr:fig:refpis}
\end{figure}

As written, the $\Pi$ term considers only a single spin transition in a single NV orientation.
As there are 4 NV orientations and two spin transitions possible in each, this requires a total of 8 interaction terms, which we number $\Pi_1$ through $\Pi_8$.
These can be incorporated into the simulation by summing all 8 together in the feedback term for the first integrator as in Fig.~\ref{vcr:fig:refpis}.

As noted previously, three integrators ($x_1$, $x_2$, and $x_3$) are required to simulate the reflection coefficient, and are labelled in Fig.~\ref{vcr:fig:refblockdiagram}.
Each corresponds to a different physical quantity as follows: $x_1$ is the number of MW photons inside the cavity, $x_2$ is the number of NV centers interacting with the MW photons (that is, undergoing a spin transition), and $x_3$ is the number of depolarized NV centers.
As increasing amounts of MW power are applied to the cavity, the number of depolarized spins increases, meaning that there are fewer NV centers available to interact with the MW photons and affect the reflection coefficient.
This effect causes the asymmetry in Fig.~\ref{vcr:fig:timeseries} as follows: when an NV orientation is addressed, the MW photons act to depolarize NV centers with that orientation, meaning the next time it is addressed, the reflection coefficient peak will be smaller.

To form a more accurate simulation, the output of $x_1$ should be used to set $n_{\text{cav}}$, but this further exacerbates the non-linearities in the system.

\subsection{Magnetometer Calibration}
\label{vcr:supp:calibration}
To perform calibration, we assume that $\tau_i$ vary linearly with the applied field, which is valid when external fields are small compared to the bias field, but breaks down for larger fields.
This non-linearity is primarily driven by the bias field slope varying with time.
For large enough applied fields, this can result in a change in the peak shape as parts of it begin to experience different bias field slopes.

\begin{figure}
    \centering
    \begin{tabular}{c}
    \includegraphics[width=3.7in]{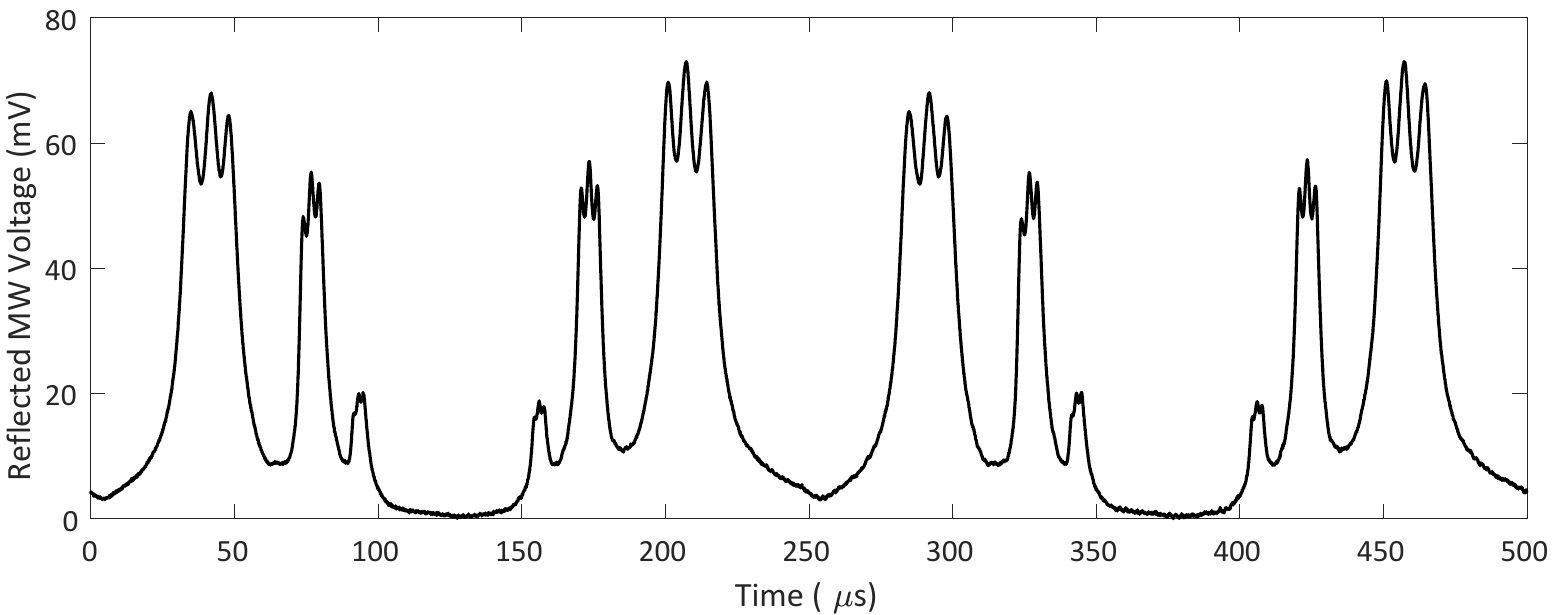} \\
    (a) Raw reflection coefficient data against time \\ 
    \includegraphics[width=3.7in]{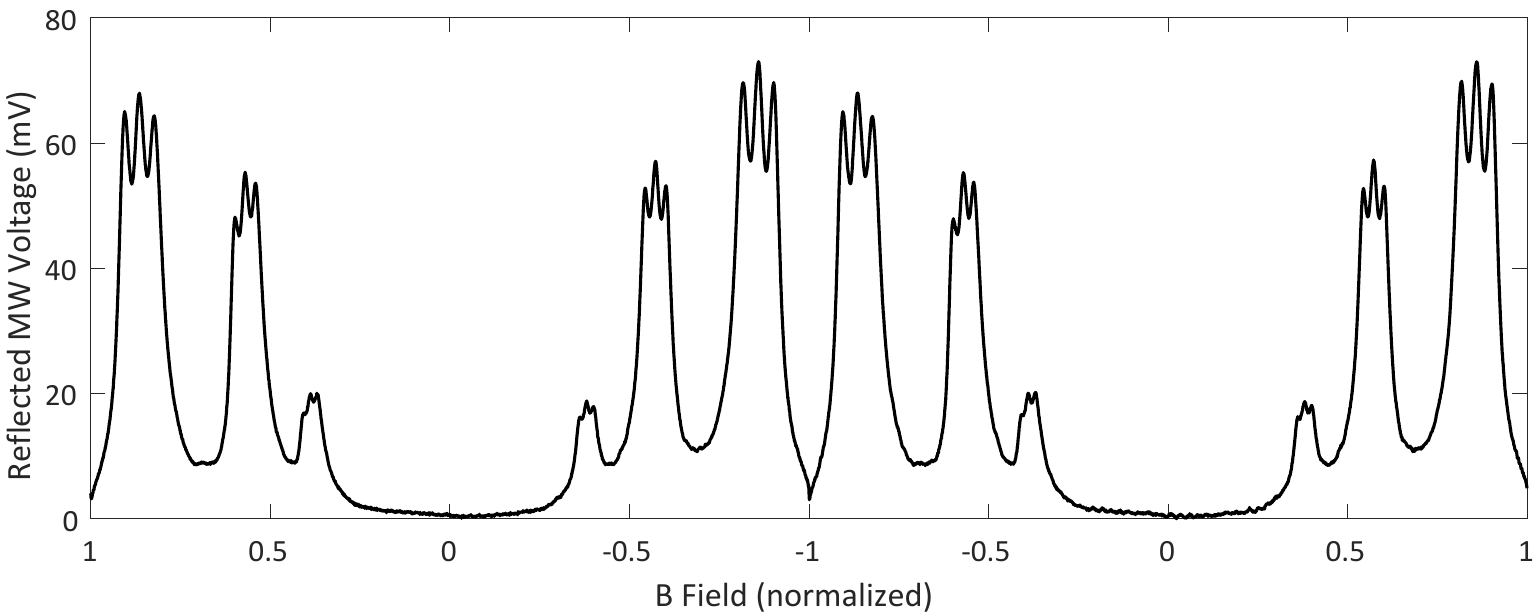} \\
    (b) Linearized reflection coefficient against bias field
\end{tabular}
    \caption{\textbf{Linearizing magnetometer response.} To linearize the response, the reflection coefficient data is resampled and plotted against the applied bias field. Compared to the non-linearized data, the main effect is shifting the locations of the peaks and making NV orientation 3 appear less broad.}
    \label{vcr:fig:linearization}
\end{figure}

However, as the applied bias field is known, the recorded reflection coefficient can be resampled and plotted against the bias field, instead of against time, as in Fig.~\ref{vcr:fig:linearization}.
Resampling has the largest effect on the NV orientation 3, which is addressed near the extrema of the bias field -- its peak width is  artificially broadened due to a reduced slope in the bias field.
The resulting linearized data is no longer affected by the non-uniformity in the bias field sweep, making it more robust for magnetometry.
Note that in Fig.~\ref{vcr:fig:linearization}, there is still a variation in peak widths across orientations -- this is because the NV spin resonances depend on the projection of the bias field along the NV axis.
Correspondingly, the NV orientations with the smallest bias field projection have the broadest peaks.

Assuming now that the magnetometer is operating within the linear regime, calibration consists of finding two matrices $A$ and $C$ defined by Eq.~\eqref{vcr:eq:A}, rewritten below for convenience
\begin{equation}
\vec{B}_T = (I + C)A\left[\begin{array}{c}\tau_1 \\ \tau_2 \\ \tau_3 \end{array} \right],
\end{equation}
where $A$ accounts for rotating the NV coordinate system to the lab coordinates and for the bias field strength, while $C$ accounts for non-idealities in the system, which are discussed in more detail later.

To calibrate the magnetometer and transform from the NV axis basis, which is not orthogonal, to the $x$-, $y$-, $z$-basis, which is orthogonal, three orthogonal magnetic fields at differing orientations and known magnitude are applied to the sensor -- $f_x = $15~Hz in the $x$-direction, $f_y = $25~Hz in the $y$-direction, and $f_z = $35~Hz in the $z$-direction.
The magnitude of the applied test fields are measured using a vector hall-effect magnetometer, and each set to be 1~\micro T$_{\text{rms}}$.
By choosing different frequencies for each direction, these fields can be differentiated and the calibration can be performed with a single measurement.

With the test magnetic fields applied to the sensor, $\tau_i(n T)$ is recorded for $i = 1,2,3$, where $T = 500$~\micro s, and integer $n$.
In practice, the calibration is performed using 10~s of data, or $20\times 10^3$ samples.
The actual duration of the signal is not important, but using a duration which permits an integer number of periods of each test field allows cleaner analysis.
A Fast-Fourier-Transform is then used to compute the amplitude spectrum with phase of each NV orientation, denoted by $R_i(\omega) = A_i(\omega)e^{i\phi_i(\omega)}$, with $A_i(\omega) > 0$ and $\phi_i(\omega)$ are both real.
Because a single external field drives the magnetometer at $f_x$, it is true, to good approximation, that $\phi_i(2\pi f_x)$ differ only by multiples of $\pi$ -- differing by odd multiples of $\pi$ correspond to negation.
We can therefore rotate all three $R_i(2\pi f_x)$ by the same constant and obtain $F_i(2\pi f_x) = \pm A_i(2\pi f_x)$, where the sign is chosen based on whether the alignment differs by an even or odd multiple of $\pi$. An analogous statement holds for $f_y$ and $f_z$.

Now form the matrix
\begin{equation}
M = \left[\begin{array}{ccc}
F_1(2\pi f_x)  & F_1(2\pi f_y) & F_1(2\pi f_z)  \\
F_2(2\pi f_x)  & F_2(2\pi f_y) & F_2(2\pi f_z)  \\
F_3(2\pi f_x)  & F_3(2\pi f_y) & F_3(2\pi f_z)
\end{array}\right]
\end{equation}
the columns of which represent the response of the magnetometer to the applied fields.

First, we solve for the $A$ matrix, which accounts for the NV geometry and applied bias field strength.
To do this, we need a functional form for $A$ so that it corresponds to a valid NV geometry.
If the peak locations, $\tau_i$, are reported in normalized, linearized units as in Fig.~\ref{vcr:fig:linearization}b, then we expect $\tau_i$ to be given by
\begin{equation}
\tau_i = \frac{\hat{n}_i^TU^T \vec{B}}{\hat{n}_i^TU^T \vec{B}_0}
\end{equation}
where $\vec{B}$ is the applied field, $\vec{B}_0$ is the bias field, $\hat{n}_i$ is NV unit vector $i$ as specified in Supplemental Material Sec.~\ref{vcr:supp:nvaxes}, and $U$ is an orthogonal matrix to transform the NV axes from the diamond coordinate from to the lab coordinate frame.
Since $A^{-1}$ will map applied magnetic fields to peak locations, this gives the representation
\begin{equation}
\label{vcr:eq:calgeom}
A^{-1} = \left[\begin{array}{ccc}\hat{n}_1^TU^T\vec{B}_0 & 0 & 0 \\ 0 & \hat{n}_2^TU^T\vec{B}_0 & 0 \\ 0 & 0 & \hat{n}_3^TU^T\vec{B}_0\end{array}\right]^{-1}N_3^T U^T,
\end{equation}
where $N_3$ is a matrix whose columns are $\hat{n}_i$.

Letting $B_{\text{app}} = 1$~\micro T$\times I$, we know that, ignoring the $C$ matrix, that we should have $B_{\text{app}} = AM$.
We can then numerically solve for the orthogonal matrix $U$ and bias field strength $B_0$ which minimizes the error $|B_{\text{app}} - AM|$.
However, the $A$ matrix does not entirely orthogonalize the system, that is $AM$ is not a diagonal matrix.
We thus use an additional calibration matrix $C$, which is computed by
\begin{equation}
C = B_{\text{app}}M^{-1}A^{-1} - I.
\end{equation}
The resulting calibration by $(I + C)A$ ensures that the orthogonal input test fields will result in orthogonal measurement outputs.
When computed, the components of $C$ have magnitude less than 0.16, which is commensurate with the estimated magnitude of effects it is used to correct, as detailed below.

The $C$ matrix accounts for two main effects which break orthogonality.
The first is that due to the overlapping peaks seen in Fig.~\ref{vcr:fig:timeseries}, the movement of one NV orientation's peak location can cause an apparent shift in the location of an adjacent orientation.
This effect will scale in proportion to the second derivative of the tail of the interfering orientation,
and is estimated to have a 10\% coupling from $\tau_2$ into $\tau_1$.
The second is that due to strong transverse components of the bias field, the NV resonant frequency becomes sensitive to transverse magnetic fields.
This effect originates from a quadratic dependence of the resonant frequency depending on the transverse magnetic fields, as detailed in Ref.~\cite{barry_sensitivity_2020}.
When the bias field is much stronger than the test field, the resonant frequency is well approximated as an affine function of the longitudinal and transverse test magnetic fields.
For NV orientation 3, where this effect is largest due to its large transverse bias field component, the resonant frequency is expected to be 10\% as sensitive to transverse fields as to longitudinal fields.
Both of the above effects are expected to be linear for small test fields, meaning that the linear correction matrix $C$ can properly correct them.



A useful diagnostic tool is treating each NV orientation as a single-axis vector sensor.
Eq.~\eqref{vcr:eq:calgeom} can easily be modified for this purpose giving
\begin{equation}
\label{vcr:eq:calsingle}
A_{\text{axis}} = \left[\begin{array}{ccc}\hat{n}_1^TU^T\vec{B}_0 & 0 & 0 \\ 0 & \hat{n}_2^TU^T\vec{B}_0 & 0 \\ 0 & 0 & \hat{n}_3^TU^T\vec{B}_0\end{array}\right].
\end{equation}
As we do not know the NV orientations in the lab frame, we cannot apply test fields along each NV orientation to compute this matrix.
Instead, we look at the hyperfine splitting for each NV orientation, which allows an estimation of $\hat{n}_i^TU^T\vec{B}_0$.

NV centers in $^{14}$N experience a hyperfine splitting between 2.12~MHz and 2.32~MHz \cite{loubser_electron_1978, he_paramagnetic_1993, felton_electron_2008, steiner_universal_2010, smeltzer_robust_2009}.
We choose the average value of the presented literature values at $A_{||} = 2\pi\times 2.22$~MHz, with corresponding magnetic field $B_{\text{HF}} = A_{||}/\gamma$.
Then we can approximate
\begin{equation}
\label{vcr:eq:deltab1}
\hat{n}_i^TU^T\vec{B}_0 = \frac{B_{\mathrm{HF}}}{\Delta B_{i, \tau}}
\end{equation}
where $\Delta B_{i, \tau}$ is the hyperfine splitting of NV orientation $i$ in normalised units as in Fig.~\ref{vcr:fig:linearization}b.
This allows computation of $A_{\text{axis}}$ from Eq.~\eqref{vcr:eq:calsingle}.
Furthermore, Eq.~\eqref{vcr:eq:deltab1} is of the form required for Eq.~\eqref{vcr:eq:3proj}, allowing computation of $B_0$, and subsequently $\hat{n}_i^TU^T\hat{B}_0$.
Using this procedure, we estimate $\hat{n}_i^TU^T\hat{B}_0$ to be $0.85$, $0.61$, and $0.44$ for NV orientations 1, 2, and 3, respectively.

Using the $A$ matrix, the bias field strength is estimated to be 23~G$_{rms}$.
When computed using the hyperfine method above with the $A_{\text{axis}}$ matrix, we estimate the bias field strength to be 27~G$_{\text{rms}}$.
These two values agree to within 17\%, which is of similar magnitude to the corrections introduced in the $C$ matrix, suggesting the discrepancy caused from the non-idealities.
The mean of these two values is used as the estimate for the bias field given in Sec.~\ref{vcr:sec:setup}.

\subsection{Magnetometer Sensitivity in an Orthogonal Basis}
\label{vcr:supp:orthnoise}
We wish to find how the sensitivity depends on orientation when we convert to an orthogonal basis. To that end, we compute the sensitivity along an arbitrary axis $\vec{v}$ with $||\vec{v}|| = 1$, then bound this sensitivity.

Let $\vec{\tau}' = [\begin{array}{ccc}\tau_1 & \tau_2 & \tau_3\end{array}]^T + \vec{e}$, where $\vec{e}$ is vector of measurement noise in each NV orientation. The sensitivity of NV orientation $i$ is then determined by $\mathbb{E}e_i^2$. Define the covariance matrix as $\sigma^2 = \mathbb{E}\vec{e}\vec{e}^T$. $\sigma^2$ can be computed empirically from recorded data.

Our noisy estimate of the magnetic field in an orthogonal matrix is given using Eq.~\eqref{vcr:eq:A} as $\vec{B}' = (I + C)A\vec{\tau}'$. The magnetic field component along $\vec{v}$ is then given by $B_v = \vec{v}^T\vec{B}'$, for any unit length vector $\vec{v}$. The variance of $B_v$ is then
\begin{equation}
\mathbb{E}(B_v - \mathbb{E}B_v)^2 = \mathbb{E}(\vec{v}^T(I + C)A\vec{e})^2 = \vec{v}^T(I + C)A\sigma^2 A^T(I + C^T)\vec{v}.
\end{equation}

Using eigenvalue decomposition, write $M \Sigma M^T = (I + C)A\sigma^2 A^T(I + C^T)$, where $\Sigma$ is a diagonal matrix of positive, real eigenvectors and $M$ is orthogonal due to the spectral theorem as $(I + C)A\sigma^2 A^T(I + C^T)$ is normal \cite{axler_linear_2024}. Then
\begin{equation}
\mathbb{E}(B_v - \mathbb{E}B_v)^2 = \vec{v}^TM\Sigma M^T\vec{v} = \vec{u}^T\Sigma \vec{u},
\end{equation}
where $\vec{u} = M^T\vec{v}$. Because $M$ is orthogonal, $\vec{u}$ is also a unit vector which can be chosen arbitrarily by setting $\vec{v} = M^T\vec{u}$.
As $\Sigma$ is diagonal, it is clear that 
\begin{equation}
\lambda_{\text{min}} \leq \mathbb{E}(B_v - \mathbb{E}B_v)^2 \leq \lambda_{\text{max}},
\end{equation}
where $\lambda_{\text{max}}$ and $\lambda_{\text{min}}$ are the maximimal and minimal eigenvalues of $(I + C)A\sigma^2 A^T(I + C^T)$.

Using the measured $\sigma^2$ from the recorded data gives sensitivity eigenvalues 5.6~nT/\rthz, 530~pT/\rthz, and 210~pT/\rthz. These are similar to the NV axis orientation sensitivities of 5.3~nT/\rthz, 580~pT/\rthz, and 240~pT/\rthz. The best orthgonal direction is slightly better than the best NV orientation because multiple NV orientations are combined to improve the measurement.
The worst orthogonal sensitivity is slightly worse than the worst NV orientation.
This worst orthogonal sensitivity direction is orthogonal to the two good NV orientations meaning it only receives information from the bad NV orientation.
It is, however, not parallel to the bad NV orientation, so its noise is increased compared to this bad orientation.



\subsection{Bias Field Noise Suppression and Magnetometer Frequency Response}
\begin{figure*}[t] 
\hspace{-2mm}
\begin{minipage}[b]{1\textwidth}
\vspace{10mm}
\end{minipage}

\begin{minipage}[b]{0.45\textwidth}
\begin{Overpic}{\put(-110, 8){\includegraphics[width=3.2in]{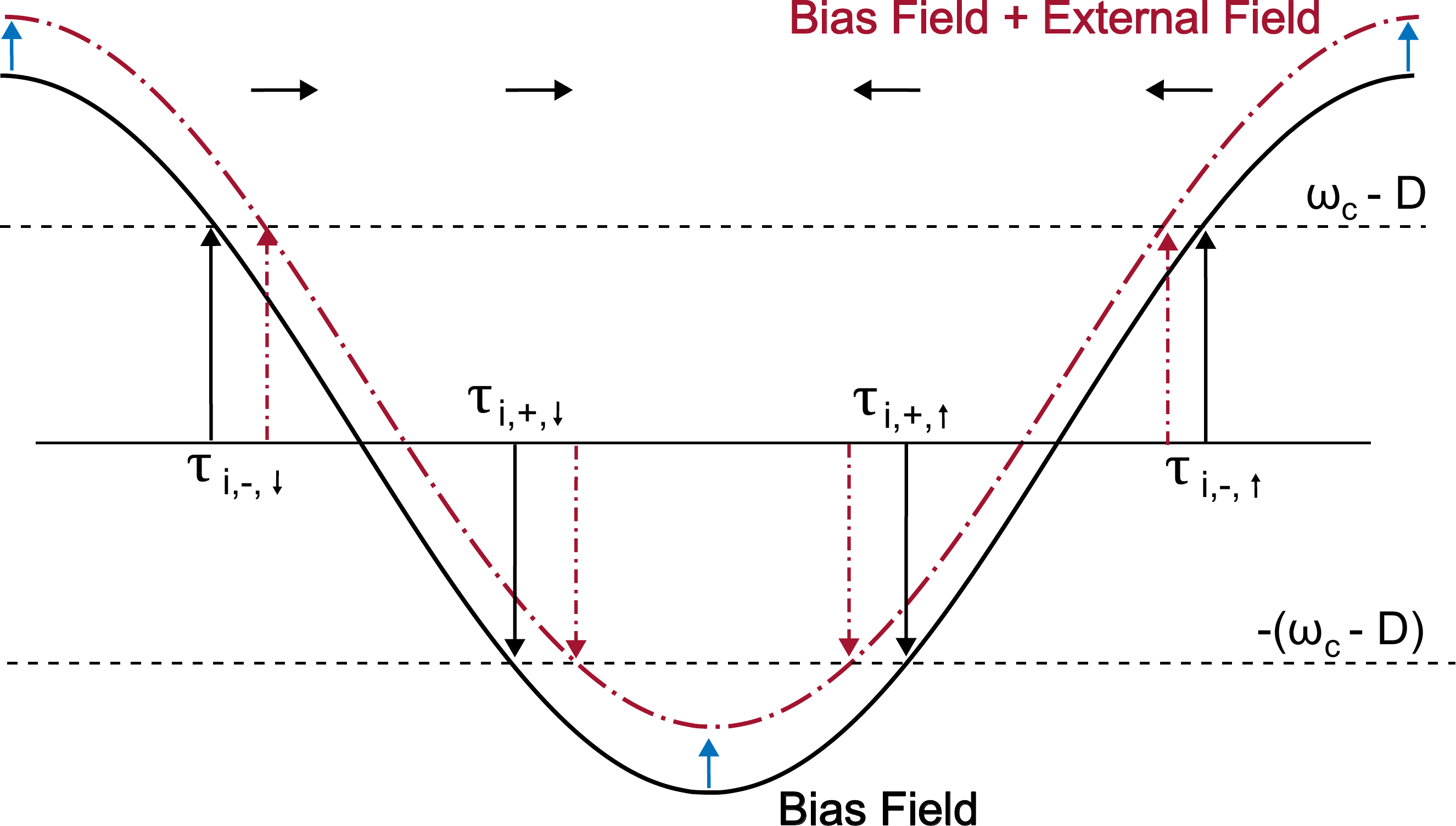}}} \put(-80,104){\textbf{a)}} 
\end{Overpic}
\end{minipage}
\;
\begin{minipage}[b]{0.45\textwidth}
\begin{Overpic}{\put(-110, 0){\includegraphics[width=3.2in]{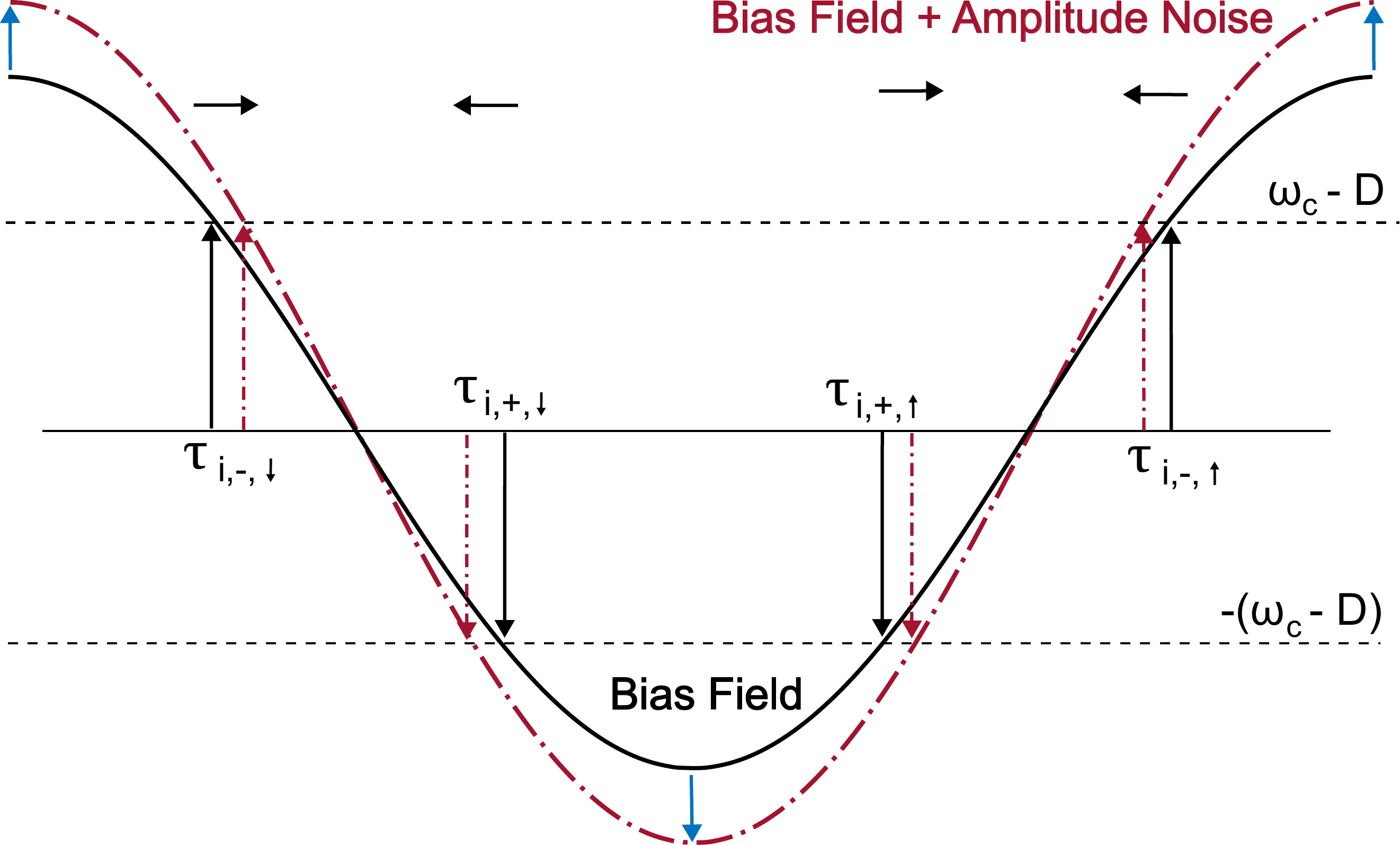}}} \put(-78,105){\textbf{b)}} 
\end{Overpic}
\end{minipage}
\;
\begin{minipage}[b]{0.45\textwidth}
\begin{Overpic}{\put(-230, 0){\includegraphics[width=3.2in]{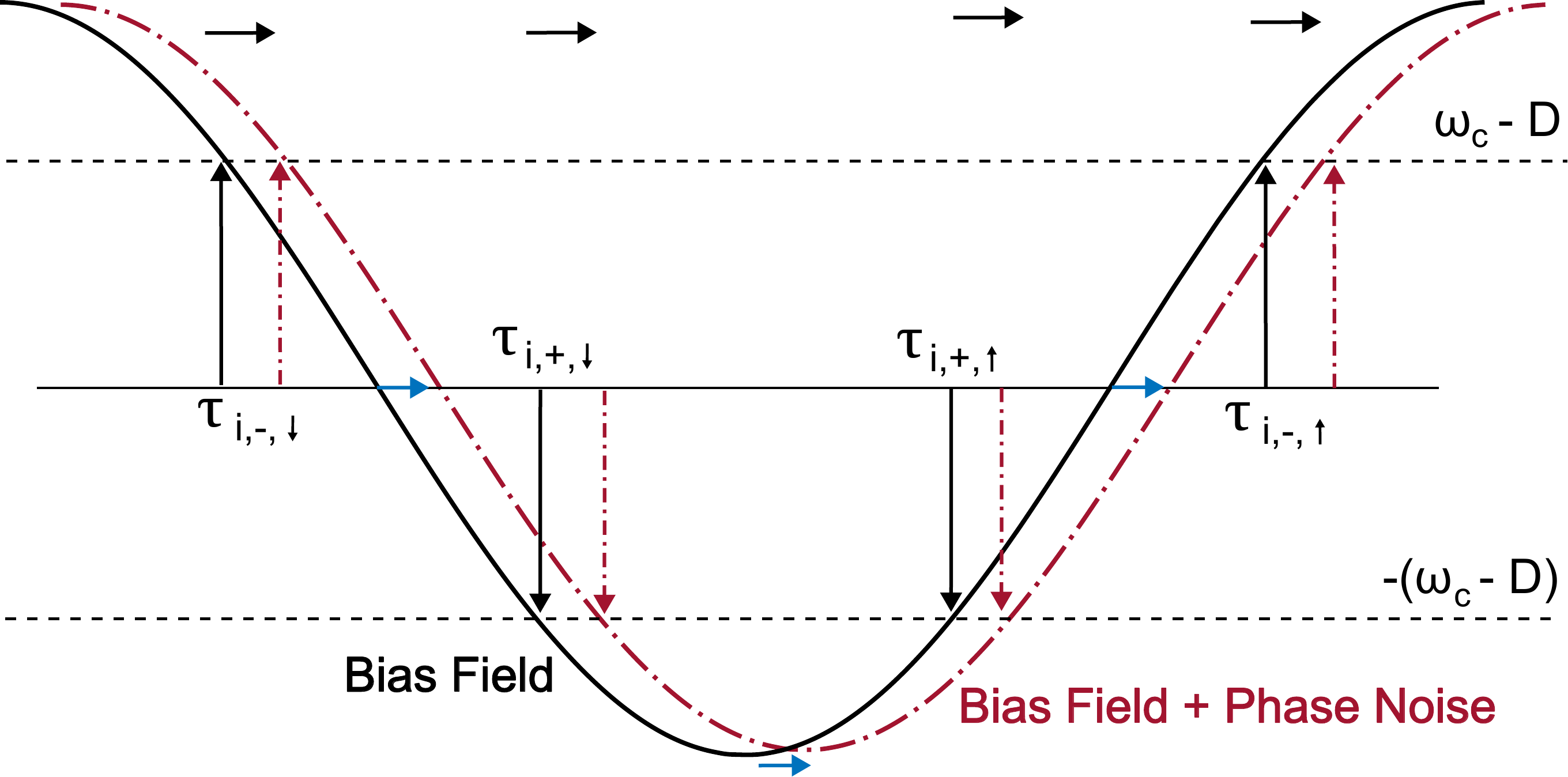}}} \put(-200,108){\textbf{c)}} 
\end{Overpic}
\end{minipage}
\caption{\textbf{Bias field noise suppression.}
\textbf{a)} External magnetic field response. When a low frequency external magnetic field is applied, this appears as a constant shift in the bias field. As a result, the times $\tau_{i, \pm, \uparrow}$ shift in the same direction, while the times $\tau_{i,\pm,\downarrow}$ shift in the opposite direction.
\textbf{b)} Bias field amplitude noise response. Amplitude noise results in an overall scaling of the bias field strength. As a result, the times $\tau_{i, -,\downarrow}$ and $\tau_{i, +,\uparrow}$ shift in the same direction, while times $\tau_{i, +,\downarrow}$ and $\tau_{i, -,\uparrow}$ shift in the opposite direction.
\textbf{c)} Bias field phase noise. Phase noise results in a shift in time of the bias field. As a result, all times shift in the same direction.}
\label{vcr:fig:biasnoise}
\end{figure*}
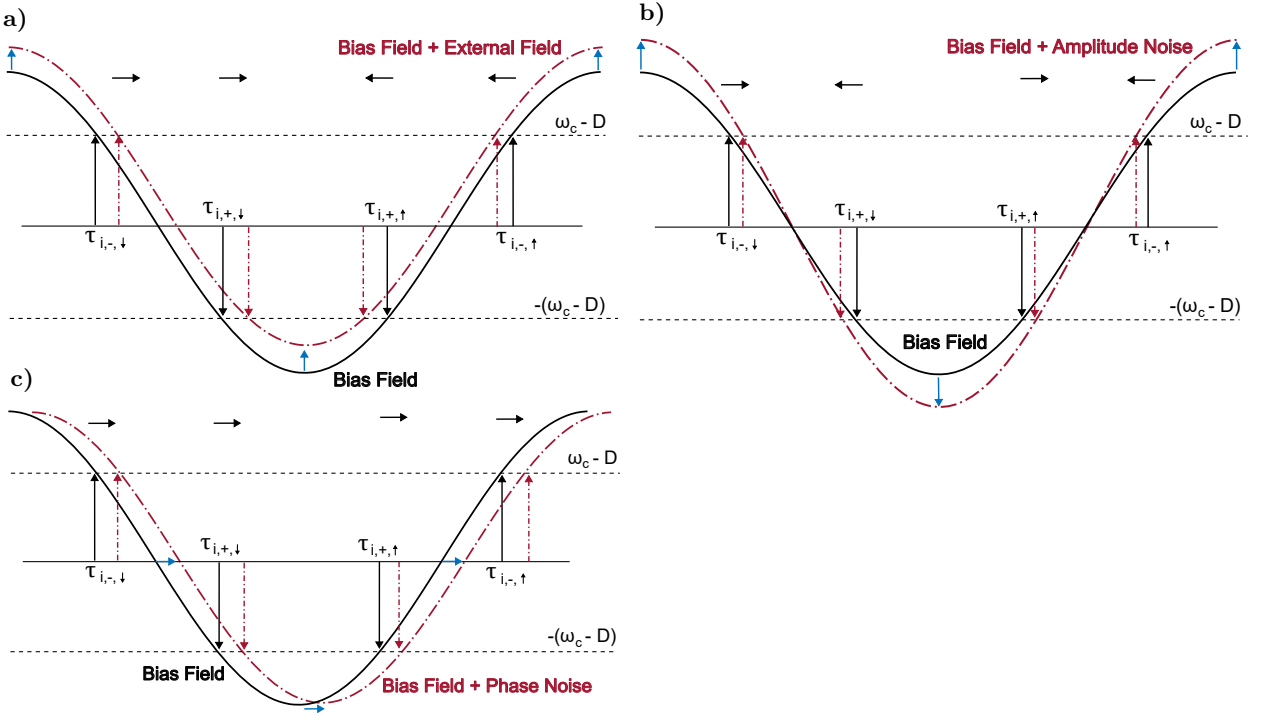

\label{vcr:supp:biasnoise}
When a noisy bias field is applied to the diamond, it behaves qualitatively differently from an external magnetic field which we want to sense.
Define $\tau_{i, \pm,\uparrow}$ as the time at which the peak for NV orientation $i$ is observed while addressing the $|m_s = 0\rangle\leftrightarrow|m_s = \pm 1\rangle$ spin transition with a positive slope bias field, and  $\tau_{i, \pm,\uparrow}$ as the time at which the peak for NV orientation $i$ is observed while addressing the $|m_s = 0\rangle\leftrightarrow|m_s = \pm 1\rangle$ spin transition with a negative slope bias field.
As seen in Fig.~\ref{vcr:fig:biasnoise}, there are three different patterns to how the resonance times $\tau_{i, \pm,\uparrow/\downarrow}$ shift, depending on whether we have an external field, amplitude noise on the bias field, or phase noise on the bias field.
Based on the shift signatures seen in Fig.\ref{vcr:fig:biasnoise}, we form the quantity
\begin{equation}
\label{vcr:eq:tau}
\tau_i = \frac{\tau_{i, -, \downarrow} + \tau_{i, +, \downarrow} - \tau_{i, +, \uparrow} - \tau_{i, -, \uparrow}}{4}
\end{equation}
which constructively combines the temporal shifts from an external magnetic field, while destructively combining both the amplitude and phase noise.
As we will show, this results in a nearly flat response to external magnetic fields up to $\approx$~400~Hz, while offering excellent phase noise and amplitude noise suppression.

We represent the bias field $\vec{B}_B(t)$ as
\begin{equation}
\label{vcr:eq:bnoisy}
\vec{B}_B(t) = \vec{B}_0\cos\omega_m t + \Delta \vec{B}(t),
\end{equation}
where $\vec{B}_0$ is the bias field amplitude and direction, and $\Delta \vec{B}(t)$ accounts for bias field noise and external magnetic fields.
Denote the measured peak times with a noise-free bias field and no external field ($\Delta\vec{B}(t) = 0$) as $\tau_{i, \pm,\uparrow/\downarrow, 0}$.
The condition that our spins are resonant and a peak occur at $\tau_{i, \pm,\uparrow/\downarrow, 0}$ is then
\begin{equation}
\label{vcr:eq:b0def}
\gamma \vec{B}_0 \cdot \hat{n}_i \cos(\omega_m \tau_{i, \pm,\uparrow/\downarrow, 0}) = \mp (\omega_c - D).
\end{equation}
Taylor expanding Eq.~\eqref{vcr:eq:bnoisy} and simplifying using Eq.~\eqref{vcr:eq:b0def} yields
\begin{equation}
\label{vcr:eq:btaylor}
\gamma \vec{B}_B(\tau_{i, \pm,\uparrow/\downarrow, 0} + \Delta\tau_{i, \pm,\uparrow/\downarrow})\cdot \hat{n}_i \approx \mp (\omega_c - D) - \Delta\tau_{i, \pm,\uparrow/\downarrow} \omega_m \gamma \vec{B}_0\cdot \hat{n}_i \sin\omega_m \tau_{i, \pm,\uparrow/\downarrow, 0} + \Delta\vec{B}(t)\cdot\hat{n}_i.
\end{equation}
Let $\tau_{i, \pm,\uparrow/\downarrow} = \tau_{i, \pm,\uparrow/\downarrow, 0} + \Delta\tau_{i, \pm,\uparrow/\downarrow}$.
Solving for $\Delta\tau_{i, \pm,\uparrow/\downarrow}$ by setting Eq.~\eqref{vcr:eq:btaylor} to $\pm(\omega_c - D)$ as was done in Eq.~\eqref{vcr:eq:b0def} yields
\begin{equation}
\Delta\tau_{i, \pm,\uparrow/\downarrow} = \frac{\Delta\vec{B}(t)\cdot\hat{n}_i}{\omega_m \vec{B}_0\cdot \hat{n}_i \sin\omega_m \tau_{i, \pm,\uparrow/\downarrow, 0}}.
\end{equation}
Define
\begin{equation}
\beta_i^2 =\left(\frac{\omega_c - D}{\gamma\vec{B}_0\cdot \hat{n}_i}\right)^2, \;\;\;\;
\alpha_i^2 = 1 - \beta_i^2
\end{equation}
with signs chosen so that
\begin{equation}
\cos(\omega_m \tau_{i, \pm,\uparrow/\downarrow, 0}) = \pm\beta_i,\;\;\;\; \sin(\omega_m \tau_{i, \pm,\uparrow/\downarrow, 0}) = s_{\uparrow/\downarrow}\alpha_i,
\label{vcr:eq:alphabetasincos}
\end{equation}
where $s_{\uparrow} = -1$ and $s_{\downarrow} = 1$.
We then arrive at the expression
\begin{align}
\Delta \tau_i &= \frac{\Delta\tau_{i,-,\downarrow} + \Delta\tau_{i,+,\downarrow} - \Delta\tau_{i,+,\uparrow} - \Delta\tau_{i,-,\uparrow}}{4}  \\
&= \frac{(\Delta\vec{B}(\tau_{i, -, \downarrow}) + \Delta\vec{B}(\tau_{i, +, \downarrow}) + \Delta\vec{B}(\tau_{i, +, \uparrow}) +  \Delta\vec{B}(\tau_{i, -, \uparrow})))\cdot\hat{n}_i}{4\omega_m \vec{B}_0\cdot\hat{n}_i\alpha_i},
\end{align}
which we can use to compute the frequency response of the system to external fields and bias field noise.
Substituting in an external field $\vec{B}_{\text{ext}}(t)$ for $\Delta\vec{B}(t)$ and Fourier transforming yields
\begin{equation}
\Delta\tau_i(\omega) = \frac{\vec{B}_{\text{ext}}(\omega)\cdot \hat{n}_i}{4\omega_m \vec{B}_0\cdot\hat{n}_i\alpha_i}(e^{i\omega\tau_{i,-,\downarrow}} + e^{i\omega\tau_{i,+,\downarrow}} + e^{i\omega\tau_{i,+,\uparrow}} + e^{i\omega\tau_{i,-,\uparrow}}).
\end{equation}
Define the response by $\Delta\tau_i(\omega) = \vec{B}_{\text{ext}}(\omega)\cdot\hat{n}_i H_{\text{ext}}(\omega)$.
$H_{\text{ext}}(\omega)$ clearly has a maximum at $\omega = 0$, and normalizing to this point to unity yields the frequency response of the system
\begin{equation}
H_{\text{ext}}(\omega) = \frac{|e^{i\omega\tau_{i,-,\downarrow}} + e^{i\omega\tau_{i,+,\downarrow}} + e^{i\omega\tau_{i,+,\uparrow}} + e^{i\omega\tau_{i,-,\uparrow}}|}{4}.
\label{vcr:eq:Hext}
\end{equation}

To compute the amplitude noise and phase noise suppression, we write our bias field as
\begin{equation}
\vec{B}_B(t) = \text{Re}\{\vec{B}_0(1 + a(t) + i\phi(t))e^{i\omega_m t}\}
\end{equation}
where $a(t)$ is the amplitude noise, and $\phi(t)$ is the phase noise. Then for amplitude noise, we have \cite{rubiola_phase_2009}
\begin{equation}
\Delta\vec{B}(t) = \vec{B}_0a(t)\cos\omega_m t
\end{equation}
and for phase noise we have
\begin{equation}
\Delta\vec{B}(t) = -\vec{B}_0 \phi(t)\sin\omega_m t.
\end{equation}
Define the amplitude and phase resonances by
\begin{align*}
\Delta\tau_i(\omega) &= A(\omega)\hat{B}_0\cdot\hat{n}_iH_{\text{amp}}(\omega)   \\
\Delta\tau_i(\omega) &= \Phi(\omega)\hat{B}_0\cdot\hat{n}_iH_{\text{phase}}(\omega),
\end{align*}
with $\Phi(\omega)$ and $A(\omega)$ the spectra of $a(t)$ and $\phi(t)$, respectively. Using the relations in Eq.~\eqref{vcr:eq:alphabetasincos} and the same normalization as $H_{\text{ext}}(\omega)$, we find
\begin{align}
H_{\text{amp}}(\omega) &= \beta_i \frac{|e^{i\omega\tau_{i,-,\downarrow}} - e^{i\omega\tau_{i,+,\downarrow}} - e^{i\omega\tau_{i,+,\uparrow}} + e^{i\omega\tau_{i,-,\uparrow}}|}{4} \label{vcr:eq:Hampnoise} \\
H_{\text{phase}}(\omega) &= \alpha_i \frac{|e^{i\omega\tau_{i,-,\downarrow}} + e^{i\omega\tau_{i,+,\downarrow}} - e^{i\omega\tau_{i,+,\uparrow}} - e^{i\omega\tau_{i,-,\uparrow}}|}{4} \label{vcr:eq:Hphasenoise}.
\end{align}
Note that $\Delta\tau_i(\omega) \propto \hat{B}_0\cdot \hat{n}_i$, which indicates that the lower the projection of the bias field onto the NV axis, the more robust it is against phase and amplitude noise.

\begin{figure}
    \centering
    \includegraphics[width=3.2in]{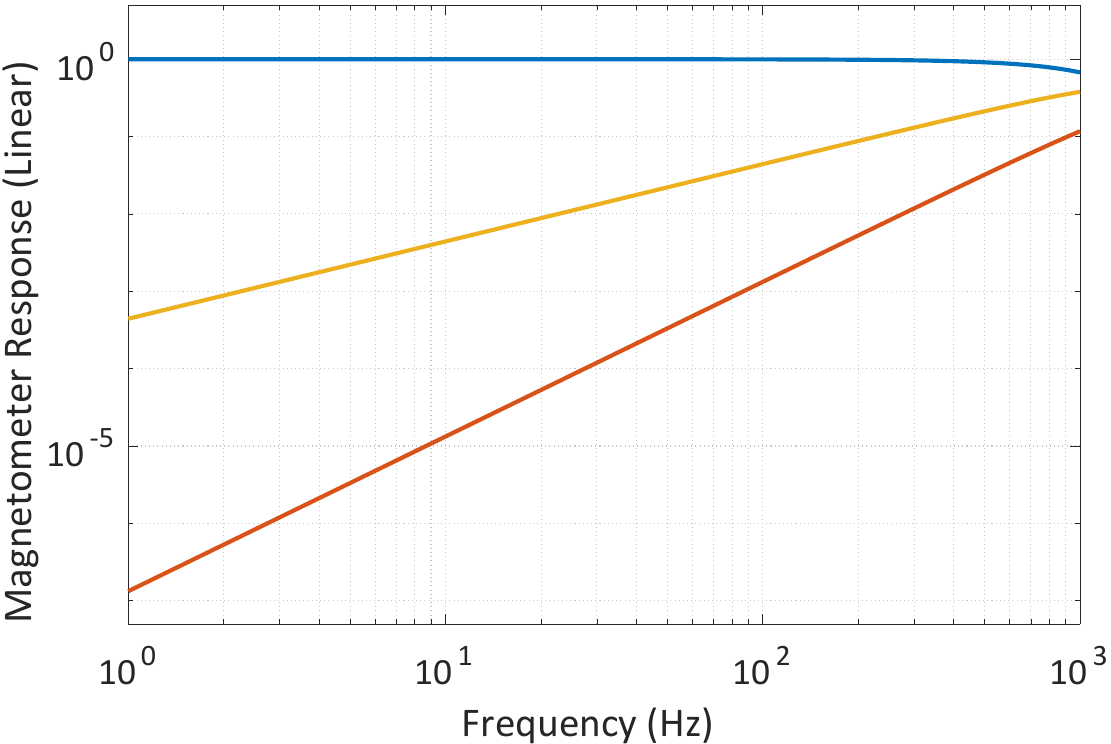}
    \caption{\textbf{Magnetometer Frequency Response.}
    The magnetometer response to external magnetic fields $H_{\text{ext}}(\omega)$ is plotted (\textcolor{matlab1}{\rule[0.75mm]{3mm}{.25mm}}), alongside its response to amplitude noise $H_{\text{amp}}(\omega)$  (\textcolor{matlab2}{\rule[0.75mm]{3mm}{.25mm}}) and phase noise $H_{\text{phase}}(\omega)$  (\textcolor{matlab3}{\rule[0.75mm]{3mm}{.25mm}}). Note that $H_{\text{ext}}(\omega)$ is nearly flat up to 400~Hz, and both $H_{\text{amp}}(\omega)$ and $H_{\text{phase}}(\omega)$ are suppressed by several orders of magnitudes at low frequencies.}
    \label{vcr:fig:magresponse}
\end{figure}

Fig.~\ref{vcr:fig:magresponse} plots each of $H(\omega)$ against frequency.
The plots are generated using measured peak location times and an estimate of $\hat{B}_0\cdot\hat{n}_2$.
$H_{\text{ext}}(\omega)$ remains nearly flat until about 400~Hz, and drops by about 30\% at 1~kHz, indicating a very flat response over the sampling band.
Both amplitude and phase noise are suppressed by several orders of magnitude at low frequencies.

\begin{figure}
    \centering
    \includegraphics[width=3.2in]{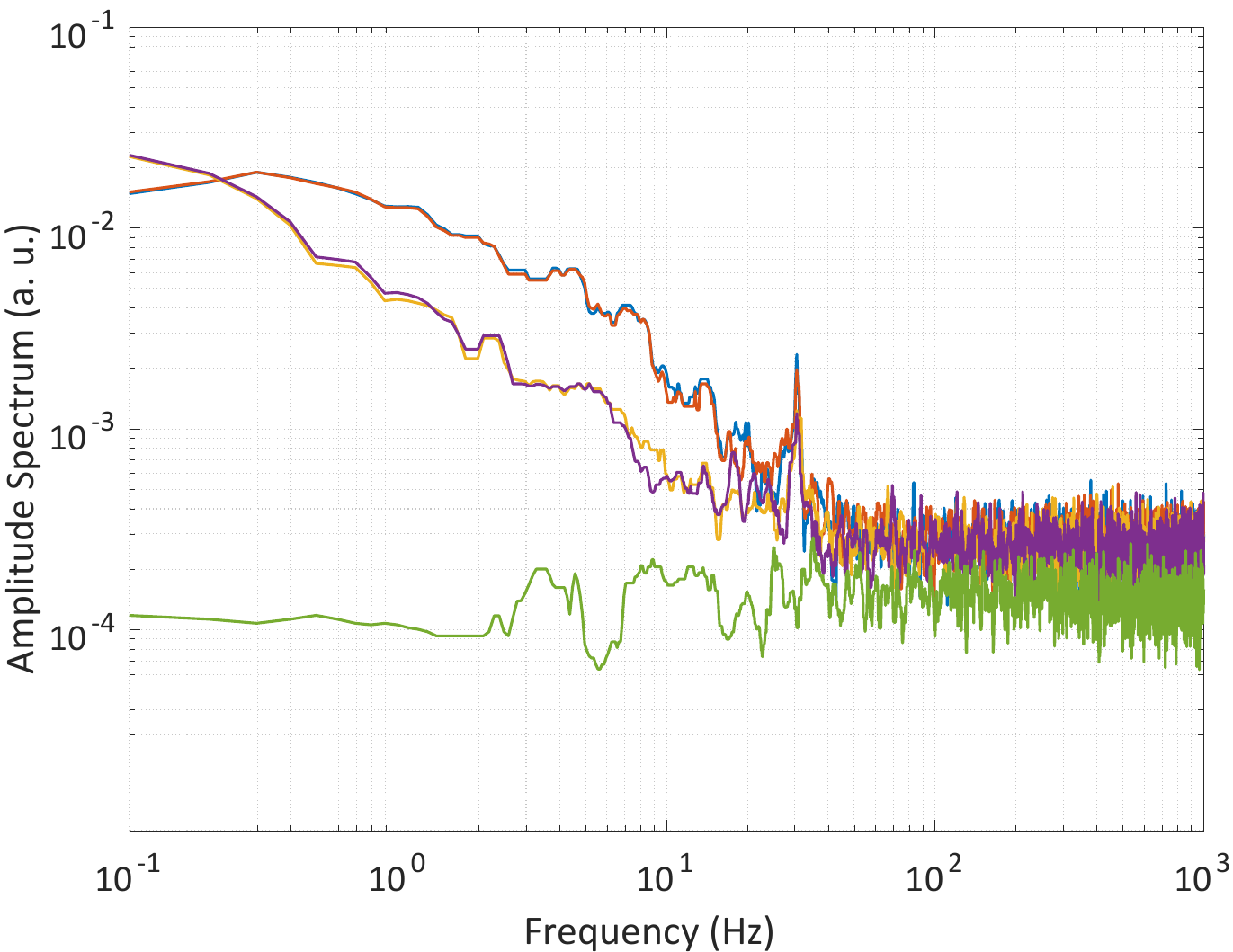}
    \caption{\textbf{Demonstration of bias field noise suppression.}
    The amplitude spectrum of any $\tau_{i,\pm,\uparrow/\downarrow}$ (\textcolor{matlab1}{\rule[0.75mm]{3mm}{.25mm}}, \textcolor{matlab2}{\rule[0.75mm]{3mm}{.25mm}} \textcolor{matlab3}{\rule[0.75mm]{3mm}{.25mm}} \textcolor{matlab4}{\rule[0.75mm]{3mm}{.25mm}}) exhibits signficant low frequency noise. The linear combination $\tau_i$,  (\textcolor{matlab5}{\rule[0.75mm]{3mm}{.25mm}}) demonstrates a significant reduction in noise at low frequencies. MW noise prevents additional suppression, seen by the flat noise floor in $\tau_i$. As seen in Fig.~\ref{vcr:fig:magresponse}, external magnetic fields are unaffected by this cancellation technique.}
    \label{vcr:fig:noisedemo}
\end{figure}

The noise suppression can be verified by looking individually at the amplitude spectra of $\tau_{i, \pm, \uparrow/\downarrow}$ and comparing it against the amplitude spectrum of $\tau_i$, as in Fig.~\ref{vcr:fig:noisedemo}. Note that the spectra for $\tau_{i,\pm,\uparrow/\downarrow}$ is dominated by bias field noise up to $\approx 100$~Hz, above which it begins to be limited by MW noise (see Supplemental Material Sec.~\ref{vcr:supp:mwnoise} for a discussion of MW noise). When it is limited by bias field noise, the noise suppression scheme eliminates nearly all noise, down to the noise floor set by the MW noise. The MW noise is independent across all peak times $\tau_{i,\pm,\uparrow/\downarrow}$ and therefore the only reduction in it is that due to averaging the peak times together.

\begin{figure*}[t] 
\hspace{-2mm}
\begin{minipage}[b]{1\textwidth}
\vspace{10mm}
\end{minipage}

\begin{minipage}[b]{0.45\textwidth}
\begin{Overpic}{\put(-110, 0){\includegraphics[width=3.2in]{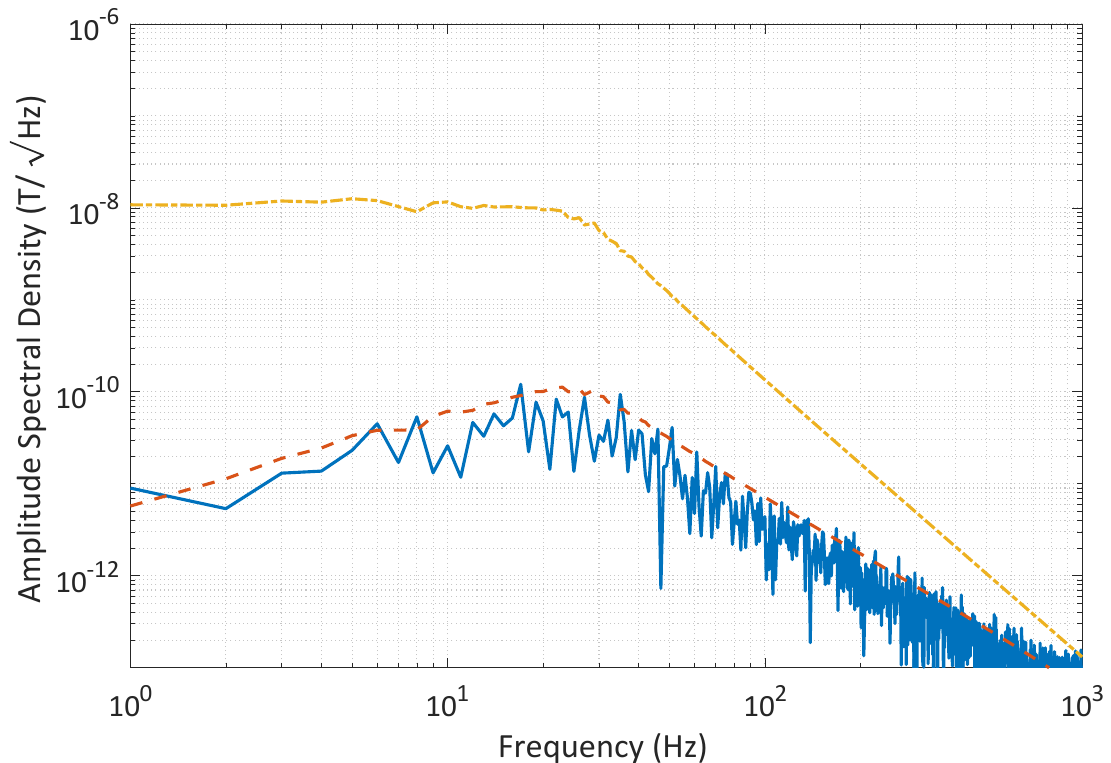}}} \put(-80,104){\textbf{a)}} 
\end{Overpic}
\end{minipage}
\;
\begin{minipage}[b]{0.45\textwidth}
\begin{Overpic}{\put(-110, 0){\includegraphics[width=3.2in]{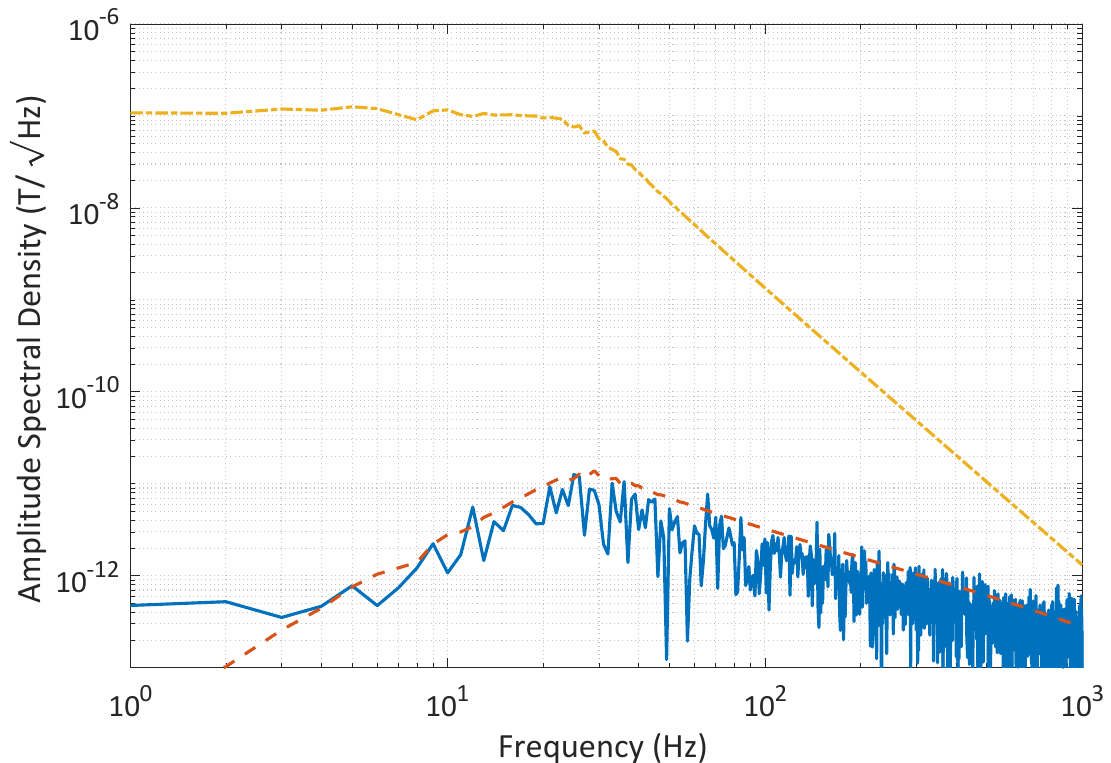}}} \put(-78,105){\textbf{b)}} 
\end{Overpic}
\end{minipage}
\caption{\textbf{Simulations of bias field phase and amplitude noise.}
\textbf{(a) Bias field phase noise simulation.} Phase noise with a known spectrum is applied to the bias field (\textcolor{matlab3}{\plotdotdashline}). When the resulting reflection coefficient data is analyzed, the noise floor (\textcolor{matlab1}{\plotline}) closely matches the noise floor prediction from Eq.~\eqref{vcr:eq:Hphasenoise} (\textcolor{matlab2}{\plotdashline}).
\textbf{(b) Bias field amplitude noise simulation.} Amplitude noise with a known spectrum is applied to the bias field (\textcolor{matlab3}{\plotdotdashline}). When the resulting reflection coefficient data is analyzed, the noise floor (\textcolor{matlab1}{\plotline}) closely matches the noise floor prediction from Eq.~\eqref{vcr:eq:Hampnoise} (\textcolor{matlab2}{\plotdashline}).}
\label{vcr:fig:biasnoise_sim}
\end{figure*}

As a further demonstration of the amplitude and phase noise suppression of the measurement method, a simulation of the system is performed.
A simulated bias field is created, which allows precise control of the amplitude and phase noise on it.
This is then used to simulate the reflection coefficient which is then processed using the same code as the real magnetometry data.
The resulting noise floor of the system closely matches the prediction from Eq.~\eqref{vcr:eq:Hampnoise} and  Eq.~\eqref{vcr:eq:Hphasenoise}, shown in Fig.~\ref{vcr:fig:biasnoise_sim}.

\subsection{Understanding the Reflection Coefficient Spectrum}
\label{vcr:supp:spectrum}
\begin{figure*}[t] 
\includegraphics[width=3.2in]{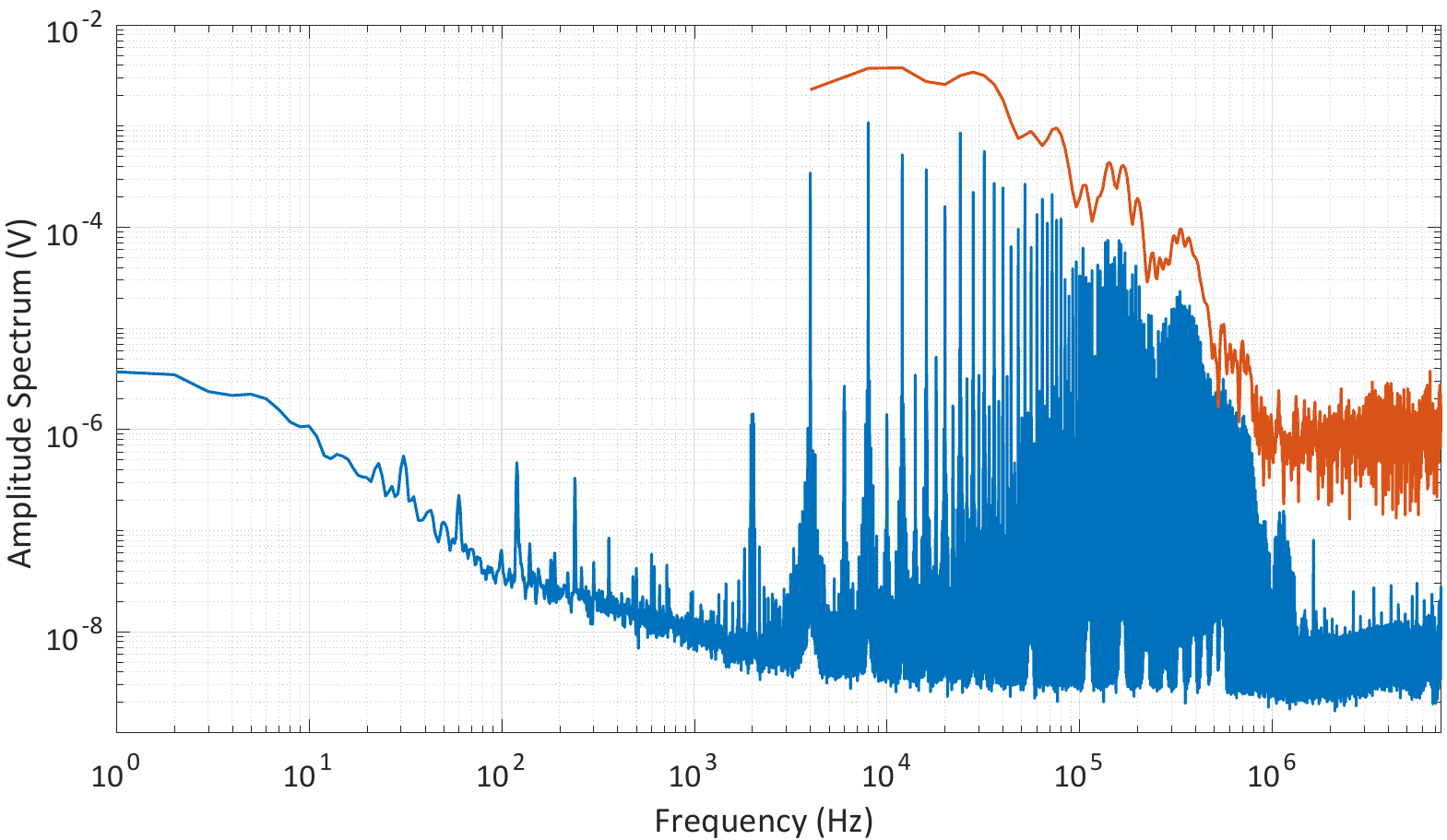} \\
\caption{\textbf{Reflection coefficient spectrum.} The reflection coefficient spectrum exhibits strong harmonics of the bias field out to 785~kHz (\textcolor{matlab1}{\plotline}). The height of the harmonics is determined by the spectrum of the reflection coefficient over a single half period of the bias field (\textcolor{matlab2}{\plotline}). Harmonics of the applied bias field show up at even multiples of $\omega_m = 2\pi\times 2$~kHz, ie. 4~kHz, 8~kHz, 12~kHz, etc. The applied test field appears at the odd harmonics of $\omega_m$, ie. 2~kHz, 6~kHz, 10~kHz, etc.}
\label{vcr:fig:harmonics}
\end{figure*}

The spectrum of the reflection coefficient is shown in Fig.~\ref{vcr:fig:harmonics}.
It consists of a series harmonics of the 2~kHz bias field, out to about 785~kHz with strength of each harmonic determined by the spin linewidth.
This indicates that the process of sweeping the bias field is an extremely non-linear process, which complicates both an understanding of the reflected power spectrum and the noise floor.

The bandwidth in Fig.~\ref{vcr:fig:harmonics} can be understood by modeling the received signal as a half period of the collected reflection data repeated indefinitely -- that is, taking the first half of Fig.~\ref{vcr:fig:timeseries} and repeating it.
Note that we take a half period because, in the absence of an external field, the first and second half periods are identical.
Letting $r(t)$ be the first half of the signal in Fig.~\ref{vcr:fig:timeseries}, $0\leq t \leq 250$~\micro s, then the full received signal $s(t)$ can be modeled as
\begin{equation}
s(t) = r(t)\star \sum_n \delta(t - n T/2)
\end{equation}
where $T = 500$~\micro s is the bias field period. The Fourier transform is then given by \cite{bracewell_fourier_1985}
\begin{equation}
S(\omega) = 2\omega_m R(\omega)\sum_n \delta\left(\omega - 2n\omega_m\right),
\end{equation}
where $R(\omega)$ is the Fourier transform of $r(t)$.
This indicates that we expect the spectrum of the reflection coefficient to be given by an impulse train every 4~kHz, with an envelope shaped by the spectrum of a single half period.
The spectrum of a half period is determined by the spin linewidth, with a narrower spin linewidth increasing the bandwidth over which the harmonics appear.

An additional signal is seen at 2~kHz in Fig.~\ref{vcr:fig:harmonics}, which corresponds to an externally applied magnetic field.
To first order, we can model an applied magnetic field as shifting $r(t)$ to the right during odd periods of 4~kHz and to the left during even period of 4~kHz, as seen in Fig.~\ref{vcr:fig:biasnoise}a -- this ignores all vector effects, but is still useful to explain the spectrum behavior.
Letting $B_n$ be the applied magnetic field during half period $n$ and $a$ the proportionality between time shifts and magnetic field, the signal becomes
\begin{equation}
s(t) = r(t)\star \left(\sum_n \delta(t - nT + a B_{2n}) + \sum_n \delta(t - nT - T/2 - a B_{2n + 1})\right).
\end{equation}
The Fourier transform is
\begin{equation}
S(\omega) = R(\omega)\left(\sum_n e^{i\omega (nT - a B_{2n})} + \sum_n e^{i\omega (nT + T/2 + a B_{2n + 1})}\right).
\end{equation}
We assume $\omega$ is small so  we can approximate $e^{i\omega a B_n}\approx 1 + i\omega a B_n$, and assume that $B_n$ changes slowly so that $B_{2n} = B_{2n + 1}$.
This allows the simplification
\begin{equation}
S(\omega) = 2\omega_m R(\omega)\left(i\omega a \left(e^{i \pi\omega/\omega_m}  - 1\right) B\left(\frac{\omega}{2\omega_m}\right) + \sum_n\delta\left(\omega - 2n\omega_m\right)\right),
\end{equation}
where
\begin{equation}
    B\left(\frac{\omega}{2 \omega_m}\right) = \sum_n e^{i 2\pi \frac{\omega}{2\omega_m}}B_n
\end{equation}
is the discrete time Fourier transform of $B_n$.
Taking the magnitude of the spectrum gives
\begin{equation}
|S(\omega)| = 2\omega_m|R(\omega)|\left(2\omega a \sqrt{2 - 2\cos(\pi\omega/\omega_m)} \left|B\left(\frac{\omega}{2\omega_m}\right)\right| + \sum_n\delta\left(\omega - 2n\omega_m\right)\right).
\end{equation}
Note that $B(\omega/(2\omega_m))$ is periodic in $\omega$ with period $2\omega_m$, but $1 - \cos(\pi\omega/\omega_m) = 0$ for $\omega = 2n\omega_m$, meaning that we will see the signal from the external magnetic field only from odd multiples of $\omega_m$, ie. 2~kHz, 6~kHz, 10~kHz, etc. as seen in Fig.~\ref{vcr:fig:harmonics}.

The spectrum in Fig.~\ref{vcr:fig:harmonics} is thus understood to be a series of impulses at multiples of $\omega_m$, with overall strength determined by the spectrum of a single half-period of the reflected signal.
On even harmonics of $\omega_m$ (ie. 4~kHz, 8~kHz, 12~kHz, etc.), we have impluses due to the periodic nature of the reflected signal.
On odd harmonics of $\omega_m$ (ie. 2~kHz, 6~kHz, 10~kHz, etc.), we see the signal from an external magnetic field.

Importantly, this means that even narrowband external magnetic fields are now encoded across a broad spectrum of $\approx$~785~kHz.
To demodulate the signal, this entire bandwidth must be used.
As a consequence, the sensor is now susceptible to MW noise across this entire spectrum -- it is not enough to lower MW noise at a single harmonic, but must be reduced over a broad band to see sensitivity improvements.
The consequences of this are discussed in more detail in Supplemental Material Sec.~\ref{vcr:supp:mwnoise}.

From a basic signals processing standpoint, the appearance of harmonics indicates the presence of a non-linearity in the system.
Specifically, we are measuring the reflected signal off the cavity, given by Eq.~\eqref{vcr:eq:ref}, which is seen to be a non-linear function of the spin transition frequency.
The spin transition frequency varies linearly with applied magnetic field, thus making the reflected signal a non-linear function of applied magnetic field.
In previous demonstrations of MW cavity readout, the applied bias field was static and the test magnetic fields were small so that a linear approximation of Eq.~\eqref{vcr:eq:ref} could be used \cite{eisenach_cavity-enhanced_2021, wilcox_thermally_2022}.
However, this work requires addressing each NV orientation separately for vector magnetometry, and the process of sweeping the magnetic field from one spin resonance to another is inherently non-linear.

\subsection{System Noise Limitations}
\label{vcr:supp:mwnoise}
\begin{figure*}[t] 
\hspace{-2mm}
\begin{minipage}[b]{1\textwidth}
\vspace{10mm}
\end{minipage}

\begin{minipage}[b]{0.45\textwidth}
\begin{Overpic}{\put(-110, 0){\includegraphics[width=3.2in]{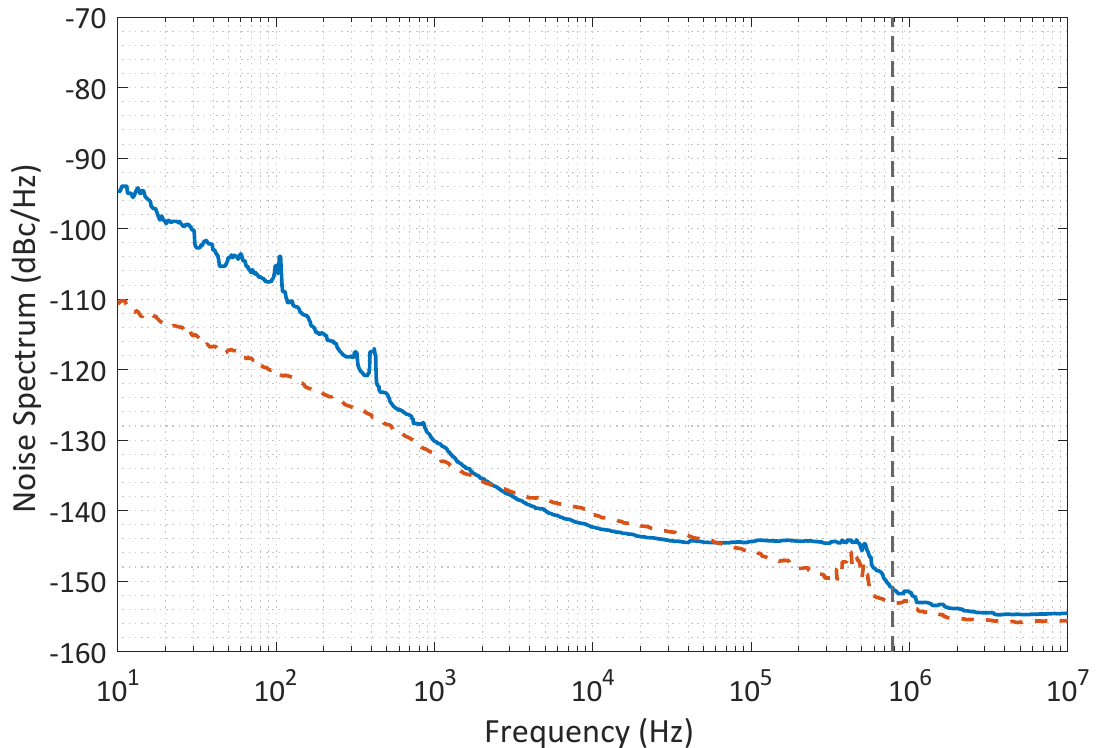}}} \put(-80,104){\textbf{a)}} 
\end{Overpic}
\end{minipage}
\;
\begin{minipage}[b]{0.45\textwidth}
\begin{Overpic}{\put(-110, 0){\includegraphics[width=3.2in]{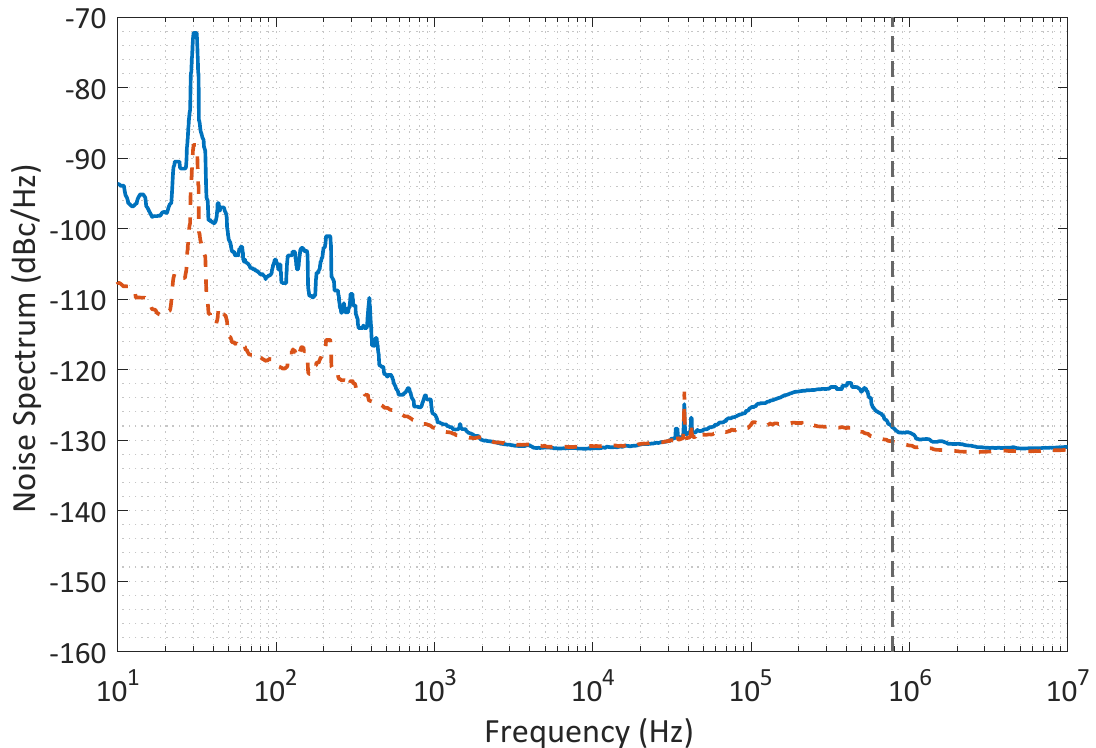}}} \put(-78,105){\textbf{b)}} 
\end{Overpic}
\end{minipage}
\caption{\textbf{MW and cavity noise.}
\textbf{(a) MW source amplitude and phase noise.} The MW source is amplified with an LNA and analyzed with a phase noise analyzer to produce phase noise (\textcolor{matlab1}{\plotline}) and amplitude noise (\textcolor{matlab2}{\plotdashline}) plots. The sensor is susceptible to noise up to 785~kHz marked by the vertical dashed line (\plotdashline).
\textbf{(b) Cavity MW amplitude and phase noise.} The MW source reflection off the cavity is measured using a phase noise analyzer to show the phase noise (\textcolor{matlab1}{\plotline}) and amplitude noise (\textcolor{matlab2}{\plotdashline}). Compared to (a), the reflection adds considerable phase noise. The sensor is susceptible to noise up to 785~kHz marked by the vertical dashed line (\plotdashline).}
\label{vcr:fig:mwnoise}
\end{figure*}
A common technique used in precision measurement applications is a lock-in measurement, whereby a signal of interest is modulated into a low-noise band of the readout electronics to lower the noise floor of the measurement \cite{horowitz_art_2024}.
However, as discussed in Supplemental Material Sec.~\ref{vcr:supp:spectrum}, the process of sweeping the bias field is extremely non-linear, resulting in strong harmonics up to to 785~kHz as seen in Fig.~\ref{vcr:fig:harmonics}.
As a result, the magnetometry information is not contained within the low-noise 2~kHz region, but instead spread across a bandwidth of 785~kHz, making our measurement vulnerable to noise across the entire system.
From the reflected signal such as Fig.~\ref{vcr:fig:timeseries}, which is sampled at 15~Msps, the final reconstructed magnetic field is produced, with a sample rate of 2~ksps, effectively downsampling the 15~Msps signal.
When the signal is downsampled to 2~ksps, the entire noise band up to 785~kHz is aliased onto our sample -- an effect which has been confirmed by simulation.
Thus, while MW noise has long been known to be detrimental to NV magnetometry \cite{berzins_impact_2024}, this system is uniquely susceptible to this kind of noise.

The MW noise of the amplified MW source is given in Fig.~\ref{vcr:fig:mwnoise}a.
When the MWs are reflected off the cavity measured as in Fig.~\ref{vcr:fig:mwnoise}b, two effects are noticeable.
First, at low frequency peaks in the noise spectra appear -- this is attributed to acoustic noise coupling into the MW through the cavity.
Second, both amplitude and phase noise increase significantly above 1~kHz.
This effect is because the MW carrier frequency, being resonant with the cavity, is absorbed, reducing the carrier power, while the higher frequency noise components, being outside the cavity linewidth, are not attenuated.
Thus, even though the noise power is not increased, the carrier power decreases, resulting in a higher relative noise power.

To confirm the limiting noise source, Fig.~\ref{vcr:fig:mwnoise}b is used to generate realistic amplitude noise, which is then used to produce a simulated reflected signal, as would be measured after the mixer in Fig.~\ref{vcr:fig:cavity}b.
This simulated reflection coefficient is then processed using the same code as the experimental data.
The resulting noise floor is 2.8~nT/\rthz{} for NV orientation 1 and 450~pT/\rthz{} for NV orientation 3, in reasonable agreement with the experimental values of 5.2~nT/\rthz{} and 250~pT/\rthz{}, suggesting that MW amplitude noise is the limiting noise source.
Simulating with no added MW noise results in a noise floor of about 1~aT/\rthz{}, demonstrating the noise floors are not a limitation of the processing software.
For comparison, using the noise spectra directly from the signal generator without the cavity, from Fig.~\ref{vcr:fig:mwnoise}a, results in a noise floor of 75~pT/\rthz{} for NV orientation 3.
A thermal-noise-limited source with a flat spectrum of -177~dBc/Hz results in a noise floor of 7.3~pT/\rthz{}, 2.1~pT/\rthz{}, and 1.2~pT/\rthz{} for NV orientations 1, 2, and 3, respectively.

\begin{figure}
    \centering
    \includegraphics[width=3.2in]{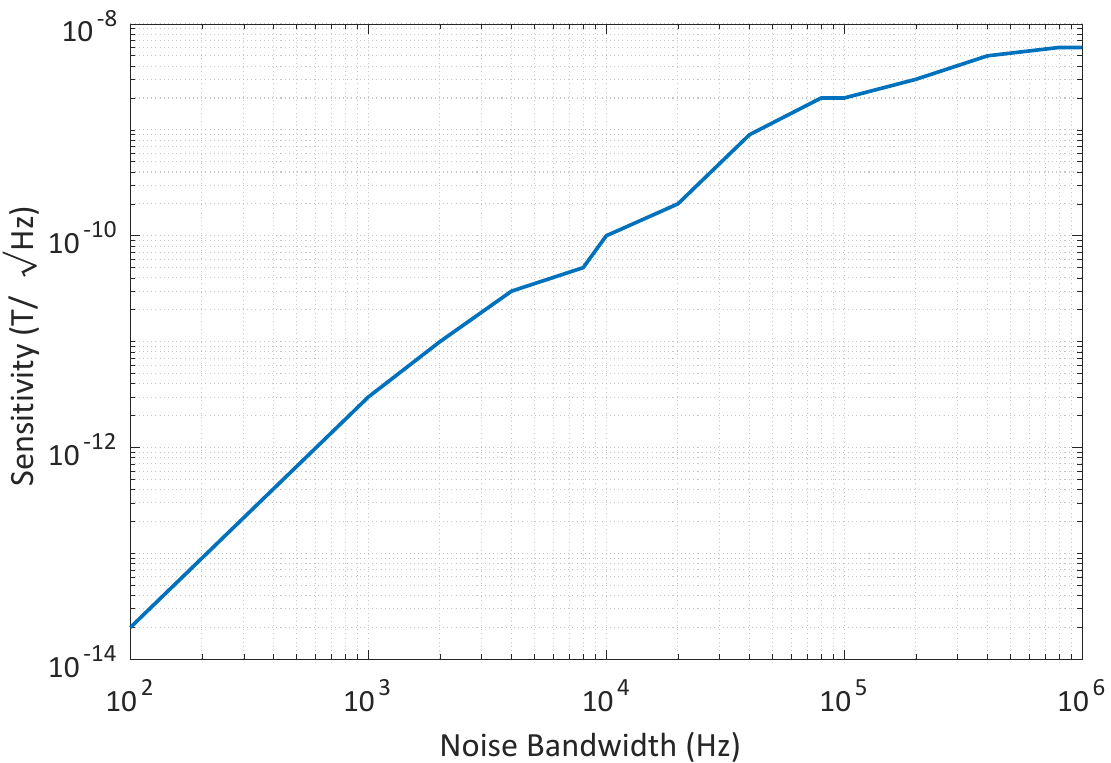}
    \caption{\textbf{Effect of noise bandwidth on sensitivity.}
    Amplitude noise with -110~dBc/Hz power spectral density and varying bandwidth is simulated, with the resulting sensitivity plotted. The sensitivity worsens as the noise bandwidth increases up to about 800~kHz, demonstrating that the sensitivity is affected by noise over the entire 785~kHz band from Fig.~\ref{vcr:fig:harmonics}.}
    \label{vcr:fig:noise_BW}
\end{figure}

The broadband noise dependence of the magnetometer is confirmed through simulation.
Amplitude noise with $-110$~dBc/Hz power spectral density is generated with varying bandwidths, then introduced to the reflection coefficient.
The resulting sensitivity is plotted against the applied noise bandwidth in Fig.~\ref{vcr:fig:noise_BW}.
As seen in the plot, the sensitivity of the magnetometer worsens with increased noise bandwidth.
Near 100~kHz, the magnetometer dependence on noise begins to decrease, consistent with the reduction in harmonics in Fig.~\ref{vcr:fig:harmonics}, with the response becoming nearly flat at 800~kHz.
Thus, even though our magnetometer outputs a 2~ksps magnetic field signal which is Nyquist limited to a 1~kHz bandwidth, its noise floor is set by all noise within the 785~kHz.

This effect is understood by an aliasing argument from signal processing.
When downsampling a signal, frequencies outside the new sampling Nyquist frequency are aliased down, meaning that higher frequency noise still appears in the downsampled spectrum.
Thus, a low-pass filter is used before signals are downsampled to prevent higher frequency noise signals from raising the system noise floor \cite{oppenheim_discrete-time_2010}.
Because this system has magnetometry information encoded throughout the entire 785~kHz bandwidth, all filtering must not affect this bandwidth (see Supplemental Material Sec.~\ref{vcr:supp:spectrum}).
Filtering the high frequency components outside the 785~kHz bandwidth offers little to no improvement to the sensitivity, indicating that sensitivity is determined by noise within the 785~kHz band.

\subsection{Thermal Drift}
\label{vcr:supp:thermal}
\begin{figure}
    \centering
    \includegraphics[width=3.2in]{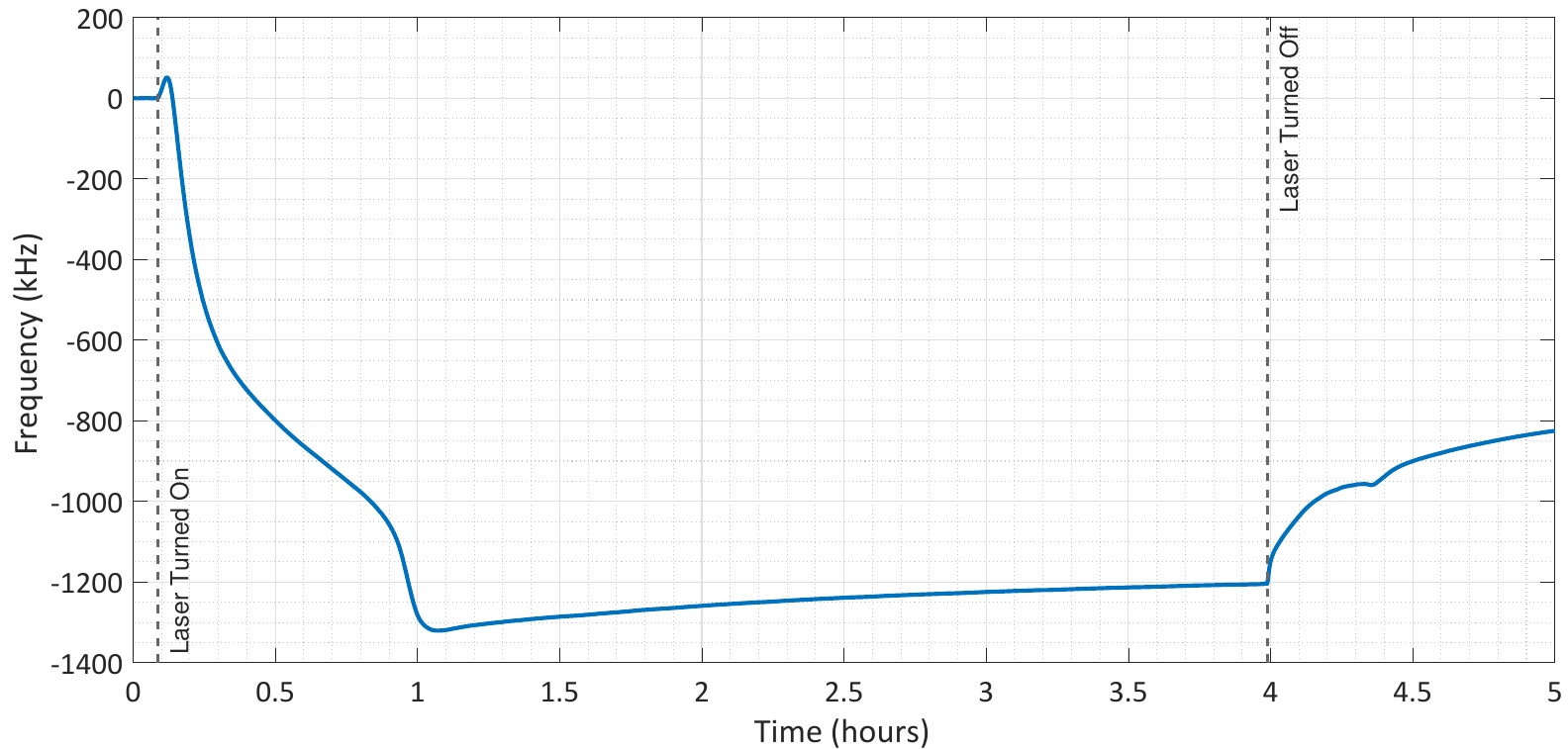}
    \caption{\textbf{Bias field frequency drift.}
    A plot showing the resonant frequency of the composite diamond and dielectric resonator over time in response to turning on and off a 3~W laser. Once the laser is turned on, the cavity undergoes a 1.3~MHz shift in frequency over the first hour before beginning to settle.}
    \label{vcr:fig:frequency_tracking}
\end{figure}

The high-Q dielectric resonator used to facilitate spin-photon coupling has a temperature dependent resonant frequency.
Fig.~\ref{vcr:fig:frequency_tracking} shows how the resonant frequency of the composite diamond and dielectric resonator cavity shifts in response to a 3~W laser.
In the first hour after the laser is turned on, the cavity undergoes a rapid shift of 1300~MHz before beginning to settle to an equilibrium position.
The measured linewidth of the dielectric resonator is 146~kHz, meaning the frequency drifts by nearly 9 linewidths under transient startup behavior.

To mitigate this behavior, a Pound-Drever-Hall (PDH) style lock \cite{pound_electronic_1946} is used to slowly track the cavity frequency with time.
A modulation frequency of 1~MHz is chosen, which is well within the bandwidth of the 15~MSps digitizer used.
As a result, the PDH feedback can be computed entirely from the digitized IQ components of the mixer, allowing for completely digital frequency tracking.
While we turn the modulation turned off when data is taken for magnetometry, the modulation and tracking could be left on during magnetometry with proper filtering and processing.

\end{document}